\documentclass{aa}  

\usepackage{graphicx,subfigure}
\usepackage{txfonts}

\usepackage[]{hyperref}
\usepackage{adjustbox}
\usepackage{lscape}
\usepackage{color}

\makeatletter
\renewcommand*\aa@pageof{, page \thepage{} of \pageref*{LastPage}}
\makeatother

\defcitealias{Carretta2009u}{C09u}
\defcitealias{Carretta2009g}{C09g}
\defcitealias{Schiappacasse-Ulloa2023}{SUL23}
\newcommand{\teff}{T$_{\mathrm{eff}}$~}
\newcommand{\vm}{\textit{v}$_{t}$}
\newcommand{\msun}{M$_{\odot}$}
\newcommand{\vr}{V$_{\rm r}$}

\begin{document}

   \title{Heavy element abundances in Galactic Globular Clusters}

   \subtitle{}

   \author{J. Schiappacasse-Ulloa
          \inst{1,}\inst{2}
          \and
          S. Lucatello\inst{2,}\inst{3}
          \and
          G. Cescutti\inst{4,}\inst{5,}\inst{6}
          \and
          E. Carretta\inst{7}}

   \institute{Dipartimento di Fisica e Astronomia, Universita’ di Padova, Vicolo dell'Osservatorio 3, I-35122, Padova, Italy.\\
              \email{joseluis.schiappacasseulloa@studenti.unipd.it}
         \and
             INAF–Osservatorio Astronomico di Padova, Vicolo dell’Osservatorio 5, 35122 Padova, Italy.
         \and
             Institute for Advanced Studies, Technische Universit{\"a}t M{\"u}nchen, Lichtenbergstra{\ss}e 2 a, 85748 Garching bei M{\"u}nchen.
        \and Dipartimento di Fisica, Sezione di               Astronomia, Università di Trieste, via G.        B. Tiepolo 11, 34143 Trieste, Italy.
        \and INAF–Osservatorio Astronomico di Trieste,        via G.B. Tiepolo 11, 34131, Trieste, Italy.
        \and INFN, Sezione di Trieste, Via A. Valerio 2,      I-34127 Trieste, Italy. 
        \and INAF-Osservatorio di Astrofisica e Scienza dello Spazio di Bologna, Via Gobetti 93/3, 40129 Bologna, Italy.
        }

   \date{Received September 15, 1996; accepted March 16, 1997}

 
  \abstract
   {Globular clusters are considered key objects for understanding the formation and evolution of the Milky Way. In this sense, their characterisation in terms of their chemical and orbital parameters can provide constraints to the chemical evolution models of the Galaxy.} 
   {We use the heavy element abundances of globular clusters to trace their overall behaviour in the Galaxy, aiming to analyse potential relations between the hot H-burning and s-process elements.}
   {We measured the content of Cu I and s- and r-process elements (Y II, Ba II, La II, and Eu II) in a sample of 210 giant stars in 18 Galactic Globular Clusters from high-quality UVES spectra. The clusters span a large metallicity range, and the sample is the largest uniformly analysed for what concerns heavy elements in Globular Clusters. }
   {Cu abundances did not show considerable spread in the sample, nor correlation with Na, meaning that the Na nucleosynthesis process does not affect the Cu abundance. Most GCs closely follow the Cu, Y, Ba, La, and Eu field stars' distribution, revealing a similar chemical evolution. The Y abundances in mid-metallicity regime GCs (-1.10 dex $<$[Fe/H]$<$-1.80 dex) display a mildly significant correlation with the Na abundance, which should be further investigated. Finally, we did not find any significant difference between the n-capture abundances among GCs with Galactic and extragalactic origin.}
 {}


   \maketitle
%

\section{Introduction}
\label{introduction}

Globular clusters (GCs) are as old as the Milky Way (MW) itself, perhaps being an important contributor to the Halo formation \citep{Martell2011}, and possibly also to that of the Bulge \citep{Lee2019}. Studying these objects, from their formation and evolution to their potential dissolution in the field, can be crucial for understanding the Galactic evolution. All the well-studies Galactic GCs show the spectroscopic and photometric evidence of multiple stellar populations (MSP) \citep[e.g.,][]{Smith1987,Kraft1994,Gratton2004,Gratton2012,Bastian2018}, revealing a star-to-star light element variation, which reflects a complex process of self-enrichment and is considered their defining signature. 

These variations are the result of the hot H-burning at the interior of polluter stars, which pollute the intra-cluster medium with material enriched in, e.g., N, Na, and Al, but depleted in C, O, and Mg \citep{Bastian2018}. In this context, a given cluster is composed of a first-generation (FG) of stars formed by the unpolluted (pristine) material and a second-generation (SG) of stars formed by a mixture of variable amounts of the pristine and polluted material \citep[e.g.,][]{Gratton2019}. While many potential sites responsible for cluster pollution have been proposed, no one can reproduce the observations. The most discussed polluter candidates are fast-rotating massive stars \citep[FRMS;][]{Decressin2007}, massive binaries \citep{DeMink2009}, and intermediate-mass ($\sim$4–8 \msun) asymptotic giant branch \citep[AGB;][]{Ventura2001} stars.

To better understand the MSP phenomenon, many studies have been carried out to constrain the nature of the polluters through detailed chemical composition. They have, however, concentrated mostly on elements lighter than Fe. On the other hand, a limited number of studies have extended the analysis to neutron- (n-) capture species. The neutron capture processes are split into two classes: rapid or r-process (neutron capture time-scale shorter than $\beta$-decay) and slow or s-process (in this case, the neutron capture time-scale is longer than $\beta$-decay). Most n-capture elements are produced by both the r- and s-process, but for some of these heavy nuclei, the production is dominated by only one process; for example, the solar system Europium is almost exclusively produced by the r-process \citep{Prantzos2020}. The main s-process takes place mainly in low-mass AGB stars ($\sim$1.2-4.0\msun; with some contribution of AGB stars up to 8\msun) during their thermal pulses \citep{Cseh2018}. Rotating massive stars can also produce s-process elements through the weak s-process, in particular, the light n-capture elements (Sr, Y, and Zr) as recently shown by \citet{Frischknecht2016} and \citet{Limongi2018}. The r-process production was thought to take place mainly in core-collapse Supernovae \citep{Cowan1991}; however, \citet[][]{Arcones2007} found that these candidates cannot efficiently host an r-process able to produce the heaviest nuclei. A possible source was proposed by \citet[e.g.,][]{Nishimura2015} with a class of supernovae, the magnetorotationally driven supernovae (MRD SNe) that can be the source of r-process; another scenario was proposed by \citet{Siegel2019} who found that collapsar can also produce neutron-rich outflows that synthesise heavy r-process nuclei. The remaining channel is a binary system of neutron stars when they merge. Neutron stars mergers are certainly a robust theoretical site \citep{Perego2021} and the only one where the production of r-process was observed \citet{Kasen2017}; however, the delay time that should be taken into account for this source is difficult to reconcile with the observations of n-capture elements at extremely low metallicity \citep[see][]{Cavallo2023}.

As mentioned earlier, n-capture elements have been the subject of limited investigations in GCs so far: studies have shown that they display quite homogeneous abundances in most clusters \citep[e.g.,][]{James2004,dorazi2010_neutron,Cohen2011}. Nevertheless, some metal-poor GCs have shown evidence of considerable spread in their abundances, e.g., NGC~7078 \citep[][]{Sobeck2011}, which shows a large spread of Eu (with a difference of Eu within the sample of about 0.55 dex) with a slight spread in Fe ($\sim$0.1 dex). In this sense, the n-capture element distribution can give us essential information for constraining the chemical enrichment of the MW. For example, the [Ba/Eu] ratio is negative at lower metallicities, indicating a prevalence of r-process products over the s-process ones, which constantly increases at higher metallicities \citet{Gratton2004}. This higher r-process domination suggests a considerable contribution of massive stars, by explosive nucleosynthesis, to the Galactic chemical enrichment at the early stages of its evolution. On the other hand, because AGB star yields of both light (l$_s$) and heavy (h$_s$) s-process elements depend strongly on the mass and metallicity of the star \citep{Busso2001,Cescutti2022}, the [l$_s$/Fe], [h$_s$/Fe], and [h$_s$/l$_s$] (e.g., [Ba/Y]) ratios can trace the s-process enrichment in GCs. Rotating massive stars can also affect these ratios at low metallicity \citep{Cescutti2014}, and their participation should be considered.

Because the stellar systems keep some information from the place they were born \citep{Geisler2007}, their chemical features \citep{Freeman-Bland2002} coupled with astrometric information, age, and orbital properties \citep{Horta2020} of GCs can be used as a tracer not only for their chemical evolution but also for their origin. According to the most accepted scenarios, all galaxies were built through the accretion of smaller stellar systems (e.g., dwarf galaxies). Then GCs in our Galaxy could have been stripped from extra-galactic bodies \citep{Arakelyan2020}. This scenario has been supported by observational evidence \citep[e.g.,][]{Massari2019,Horta2020} extracted from high-quality data from the Gaia mission \citep{GaiaDR3_2022}, which provide parallaxes and proper motions allowing to compute the orbital properties of the systems. Therefore, the complete characterisation of the different stellar systems in the MW is crucial for understanding its formation and past mergers (e.g., Sequoia and Gaia-Sausage-Enceladus). In the literature, attempts have been made to distinguish GCs born in situ and accreted, taking advantage of their different chemical signatures. For example, \citet{Fernandez-Alvar2018} and \citet{RecioBlanco2018} argued that the $\alpha$-element abundances and [Si, Ca/Fe], respectively, can distinguish populations with different origins. On the other hand, \citet{Carretta2022} claimed that iron-peak elements may efficiently identify only the GCs associated with the Sagittarius dwarf galaxy.

In the present article, we characterised a large sample of GCs in terms of Cu and n-capture elements, aiming to study their homogeneity and relation with lighter elements. Moreover, we analysed the chemical signatures of the GCs in our sample and their connection to potential galactic or extra-galactic origin. In sections \S\ref{Sec:Obs_Data}, \S\ref{Sec:Ste_Par}, \S\ref{Sec:AbunDet}, and \S\ref{Sec:Obs_Unc}, we describe the sample, the stellar parameters, abundances, and observational uncertainties determination, respectively. Sections \S\ref{Sec:Cu_abund} and \S\ref{Sec:NCap_abun} show the distribution of Cu and the n-capture elements and their relation to O, Na, and Mg. Finally, in sections \S\ref{Sec:Origin} and \S\ref{Sec:Cl_Mass}, we analysed our results regarding the origin and the cluster mass.

\section{Observational data}
\label{Sec:Obs_Data}

The present sample includes data from \citet[][hereafter C09u]{Carretta2009u} plus NGC~5634 from \citet{Carretta2017}, where p-capture element abundances for a large number of GCs were presented. The data is based on VLT FLAMES/UVES spectrograph observations under programmes 072.D-507, and  073.D-0211. The spectra have a resolution of $\sim$40,000 and a wavelength coverage of 4800-6800\AA. 

The sample includes GCs with a wide star distribution on their horizontal branch (HB), ranging from stubby red HB to blue ones with long tails. The sample includes the less massive to the more massive GCs, covering different ages. On the other hand, the star selection considered members without a close companion brighter (fainter) than -2 (+2) mag. of the target star. Moreover, the authors preferred stars near the red giant branch (RGB) ridge over the ones close to the RGB tip to reduce problems with model atmospheres. We refer to the source for a more detailed description of the cluster and star member selection. A total of 210 stars in 18 clusters are included in the dataset.

\citetalias{Carretta2009u} and \citet{Carretta2017} kindly provided the reduced spectra. The same authors reduced the spectra for their respective samples, shifted them to rest-frame, and co-added them for each star as described in the cited articles. Briefly, they reduced spectra using the ESO UVES-FLAMES pipeline (uves/2.1.1 version). They measured each spectrum's radial velocity (\vr) using the \texttt{IRAF} task called \texttt{rvidlines}. For the correspondent \vr, we refer the reader to the mentioned articles. For the present article, we only performed the continuum normalization using the \texttt{continuum} task from \texttt{IRAF}.

\section{Stellar Parameters}
\label{Sec:Ste_Par}

For homogeneity with the abundances reported by \citetalias{Carretta2009u}, we use the same stellar parameters derived by them. The procedure adopted by the author for the atmospheric parameters determination in the survey sample is exhaustively described in the cited paper. We provide a summary of the method here. The interested reader is referred to \citetalias{Carretta2009u} for more details.

2MASS \citep{Skrutskie2006} photometry was used, \textit{J} and \textit{K} filters, which were transformed into the TCS system as was indicated in \citet{Alonso1999}. Using the relations for \textit{V-K} colours given in the mentioned article, the authors then computed the \teff and the bolometric corrections (B.C.). The final \teff was computed with a relation between the former \teff and the \textit{V} mag (or \textit{K} mag for GCs with high reddening), which was built based on a sub-sample of \textit{well-behaved} stars. It is worth noticing that these stars were defined as well-behaved if they have magnitudes in the \textit{J}, \textit{K}, \textit{B}, and \textit{V} filters and they lay on the RGB. The $\log g$ was obtained using the \teff and B.C. for a stellar mass of 0.85\msun~ and an M$_{bol,\odot}$=4.75. On the other hand, the authors determined the microturbulence velocity (\vm~) by removing the dependency of the Fe I abundances with the strength of the lines measured. They preferred this method instead of the classic functions of \vm(\teff, $\log g$) to reduce the scatter on the obtained abundances. Finally, the metallicities were derived after interpolation of \citet{Kurucz1993} model atmospheres grid with overshooting. The selected model was the one with the proper stellar parameters whose abundance was the same as the ones derived from the Fe I lines.

\section{Abundance Determination}
\label{Sec:AbunDet}

\begin{table}
   \centering
   \caption{Lines used for the abundance determination of heavier elements in the present extended survey. We adopted solar abundances reported by \citet{Asplund_isoratios} for Fe, Cu, Y, Ba, La, and Eu.}
   \label{tab:lines_used}
\begin{tabular}{lcr}
\hline
\hline
Element & n & $\lambda$ (\AA) \\
\hline
Cu I    & 1 & 5105            \\
Y  II   & 4 & 4883; 5087;     \\
        &   & 5200; 5509      \\
Ba II   & 3 & 5853; 6141;     \\
        &   & 6496            \\
La II   & 2 & 6262; 6390      \\
Eu II   & 1 & 6645            \\
\hline
\end{tabular}
\end{table}

For the present article, we extended the analysis done by \citetalias{Carretta2009u} to the heavier elements Cu, Y, Ba, La, and Eu\footnote{When using the notation [X/Fe], abundances of the neutral species are indexed to Fe I, while those for ionised species are indexed to Fe II.}. Although the number of lines used by the abundance determination can vary due to specific features of the spectra (e.g., SNR), in general, the lines considered for abundance determination can be found in Table \ref{tab:lines_used}. The abundance derivation for Cu, Y, Ba, and Eu was done through spectral synthesis using \texttt{MOOG} with its driver \texttt{synth}, which is a 1D LTE\footnote{local thermodynamic equilibrium} line analysis code. The line lists for this method were generated with \texttt{linemake} code\footnote{Github site: \url{https://github.com/vmplacco/linemake}} \citep{Placco2021}, which considers hyperfine splitting for Ba II \citep{Gallagher1967}, Cu I\footnote{\url{http://kurucz.harvard.edu/atoms.html}} \citep{Kurucz1995}, and Eu II \citep{Lawler2001}. We assumed solar isotopic ratios from \citet{Asplund_isoratios} for Cu, Y, Ba, and Eu. Although the solar isotopic ratios for these elements are not necessarily appropriate for Population II stars, we note that this has negligible impact on the results at the spectral resolution under discussion. 

\begin{figure*}
    \centering
     \begin{subfigure}
         \centering
         \includegraphics[width=0.30\textwidth,height=5.7cm]{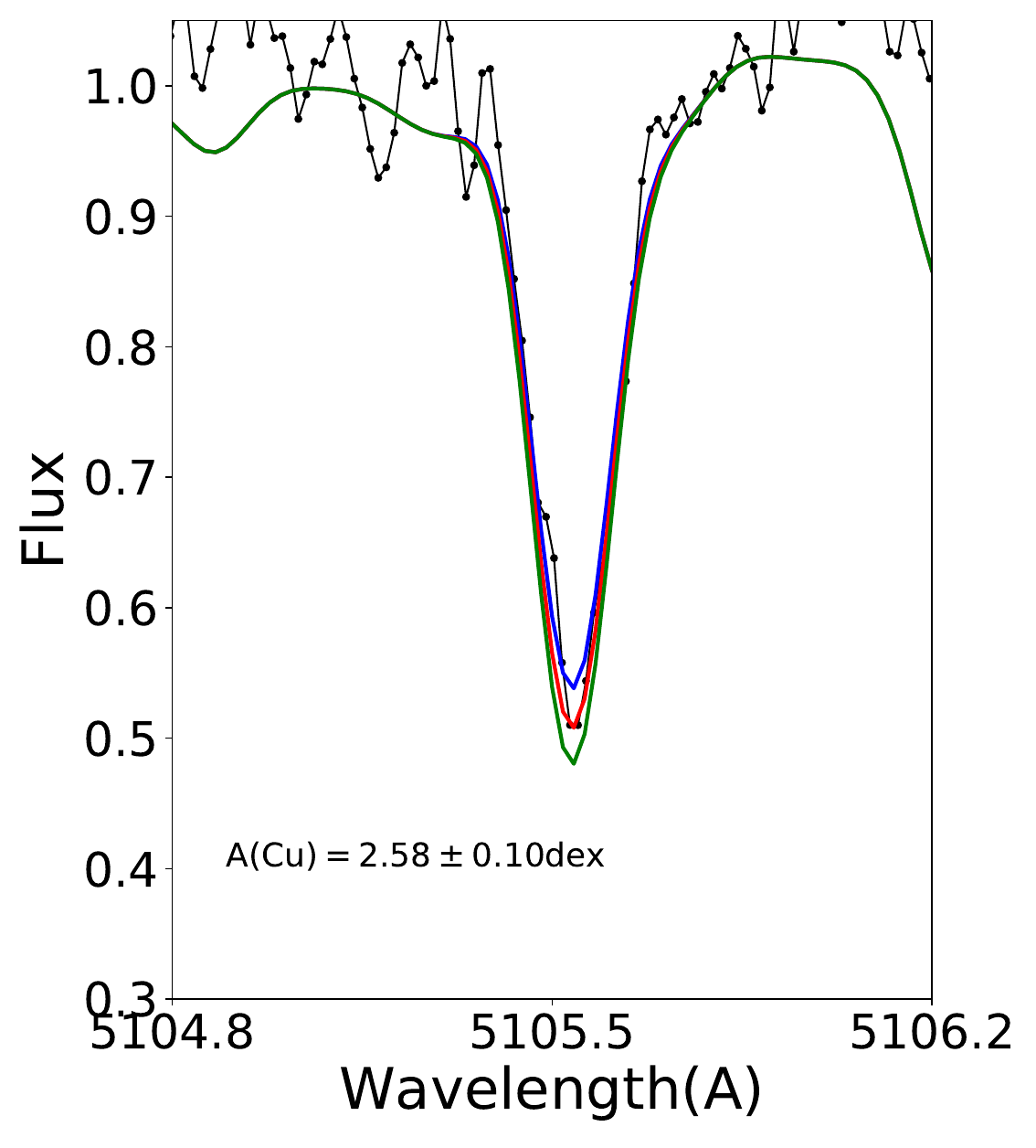}
     \end{subfigure}
     \hfill
     \begin{subfigure}
         \centering
         \includegraphics[width=0.30\textwidth,height=5.7cm]{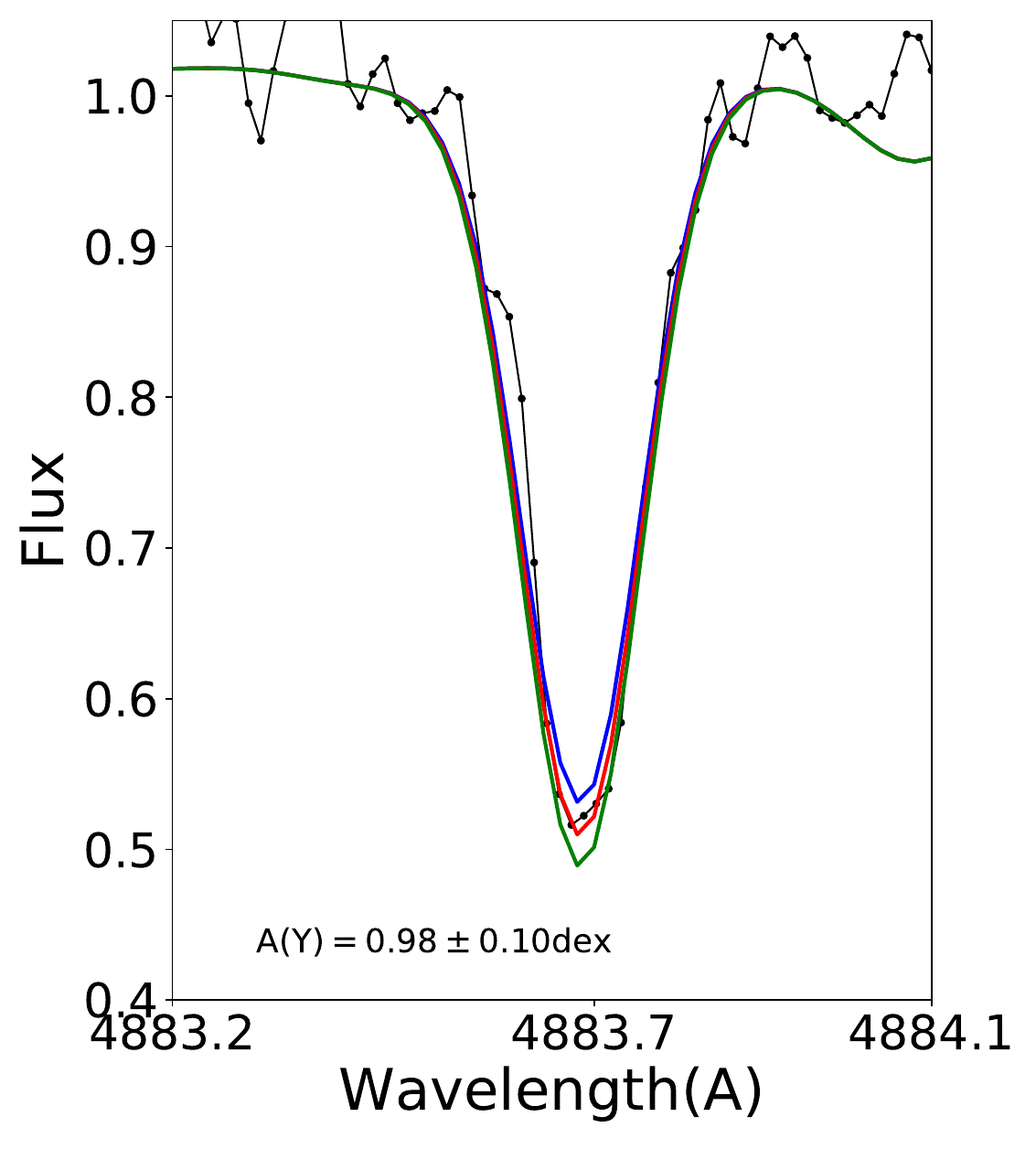}
     \end{subfigure}
   \hfill
     \begin{subfigure}
         \centering
         \includegraphics[width=0.30\textwidth,height=5.7cm]{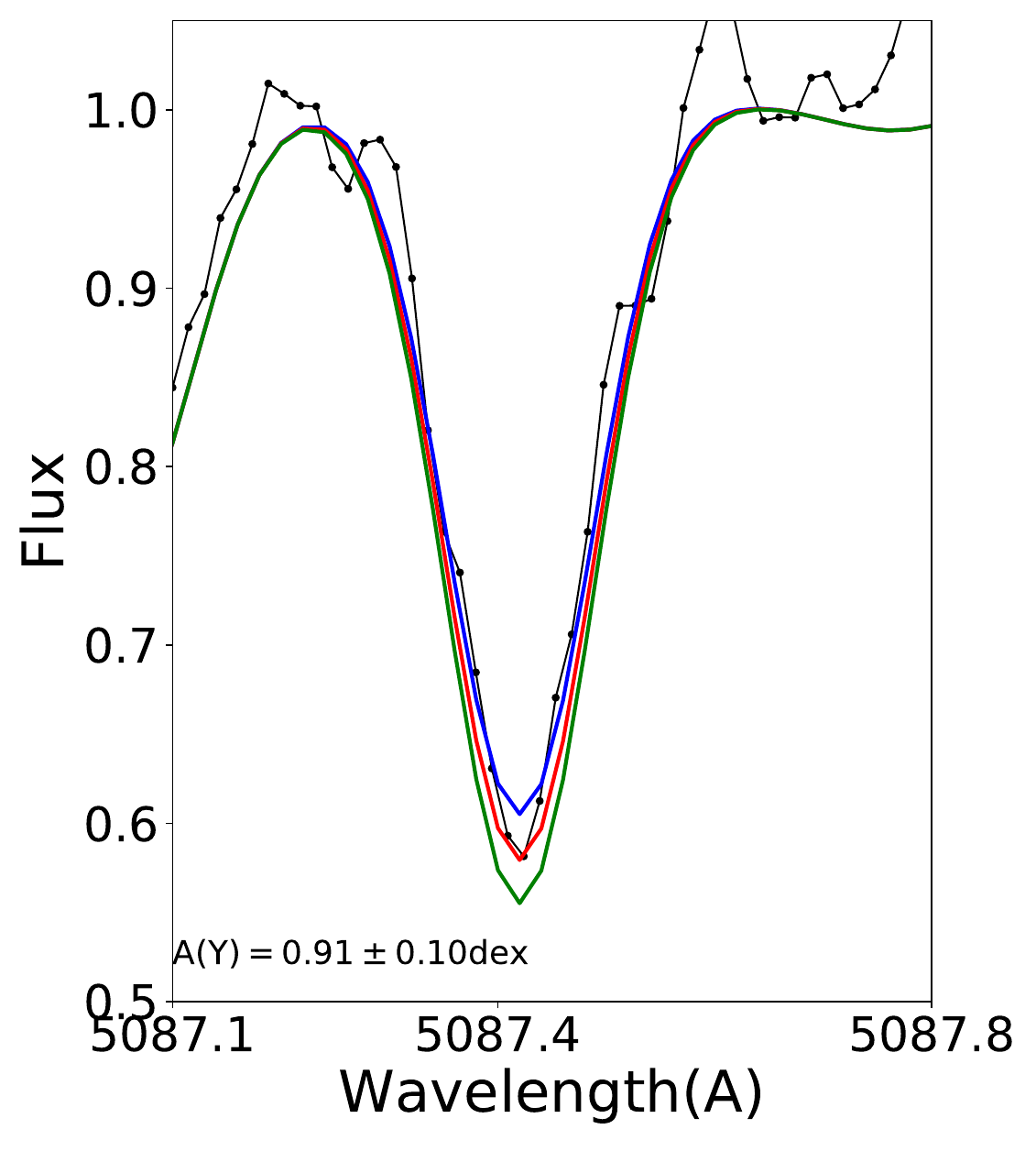}
     \end{subfigure}
   \vfill
     \begin{subfigure}
         \centering
         \includegraphics[width=0.30\textwidth,height=5.7cm]{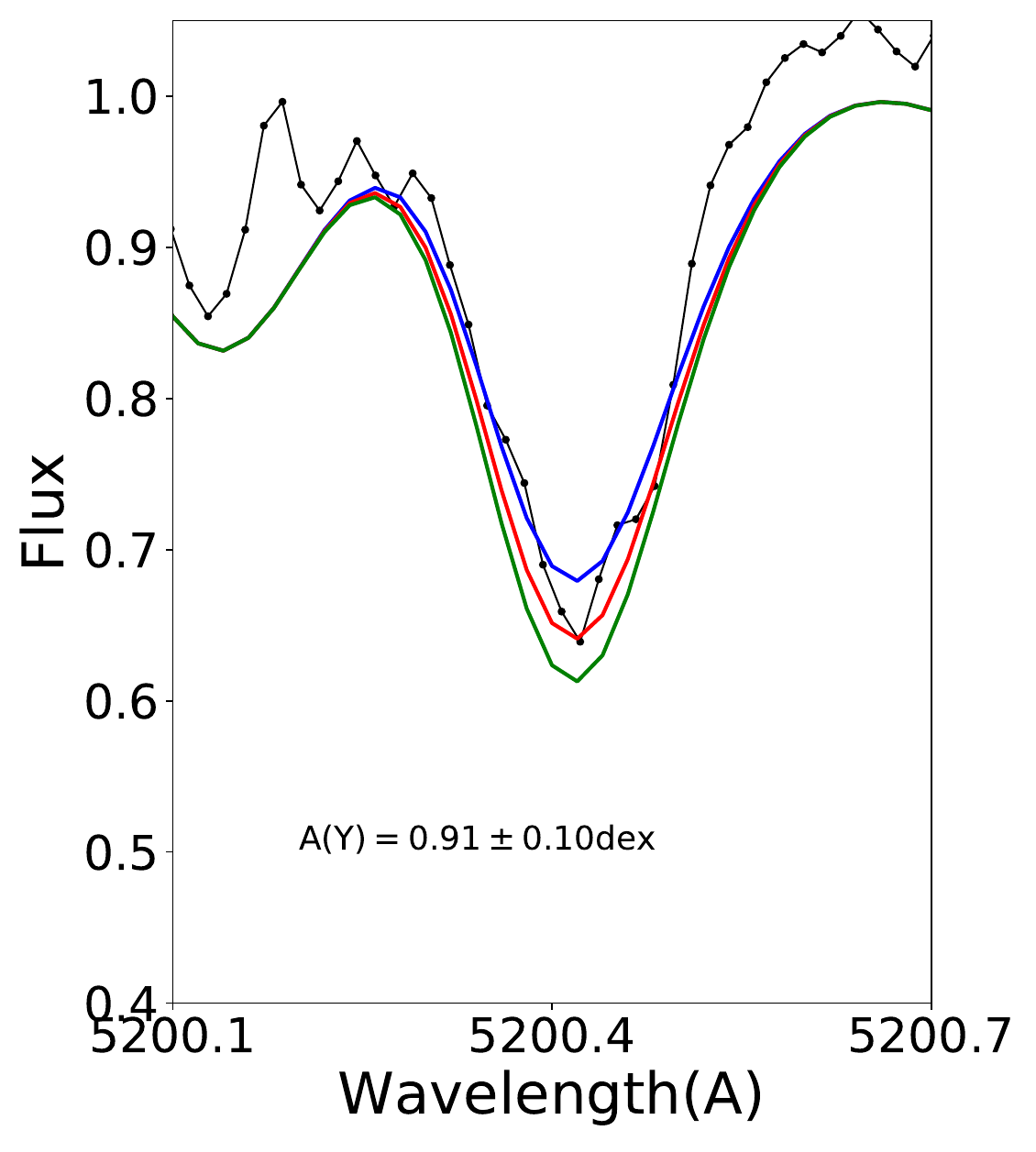}
     \end{subfigure}
   \hfill
     \begin{subfigure}
         \centering
         \includegraphics[width=0.30\textwidth,height=5.7cm]{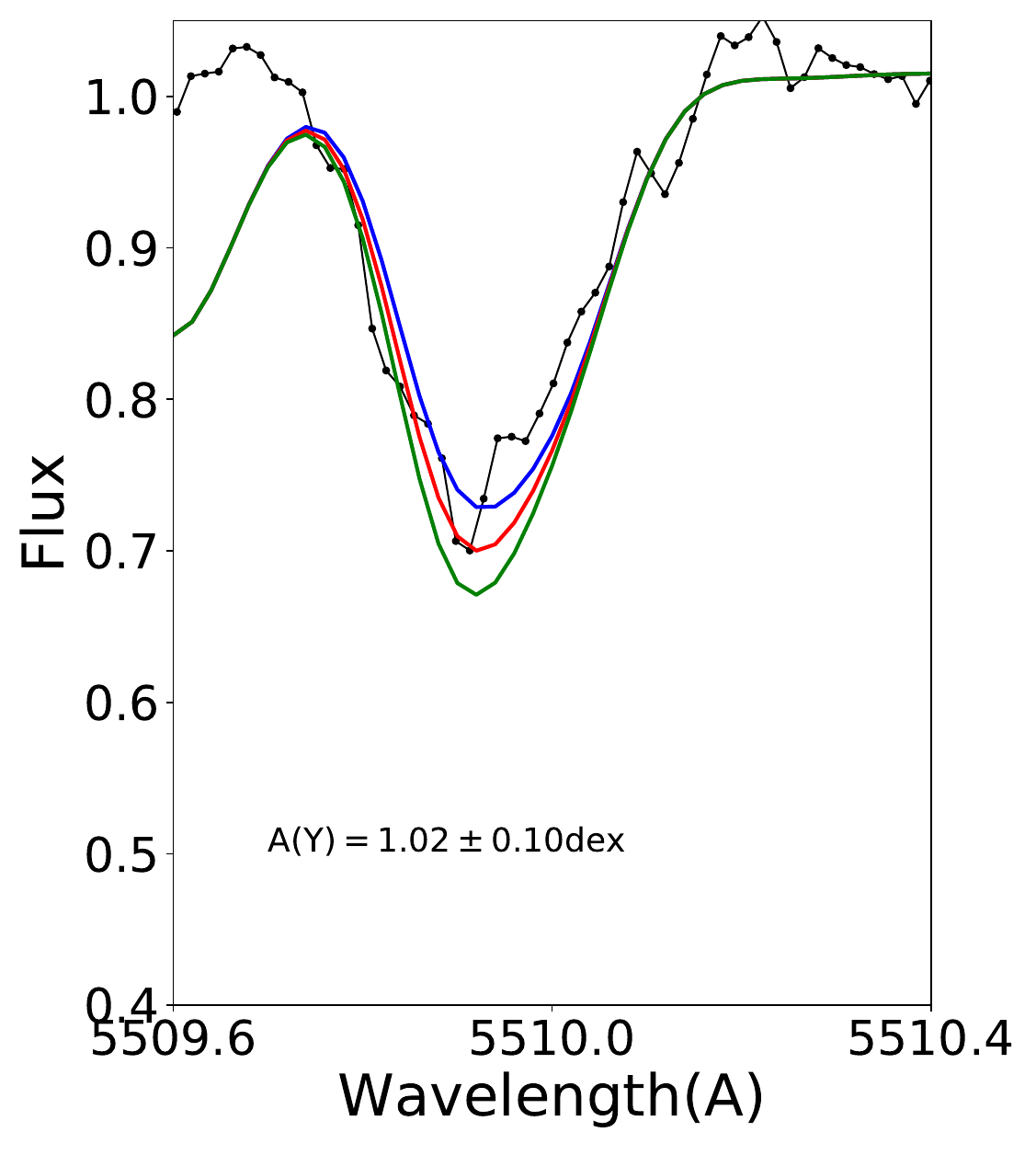}
     \end{subfigure}
   \hfill
     \begin{subfigure}
         \centering
         \includegraphics[width=0.30\textwidth,height=5.7cm]{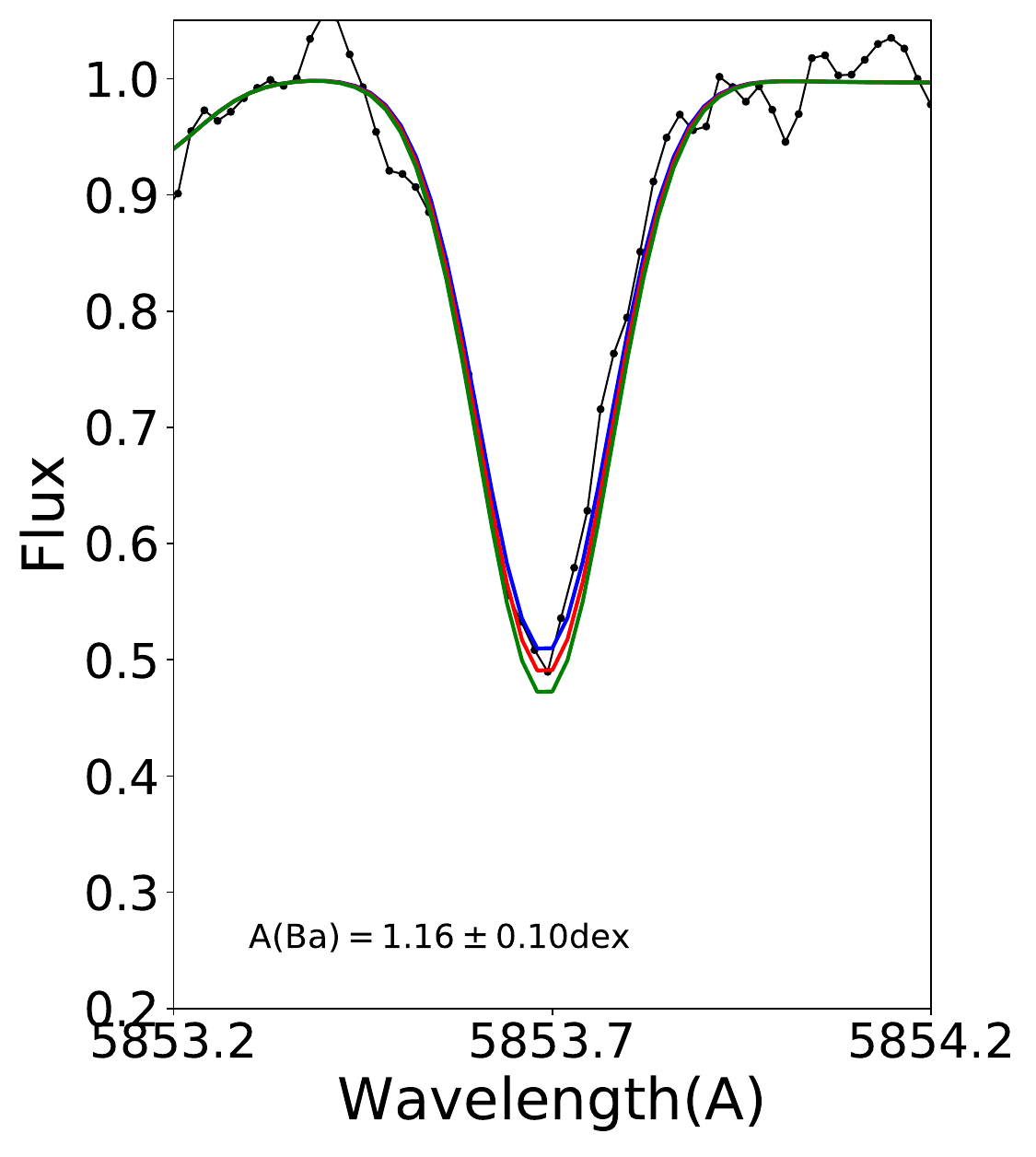}
     \end{subfigure}
   \vfill
     \begin{subfigure}
         \centering
         \includegraphics[width=0.30\textwidth,height=5.7cm]{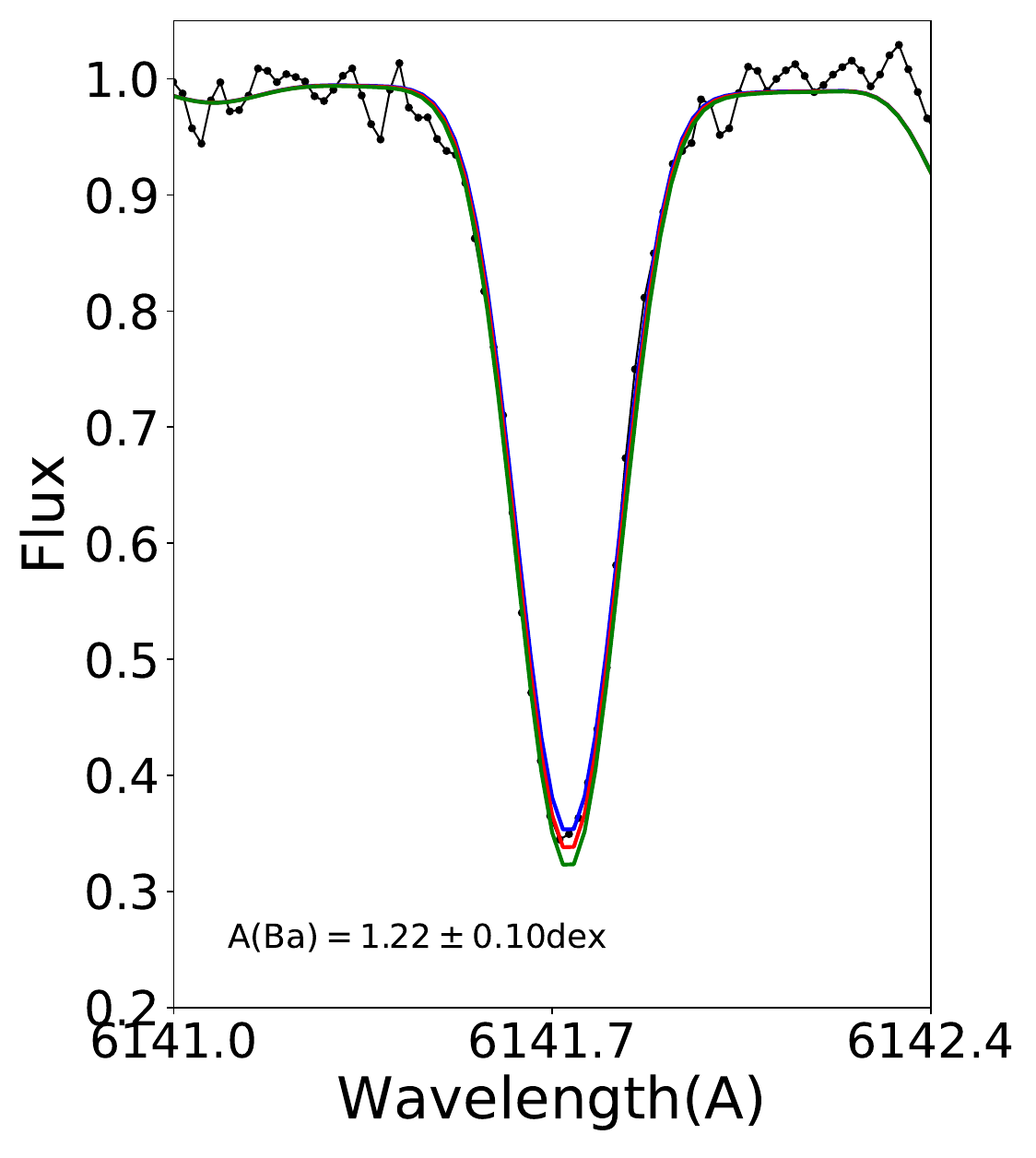}
     \end{subfigure}
   \hfill
     \begin{subfigure}
         \centering
         \includegraphics[width=0.30\textwidth,height=5.7cm]{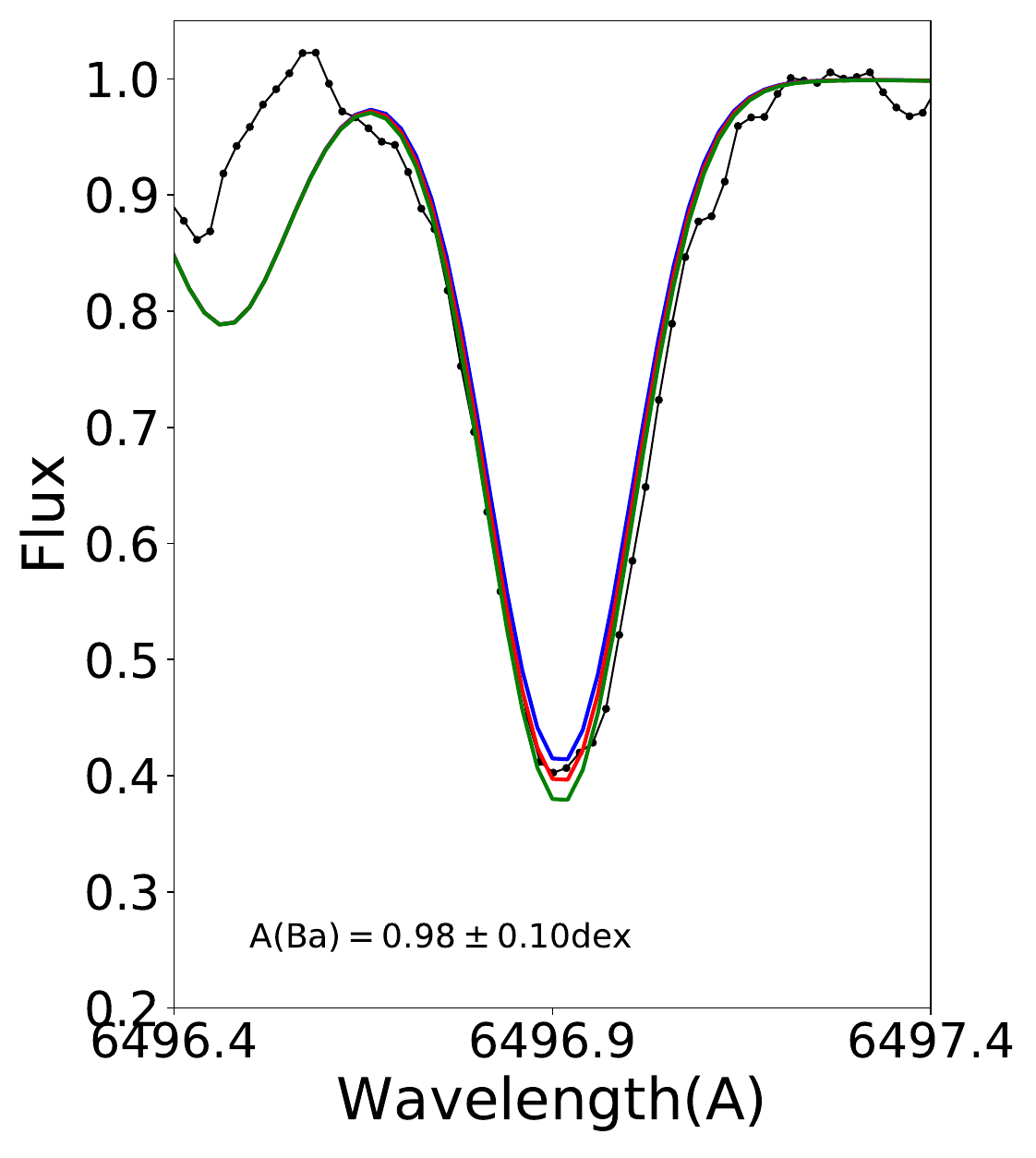}
     \end{subfigure}
   \hfill
     \begin{subfigure}
         \centering
         \includegraphics[width=0.30\textwidth,height=5.7cm]{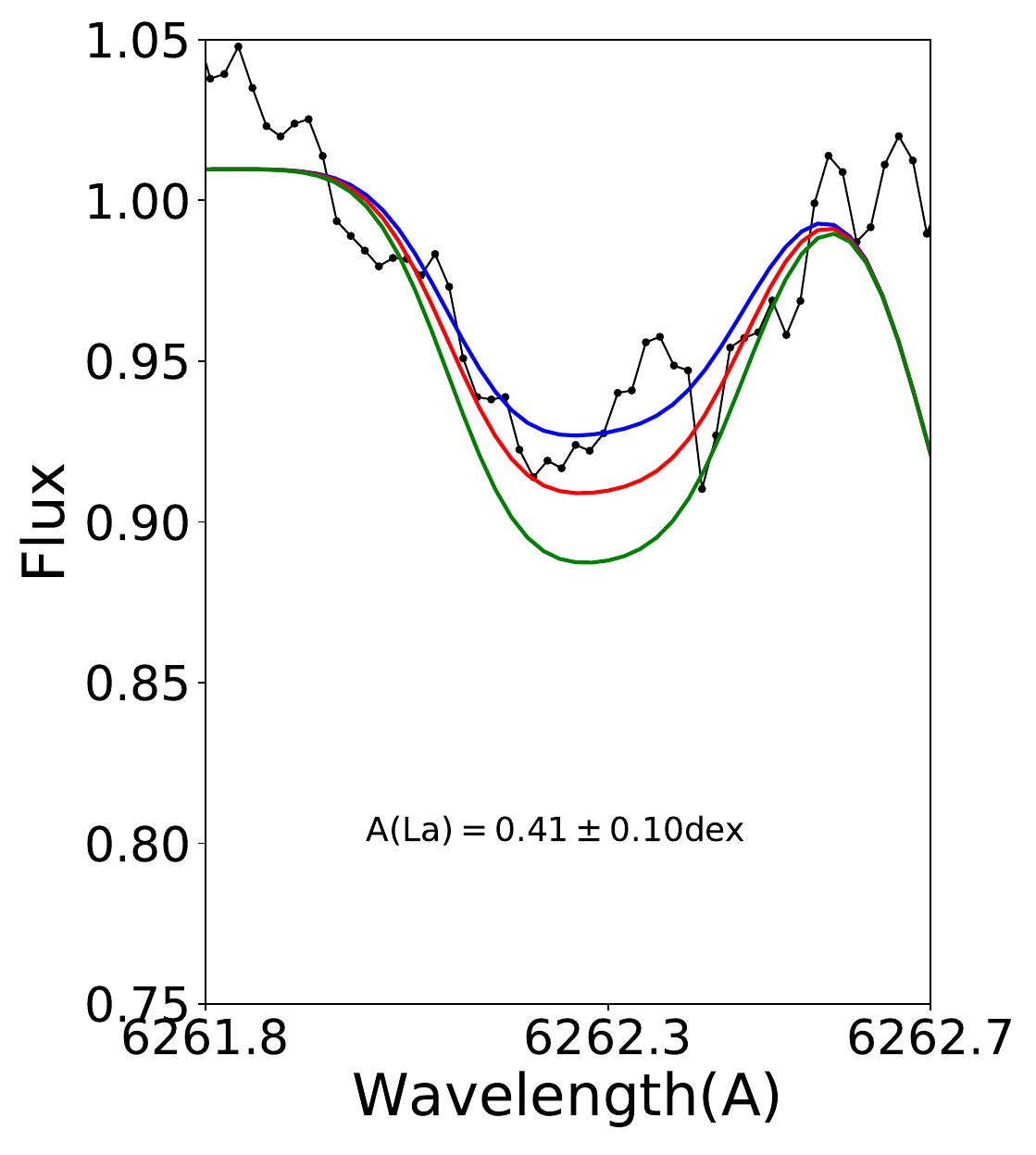}
     \end{subfigure}
   \vfill
     \begin{subfigure}
         \centering
         \includegraphics[width=0.30\textwidth,height=5.7cm]{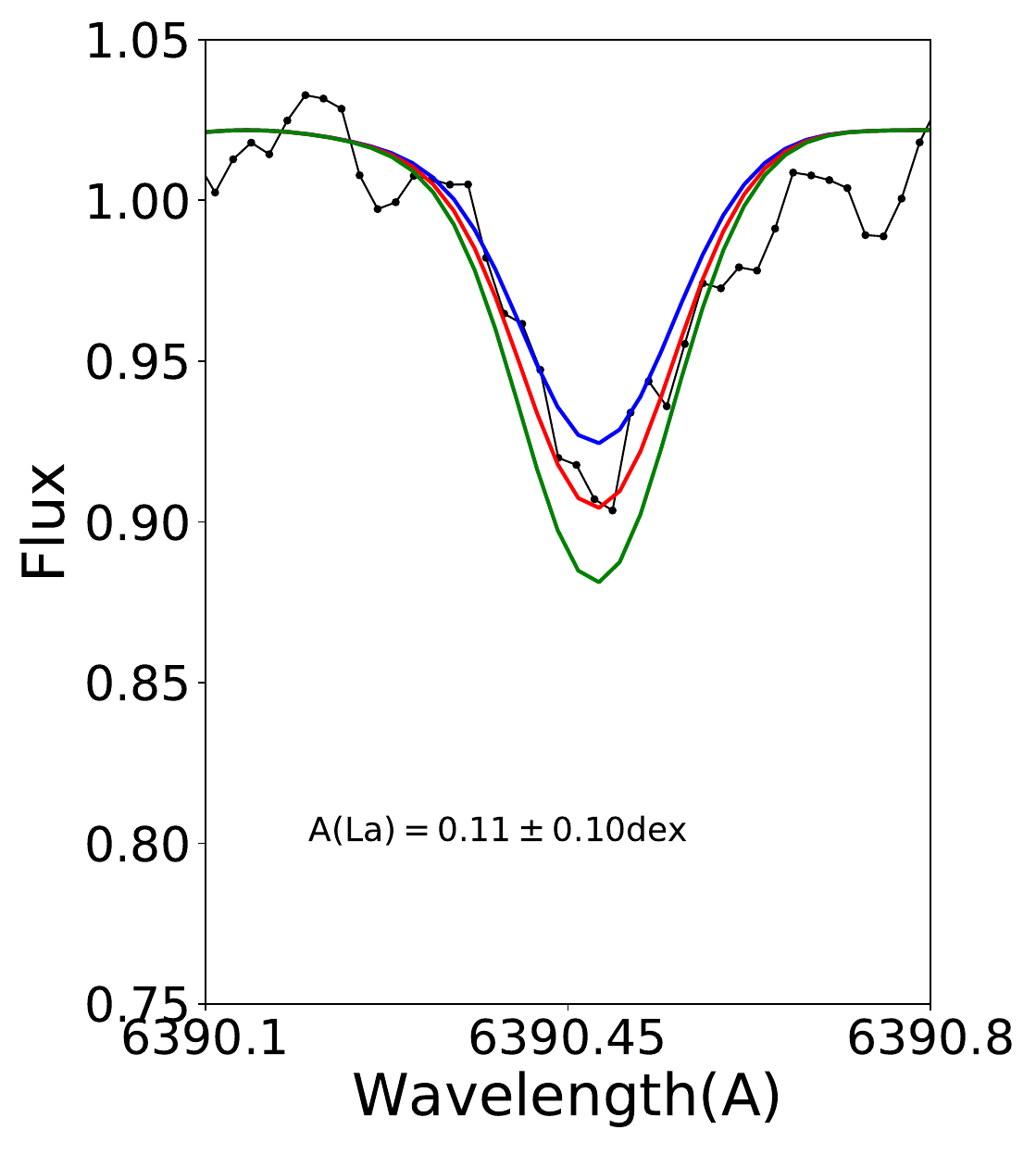}
     \end{subfigure}
   \hfill
     \begin{subfigure}
         \centering
         \includegraphics[width=0.30\textwidth,height=5.7cm]{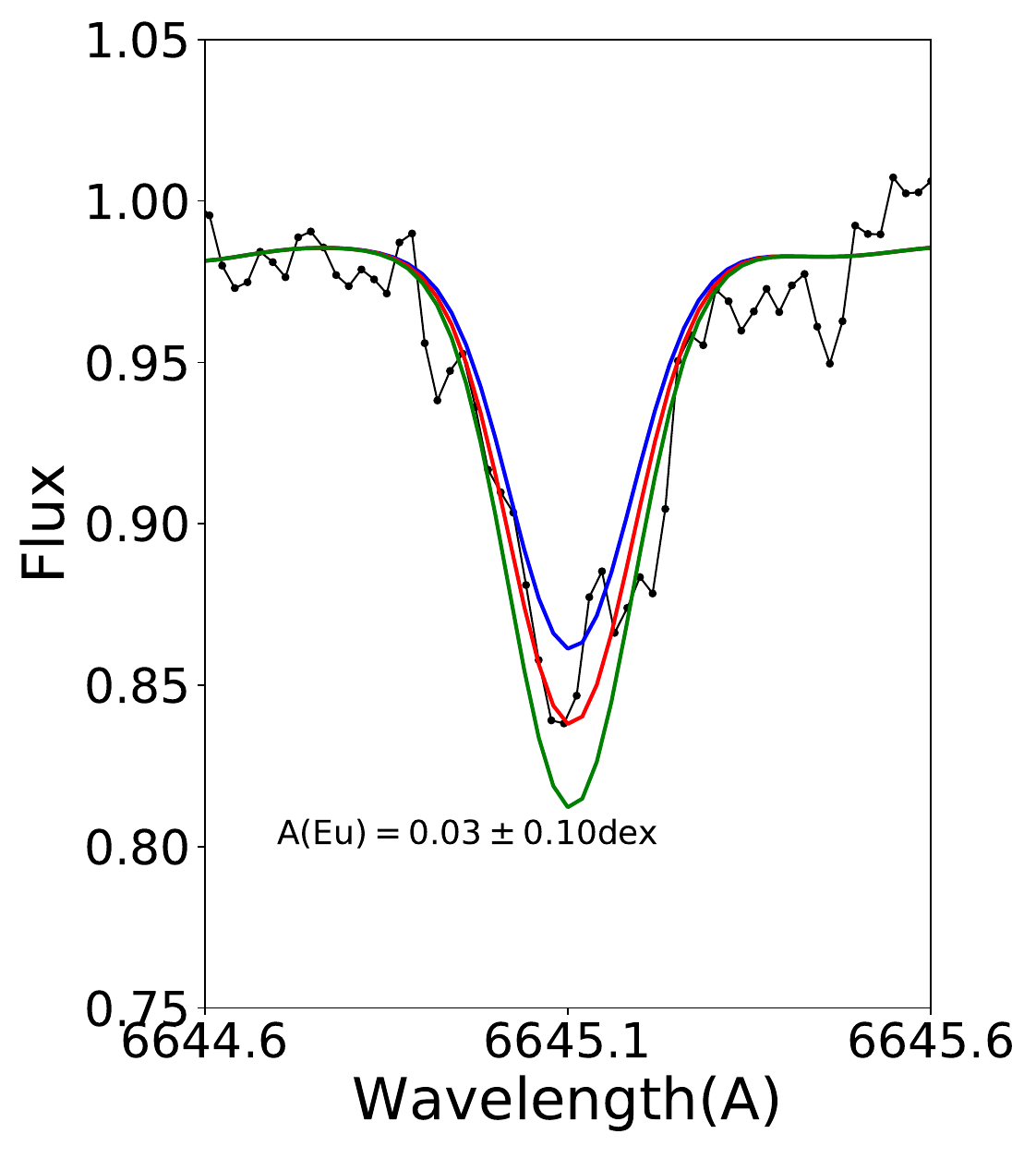}
     \end{subfigure}
   \caption{Example of the synthesised lines for one star of our sample (NGC~2808-49743). The red line represents the best fit. Blue and green fits correspond to the best fit of each element $\pm$0.1 dex, respectively.}
        \label{fig:line_fit}
\end{figure*}

We decided to synthesise La lines automatically. We took that decision because La lines are weak and have a well-behaved shape. Moreover, although La lines are affected by hyperfine splitting, this effect is negligible for these lines, considering the associated errors. We used the 1D-LTE code \texttt{PySME}\footnote{webpage:\url{pysme-astro.readthedocs.io/}} \citep{Wehrhahn2021}, considering the solar isotopic ratios cited before and the hyperfine splitting derived by \citet{Hohle1982}. We synthesised the same lines in Arcturus with both codes to confirm that it does not introduce a systematic offset with our result obtained with \texttt{MOOG}. We found an abundance of A(La)\footnote{A(X) = log(N$_X$/N$_H$ )+12, where N$_X$ is the number density of a given element.} =0.50$\pm$0.06 and A(La)=0.48$\pm$0.07 dex when we used \texttt{PySME} and \texttt{MOOG}, respectively. Using the approaches mentioned before, we analysed the Solar spectrum and obtained A(Cu) = 4.24$\pm$0.06, A(Y)= 2.19$\pm$0.06, A(Ba)= 2.40$\pm$0.06, A(La)= 1.18$\pm$0.07, and A(Eu)= 0.45$\pm$0.05 dex. Although we had good agreements with \citet{Asplund_isoratios}, we decided to use the latter as a reference in our results. As it is standard practice, \citep[see, e.g.][]{Mucciarelli2011}, we considered upper limits the abundances obtained from lines with equivalent widths (EW) smaller than three times the uncertainty associated with the EW determination. That uncertainty follows the relation defined by \citet[][eq. 7]{cayrel1988proc}.  Fig. \ref{fig:line_fit} shows an example of the lines used in the present article. Table \ref{tab:final_resulst} displays the abundances obtained for each star analysed in the present article.

\begin{table*}
\centering
   \caption{Abundances obtained for each element analysed in each GCs. Stars for which we reported actual measurement and upper limits were flagged with 0 and 1, respectively. The star IDs were taken from \citealt{Carretta2009u} and \citet{Carretta2017}. The full Table is only available at the CDS.}
   \label{tab:final_resulst}
\begin{tabular}{cccccccccc}
\hline
\hline
Cluster  & star & flag Cu & [Cu/Fe] & [Y/Fe] & [Ba/Fe] & flag La & [La/Fe] & flag Eu & [Eu/Fe] \\
\hline
NGC~7099 & 954   & 0      & -0.65   & -0.20  & -0.32   & 1       & 0.19    & 0       & 0.79    \\
NGC~7099 & 3399  & 1      & -0.67   & -0.04  & -0.13   & 1       & 1.20    & 0       & 0.98    \\
NGC~7099 & 7414  & 0      & -0.56   & -0.13  & -0.17   & 1       & 0.28    & 1       & 0.66    \\
NGC~7099 & 9817  & 0      & -0.78   & -0.21  &  0.17   & 1       & 0.51    & 0       & 0.55    \\
NGC~7099 & 9956  & 0      & -0.57   & -0.25  & -0.20   & 1       & 0.46    & 1       & 0.72    \\
NGC~7099 & 10200 & 0      & -0.78   & -0.26  & -0.32   & 1       & 0.29    & 1       & 0.67    \\
...      & ...   & ...    &  ...    &  ...   & ...     & ...     & ...     & ...     & ...     \\
\hline
\end{tabular}
\end{table*}

\section{Observational Uncertainties}
\label{Sec:Obs_Unc}

The uncertainty associated with the measurements combines the uncertainties of the best-fit determination and those associated with the uncertainties in the adopted atmospheric parameters. 
As we are adopting parameters from \citetalias{Carretta2009u} and \citet{Carretta2017}, we are also adopting the errors associated with the atmospheric parameters described there.
It is worth noticing that, for the species analysed in \citetalias{Carretta2009u} and \citet{Carretta2017}, the error associated with $\log g$ and [Fe/H] have generally a quite limited influence on the budget of the total error. Heavy elements, whose abundance is generally measured from transitions of ionised species, are more sensitive to $\log g$ variations. 

Because this sample will be compared to different GCs, the observational uncertainties should consider both the individual star-to-star errors (arising from, e.g. stochastic uncertainties in the photometry associated with the line-to-line scatter, etc) and the cluster systematic error associated with overall cluster characteristics (e.g. overall reddening). A full table with the errors computed by \citetalias{Carretta2009u} can be found in their Table 7.

\begin{table}
\footnotesize
\centering
\caption{Element sensitivity to the parameter variations ($\Delta$\teff=+50K, $\Delta$$\log g$=+0.2 dex, $\Delta$[Fe/H]=+0.10, and $\Delta$\vm=+0.10). The full table is available online.} 
\label{tab:sensitivity_survey} 
\begin{tabular}{ccccccc}
\hline
\hline 
{Cluster}                     & {Element}   & {\teff}  & {$\log g$}     & {[Fe/H]}      & {\vm}   & {star}                    \\
\hline
{NGC 7099}                    & { A(Cu)}   & { 0.08}  & {-0.02}     & {-0.01}       & { 0.00}    & {9956}                    \\
{ }                           & { A(Y)}    & { 0.04}  & { 0.06}     & { 0.01}       & {-0.01}    & { }                    \\
{ }                           & { A(Ba)}   & { 0.06}  & { 0.08}     & { 0.00}       & {-0.04}    & { }                    \\
{ }                           & { A(La)}   & { 0.06}  & { 0.09}     & { 0.01}       & { 0.04}    & { }                    \\
{ }                           & { A(Eu)}   & { 0.04}  & { 0.08}     & { 0.03}       & { 0.04}    & { }                    \\
\hline
{NGC 7078}                    & { A(Cu)}   & { 0.09}  & {-0.01}     & { 0.00}       & { 0.02}    & {29401}                    \\
{ }                           & { A(Y)}    & { 0.03}  & { 0.05}     & { 0.01}       & { 0.00}    & { }                    \\
{ }                           & { A(Ba)}   & { 0.04}  & { 0.07}     & { 0.00}       & {-0.06}    & { }                    \\
{ }                           & { A(La)}   & { 0.04}  & { 0.07}     & { 0.02}       & { 0.02}    & { }                    \\
{ }                           & { A(Eu)}   & { 0.00}  & { 0.03}     & { 0.01}       & {-0.01}    & { }                    \\
{...}                         & {...}   & {...}  & {...}     & {...}       & {...}    & {... }                    \\
\hline
\end{tabular}
\end{table}

\subsection{Individual star error}
\label{subs:IndError}

To determine individual star error, we followed the approach described by \citet[][hereafter SUL23]{Schiappacasse-Ulloa2023}. That error is associated with the abundance determinations and combines both the uncertainties of the best-fit determination and the uncertainties in the assumed stellar parameters. For abundances derived via synthesis, the first one comes from the error on the best-fit determination (e.g. the continuum position). The second is derived by evaluating the variation of the abundances to the change in each of the parameters (T$_{\mathrm{eff}}$, $\log g$, \vm, and [Fe/H]), keeping fixed the remaining ones. We selected one star of each cluster as a representative, trying to use the one with median stellar parameters. The variations in stellar parameters assumed to compute the sensitivity matrix (Table \ref{tab:sensitivity_survey}) are: $\Delta$T$_{\mathrm{eff}} =  50$ K, $\Delta$$\log g=  0.2$ dex, $\Delta$\vm~$= 0.1$ km s$^{-1}$, and $\Delta$[Fe/H]$ =  0.1$ dex. The final estimated error ($\sigma$) derived from the variation of stellar parameters is listed in Table \ref{tab:mean_spread}. Moreover, we listed the rms error defined as the standard deviation divided by the squared root of the stars with actual measurements minus one.

\subsection{Cluster systematic error}

The error coupled to \teff comes from the empirical relation between \teff and the (V-K) colour given by \citep{Alonso1999}. Since the V-K are dereddened, \citetalias{Carretta2009u} and \citet{Carretta2017} estimated the error from the reddening adopted, affecting their \teff. To get the internal error of the $\log g$, they propagate the uncertainties in distance modulus, the star's mass, and the error associated with \teff. The one associated with \vm~ is given by its internal error divided by the square root of the star number. Finally, the error coupled to the metallicity was given by the quadratic sum of the systematic error contribution of the systematic contribution \teff, $\log g$, and \vm~ multiplied with their correspondent abundance sensitivity. The last term was given by the rms scatter in a given element divided by the square root of the star number of a given cluster.

\subsection{Data interpretation}

To determine the strength of a given relationship between two abundances, we used the so-called \textit{'Spearman coefficient' (or rank)} \citep{Spearman}. To characterise a correlation, we consider p-values lower (higher) than 0.01 (0.05) as highly (poorly) significant. Moreover, a p-value between 0.01 and 0.05 is considered mildly significant. The Spearman rank and its p-value will be indicated when corresponding along the text and figures.

In addition, we quantified the variation of any correlation in the present article by simply using the slope of a one-degree fit to the two elements considered.

\section{Chemical Abundances Distribution: Cu}
\label{Sec:Cu_abund}

\subsection{Internal spread}

\begin{figure*}
        \centering
        \includegraphics[width=\textwidth]{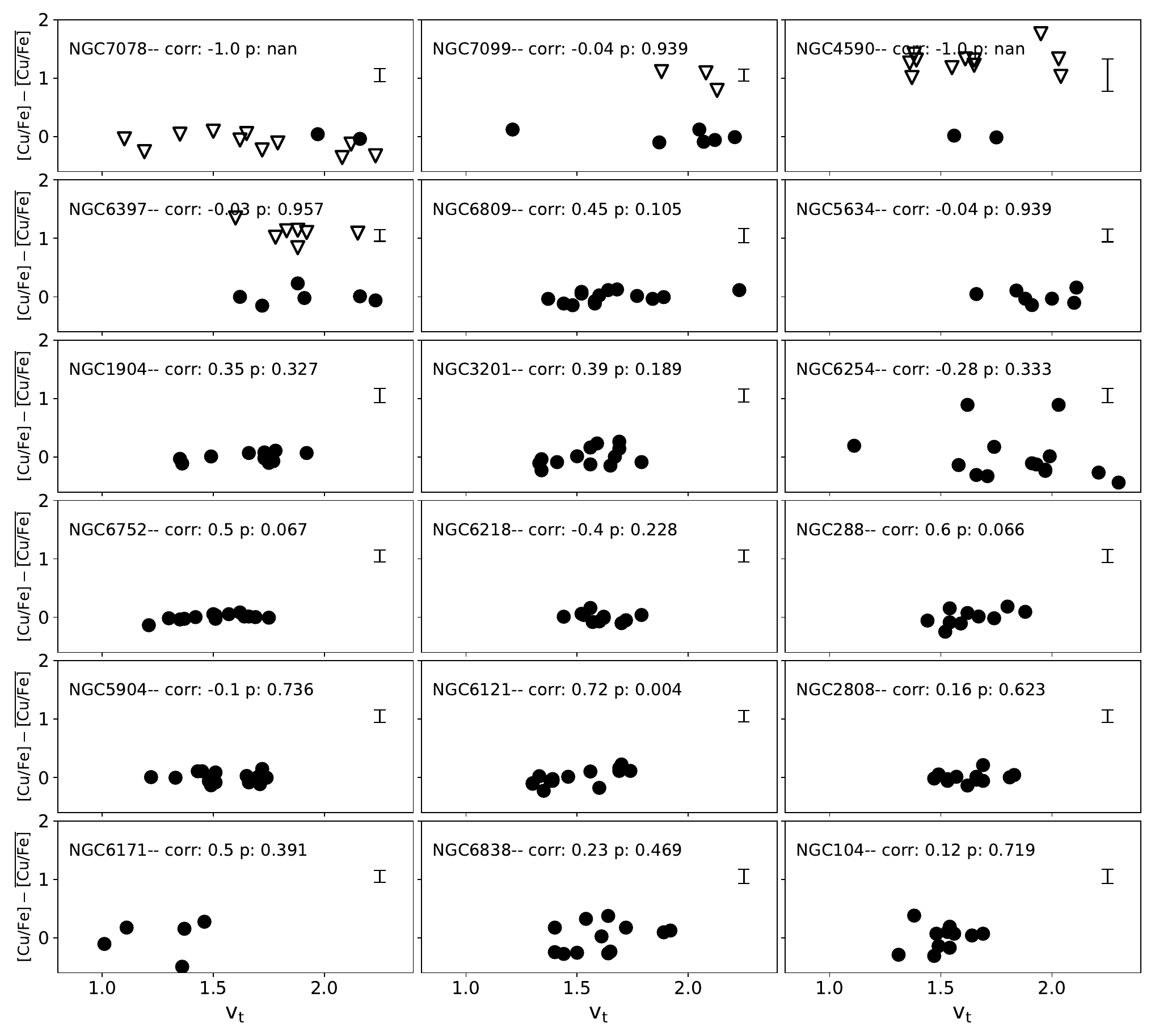}
        \caption{$\Delta$(Cu)$_{MEAN}$ along with \vm~ for each GC of the sample. The respective Spearman corr. and p-values are indicated on each panel. Filled circles and empty triangles represent actual Cu measurements and upper limits, respectively.}
        \label{fig:Cu_Mean}
\end{figure*}

Just a few studies \citep[e.g.,][]{Cunha2002,Simmerer2003} have analysed Copper in GCs, however, finding no evidence of internal variation. As Cu abundances are derived from relatively strong lines, we tested whether there is any dependency of the derived values on \vm. In the more metal-poor clusters, the Cu abundances are dominated by upper limits. Figure \ref{fig:Cu_Mean} displays the behaviour of Cu with respect to \vm, (ordered by increasing metallicity\footnote{This order will keep fixed for the upcoming Figures and Tables.}) showing no clear trend, except for NGC~6121, which has a positive correlation highly significant. Note that we use the Cu abundance obtained by re-scaling to the mean Cu within each cluster to better visualise the sample. In most cases, the Cu results seem to be, within the errors, quite flat and without spread. However, the most metal-rich GCs (NGC~6171, NGC~6838, and NGC~104) display a spread larger than the associated error. On the other hand, the GC NGC~6254 has two stars with slightly higher Cu abundances, considering the associated errors.

\begin{figure}
        \centering
        \includegraphics[width=\columnwidth]{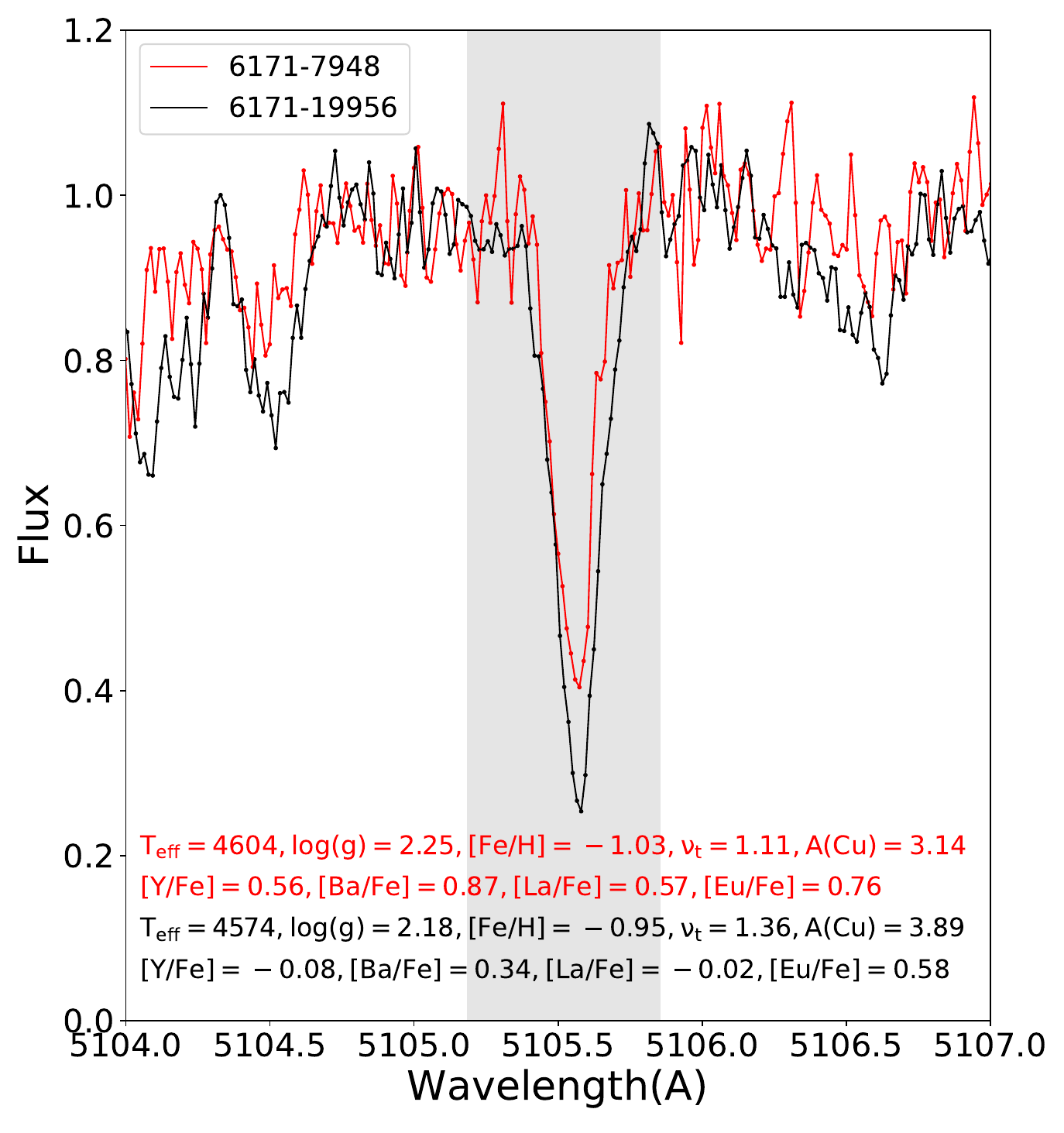}
        \caption{Pair of stars of the GC NGC~6171 with similar stellar parameters as reported by \citet{Carretta2009u}, a different Cu abundance. Black and red lines represent the spectra of ID=19956 and ID=7948, respectively.}
        \label{fig:Comp:Cu}
\end{figure}

To further analyse if this discrepancy is real, Fig. \ref{fig:Comp:Cu} shows the comparison of two stars with similar stellar parameters of the GC NGC~6171. The difference in A(Cu) is about 0.75 dex, which goes beyond the associated errors, is consistent with the difference observed in the lines and cannot be explained by the slight difference in \vm. It is worth noticing that the Cu enrichment goes in the opposite direction of the n-capture enrichment for the pair. This could suggest that in this pair, the nucleosynthesis process(s) responsible for the n-capture production is(are) not linked to the one responsible for the Cu production. 
Some authors \citep[e.g.,][]{Pignatari2010} claim that it can be related to the s-process production in massive stars or AGB stars, which will investigated in the later sections of this manuscript.

\begin{figure*}
        \centering
        \includegraphics[width=\textwidth]{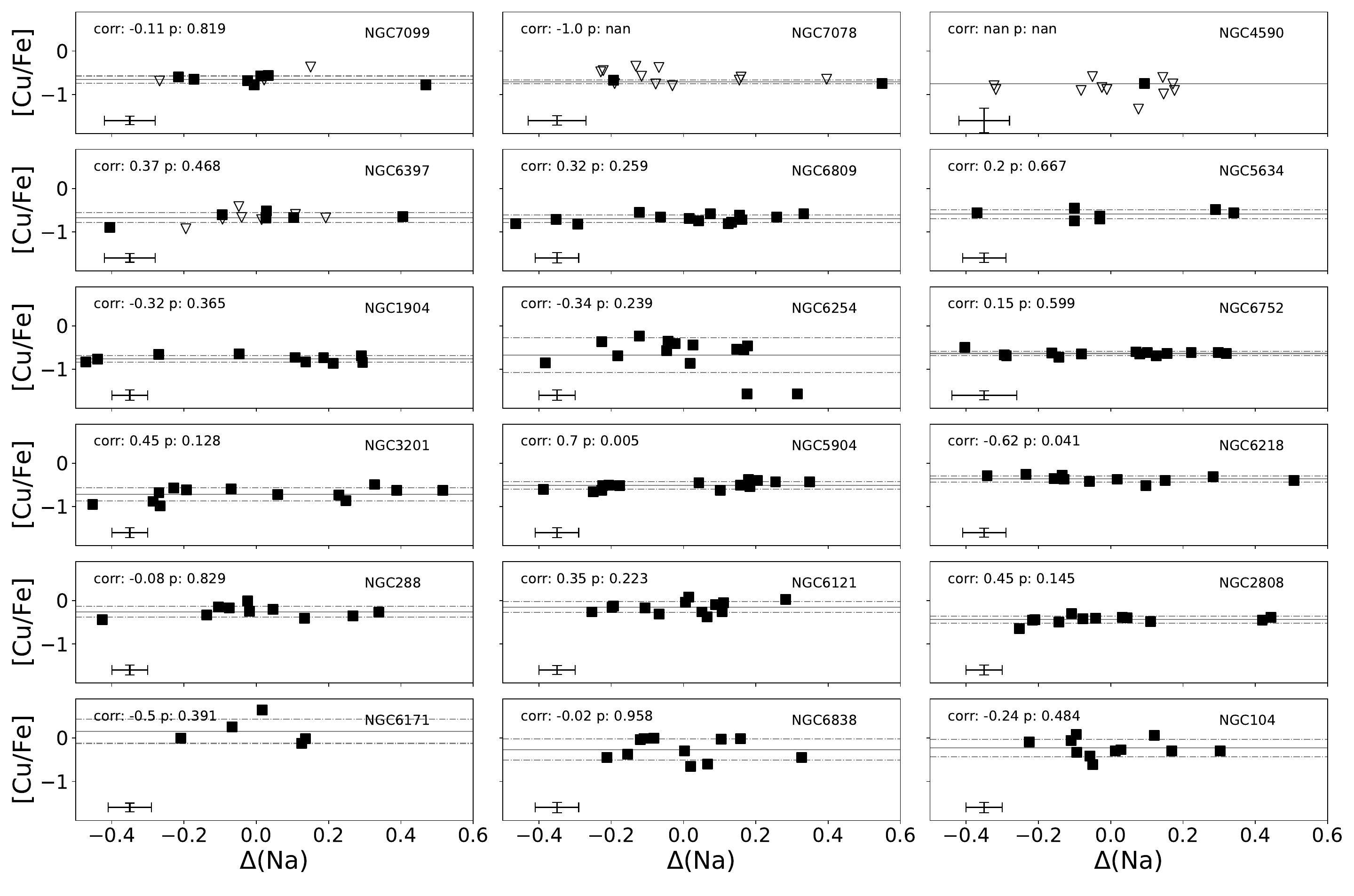}
        \caption{[Cu/Fe] distribution along with $\Delta$(Na) for each GC analysed. The respective average and standard deviation Cu abundance are indicated in solid and dashed lines, respectively. Symbols follow the same description as in Fig. \ref{fig:Cu_Mean}.}
        \label{fig:Cu_DeltaNa}
\end{figure*}

In this context, it is interesting to consider Cu relationship with Na. Fig. \ref{fig:Cu_DeltaNa} shows the distribution of Cu abundances as a function of the $\Delta$Na content in each cluster. This value has been used to eliminate any possible spurious dependencies of abundances from the adopted \vm, an effect that affects elemental abundances derived from strong lines. $\Delta_s$ were defined as follows: for a given element X, $\Delta$(X) is defined as the difference between the reported [X/Fe] abundance and a linear fit between the [X/Fe] and \vm. The distribution seems to be quite flat along with Na, meaning that there is no obvious link in the production between these two species. The only exceptions are the GCs NGC~6218 and NGC~5904, with a significantly high Spearman correlation. 

\subsection{Cu overall distribution}

\begin{figure*}
        \centering
        \includegraphics[width=\textwidth]{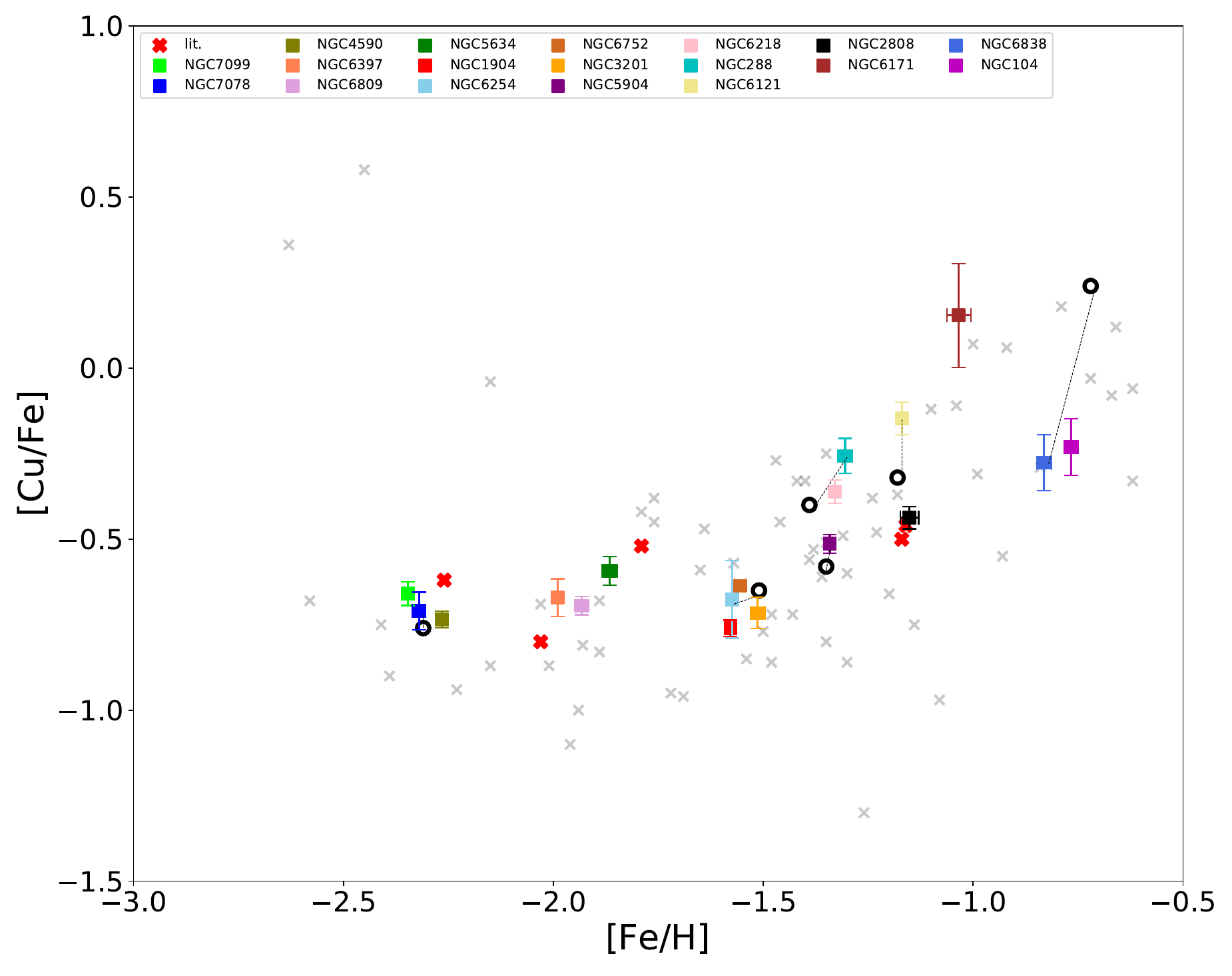}
        \caption{[Cu/Fe] distribution along the [Fe/H] for the whole sample. The analysed GCs are shown with coloured squares. Grey crosses show the field star abundances from \citet{Ishigaki2013}. Red-filled crosses display the reported abundance of Cu in different GCs in the literature. Black circles represent results reported by \citet{Simmerer2003} for our in-common GCs (linked with a black dashed line).}
        \label{fig:CuFe_Fe}
\end{figure*}

Fig. \ref{fig:CuFe_Fe} shows the Cu distribution along the [Fe/H] in the Galactic field and GCs. In the present figure, grey crosses represent the Cu abundances for mostly halo field stars (with a handful from the thick disk) reported by \citet{Ishigaki2013}. In addition, we complement our results with GC abundances from the literature when possible: NGC~1851 \citep{ngc1851}, NGC~362 \citep{ngc362}, Terzan 8 \citep{Terzan8}, NGC~4833 \citep{ngc4833}, and NGC~6093 \citep{ngc6093} all of them represented with red-filled crosses. It is worth noticing that the literature GCs added were analysed using analogous techniques (stellar parameter determination and abundance analysis). The colours were assigned to each solid symbol to represent the different GCs present in this sample. We linked the GCs in common with \citet{Simmerer2003} with a dashed line for an easy comparison. GCs display a steep increase for metallicities higher than -2.0 dex, however, most GCs closely follow the field star distribution, meaning they do not experience a particular Cu enrichment. 

\citet{Simmerer2003} analysed Cu abundance in a large sample of GC using the Cu lines at 5105\AA~ and 5787\AA. It is worth noticing that the latter line is a better Cu indicator, which is neither saturated nor crowded by other species. Unfortunately, the mentioned line is located in the gap of the spectra analysed here. While there is good agreement among in-common GCs with lower metallicity (NGC~6254 and NGC~7078), for the GCs with higher metallicities (NGC~6121, NGC~5904, and NGC~288), the cited article reported considerably lower (except for NGC~6838) Cu abundances --with differences ranging from 0.05 to 0.50 dex-- than the ones reported in the present article.

\begin{figure}
        \centering
        \includegraphics[width=\columnwidth]{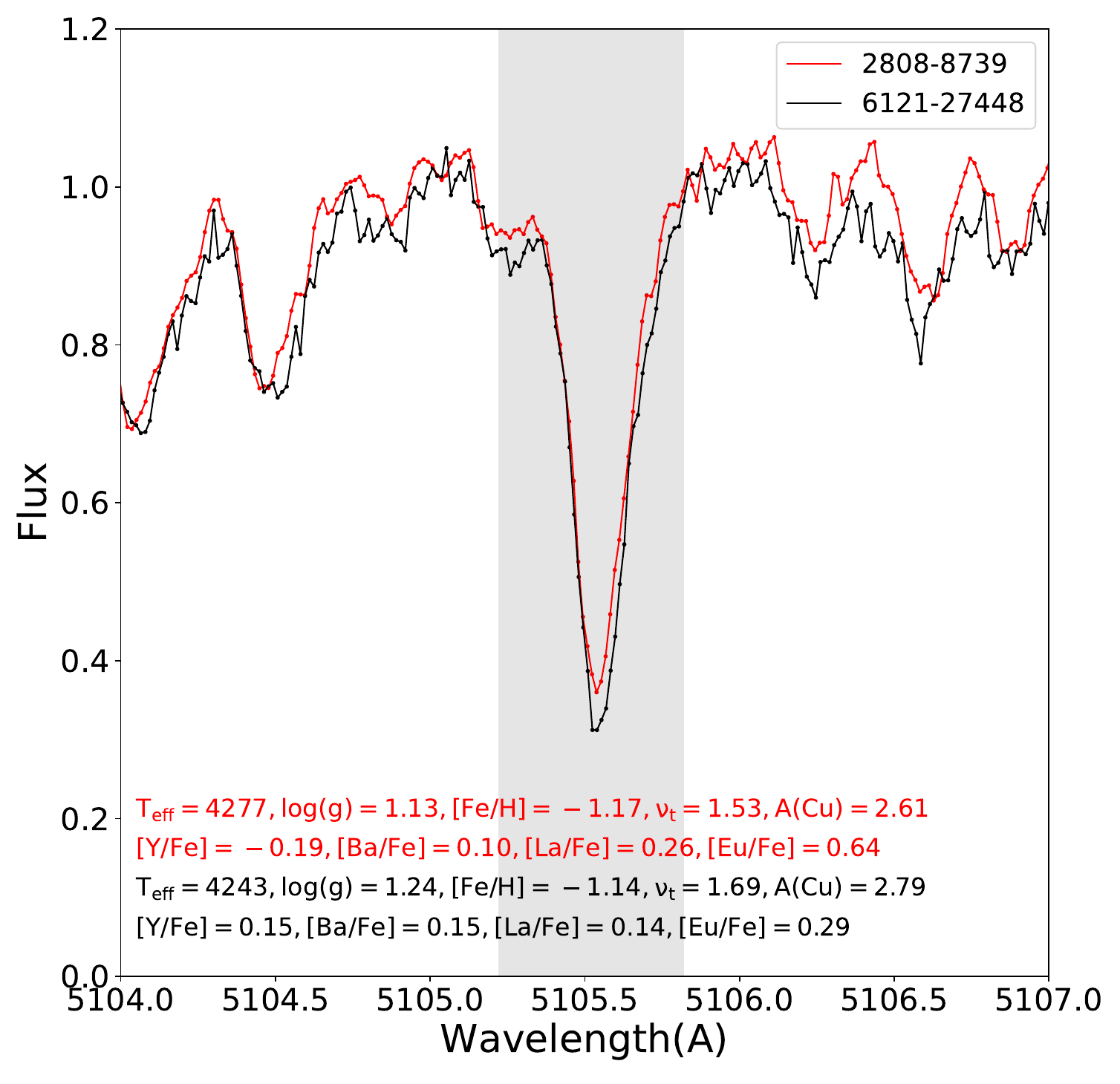}
        \caption{Pair of stars of the GCs NGC~2808 (ID=8739; red line) and NGC~6121 (ID=27448; black line) with similar stellar parameters and different Cu abundance.}
        \label{fig:Comp:Cu_difcluster}
\end{figure}

This discrepancy can be partially explained by the difference in the metallicity adopted, meaning that a model atmosphere with high metallicity reproduces a stronger Cu line than a model with a lower one. In addition, the sensitivity of the line at 5105\AA~to the change in \vm~, plus the presence of MgH lines in the more metal-rich regime, could also play a role in this difference. For the stars with these problems, \citet{Simmerer2003} determined the Cu abundance using the line at 5787\AA. Although we have stars in common with \citet{Simmerer2003}, there is only one for which they determined the abundance from the line at 5105\AA. For those stars, the stellar parameters used in both our and their analyses are practically the same, and the Cu obtained is -0.27$\pm$0.10 and -0.30$\pm$0.10 dex, respectively. In particular, the large spread found in the present article for NGC~6254 was also reported by \citet{Simmerer2003}. On the other hand, they also reported a particularly high Cu content in NGC~6121 compared with other GCs with similar metallicities.  NGC~2808 has similar metallicity as NGC~6121, but they display quite different Cu content in our analysis. Given that there is a pair of stars with similar parameters, one in each of the two clusters, it is possible to assess the existence of such a difference directly. Figure \ref{fig:Comp:Cu_difcluster} shows such a comparison for the Cu line. The figure reinforces that the difference is real and is not due to any dependency on stellar parameters. In the case of NGC~6171, the trend with \vm~ does not seem to be present, but it displays a particular Cu enrichment.

\section{Chemical Abundances Distribution: Y, Ba, La, and Eu}
\label{Sec:NCap_abun}

\subsection{Ba-Y dependency with \vm}

Based on three rather strong lines, Ba abundances show considerable sensitivity to the adopted \vm. This is a common finding in cool giants, as discussed, e.g., by \citet{Worley2013}. It is worth noticing that the sensitivity of these species to \vm~ is independent of the method used for the \vm~ derivation. We explored averaging Ba abundances weighted by their respective errors using the different combinations of lines to minimise this effect and concluded that the best combination is, indeed, the use of all three available to us. We opted to use from hereinafter all three lines for our final abundance due to the reduction of both the spread and the lessening of the \vm~ dependence. Similar considerations apply to the Y II lines used to derive [Y/Fe]II abundances. 

\begin{figure*}
        \centering
        \includegraphics[width=\textwidth]{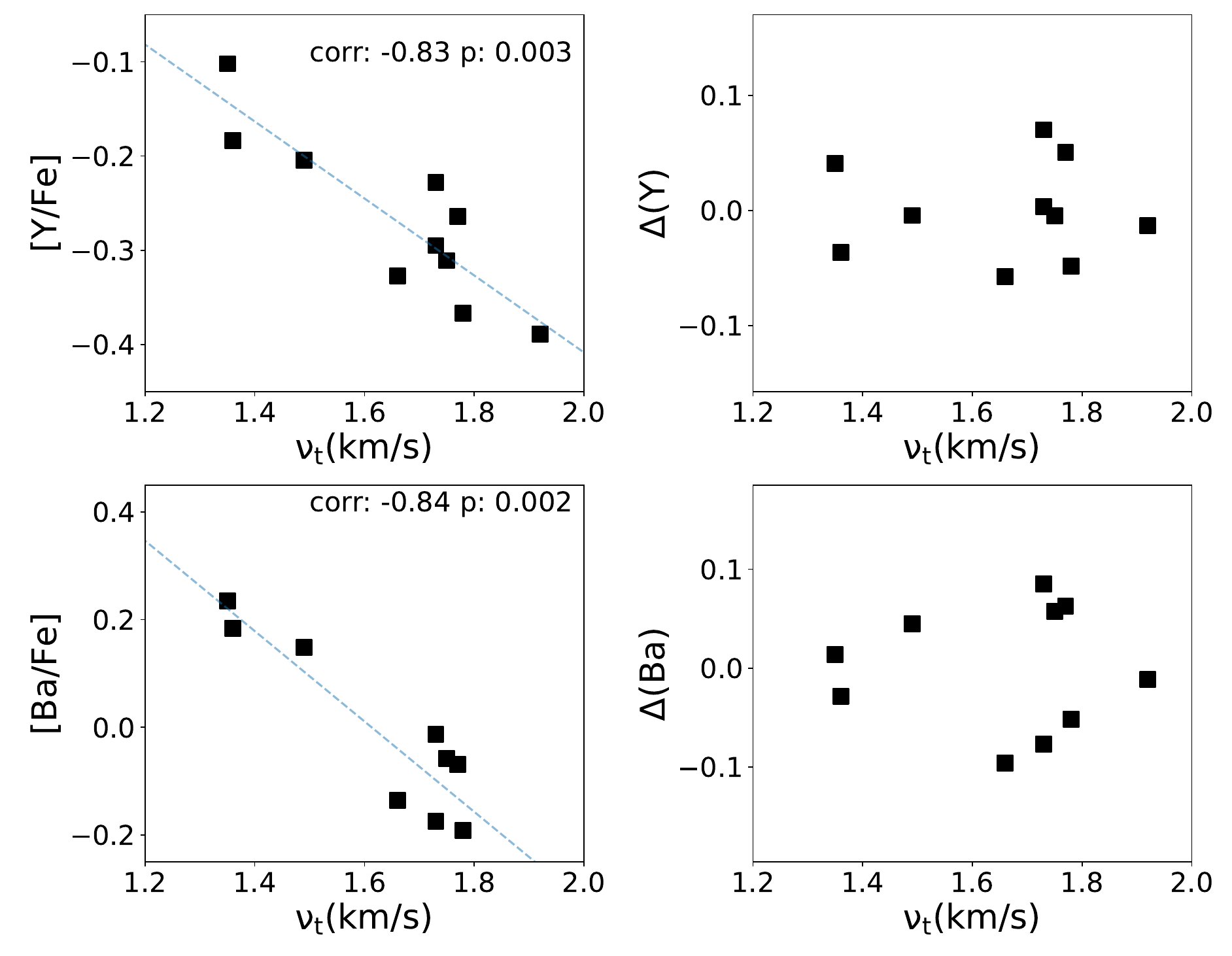}
        \caption{The upper (lower) left and right panels show the [Y/Fe] ([Ba/Fe]) abundances and $\Delta$(Y) ($\Delta$(Ba)), respectively, as a function of \vm~ for the GC NGC~1904. The blue dotted line shows the linear fit.}
        \label{fig:ngc1904_trendY}
\end{figure*}

In addition, we computed $\Delta$X for Y and Ba to get rid of the trend given by \vm~ in the whole sample. Fig. \ref{fig:ngc1904_trendY} shows an illustrative example for the GC NGC~1904. A strong negative Spearman correlation (about -0.80) is clearly shown in the left panel of both figures. The right panels show how the trend is avoided by using the $\Delta_{s}$ (defined in Sec. \ref{Sec:Cu_abund}).

\subsection{Internal n-capture spread}

For the sake of this section, we remind the reader of the effects of \vm~ on Y and Ba (discussed in Sec. \ref{Sec:AbunDet}). Because of this effect, in general, the larger the range covered by \vm~, the larger the dispersion driven by this parameter. 

As can be seen from Table \ref{tab:rms_deltas_IQR}, the GCs NGC~6171 ([Fe/H]=-1.03 dex) and NGC~7078 ([Fe/H]=-2.32 dex) display both a large rms error and IQR in Y and Ba. The Ba dispersion plus the constant Y found in NGC~7078 is in good agreement with previous results in the literature, where NGC~7078 has been reported as an r-process enriched cluster by several literature sources \citep[e.g.][]{Kirby2020}. At the cluster metallicity, Ba is mostly produced by the r-process. On the other hand, NGC~6171 shows a large spread in all the elements analysed in the present article. This mildly significant spread agrees with \citet{OConnell2011}, who speculated about a potential early r-process enrichment in the cluster due to the evidence of La and Eu spread. However, because of the small number of stars it is based on, this spread should be taken with caution.

\begingroup

\begin{table*}
\scriptsize
\caption{Summary table with the mean abundances for each cluster with their respective errors (see Section \ref{subs:IndError}). Note that the ratios of ionised species are indexed to Fe II. The letter 'n' next to each element abundance refers to the number of stars used for the mean value. The [Fe/H] reported by \citealt{Carretta2009u} and \citet{Carretta2017}.}
\label{tab:mean_spread}
\setlength{\tabcolsep}{4.8pt}
\begin{tabular}{lccccccccccccccccccccr}
\hline
\hline
Clusters & [Fe/H] & [Cu/Fe] & n & $\sigma$ & rms &  [Ba/Fe] & n & $\sigma$ & rms  & [Y/Fe] & n & $\sigma$ & rms & [La/Fe] & n & $\sigma$ & rms  & [Eu/Fe] & n & $\sigma$ & rms \\
\hline
NGC~7099 & -2.34 & -0.66    & 7  & 0.10     & 0.04  & -0.22    & 10 & 0.11 & 0.06     & -0.16     & 10 & 0.10 &  0.03     &  ---        & 0  &  ---  & ---      & ---      &  0  & ---  &  ---   \\
NGC~7078 & -2.32 & -0.71    & 2  & 0.11     & 0.06  &  0.19    & 13 & 0.19 & 0.11     & -0.06     & 13 & 0.10 &  0.03     &  0.32       & 1  &  ---  & ---      & 0.90     &  6  & 0.12 &  0.09  \\
NGC~4590 & -2.27 & -0.74    & 2  & 0.11     & 0.02  & -0.17    & 13 & 0.22 & 0.05     & -0.37     & 13 & 0.13 &  0.03     &  ---        & 0  &  ---  & ---      & ---      &  0  & ---  &  ---   \\
NGC~6397 & -1.99 & -0.67    & 6  & 0.10     & 0.06  & -0.12    & 13 & 0.11 & 0.03     & -0.21     & 13 & 0.10 &  0.01     &  ---        & 0  &  ---  & ---      & 0.63     &  5  & 0.10 &  0.05  \\
NGC~6809 & -1.93 & -0.69    & 14 & 0.10     & 0.03  &  0.20    & 14 & 0.16 & 0.06     & -0.08     & 14 & 0.11 &  0.04     &  0.42       & 7  &  0.10 & 0.04     & 0.74     &  13 & 0.12 &  0.03  \\
NGC~5634 & -1.87 & -0.59    & 7  & 0.11     & 0.04  & -0.02    & 7  & 0.11 & 0.06     & -0.10     & 7  & 0.10 &  0.08     &  0.42       & 6  &  0.12 & 0.04     & 0.56     &  7  & 0.10 &  0.03  \\
NGC~1904 & -1.58 & -0.76    & 10 & 0.10     & 0.03  & -0.03    & 10 & 0.13 & 0.06     & -0.27     & 10 & 0.10 &  0.03     &  0.30       &  2 &  0.11 & 0.11     & 0.49     &  7  & 0.10 &  0.04  \\
NGC~6254 & -1.58 & -0.68    & 14 & 0.10     & 0.12  &  0.06    & 14 & 0.12 & 0.07     & -0.08     & 14 & 0.11 &  0.03     &  0.33       & 11 &  0.11 & 0.03     & 0.52     &  11 & 0.10 &  0.03  \\
NGC~6752 & -1.56 & -0.64    & 14 & 0.10     & 0.01  &  0.17    & 14 & 0.11 & 0.06     & -0.08     & 14 & 0.11 &  0.03     &  0.14       & 14 &  0.10 & 0.02     & 0.43     &  14 & 0.12 &  0.02  \\
NGC~3201 & -1.51 & -0.72    & 13 & 0.10     & 0.04  & -0.01    & 13 & 0.11 & 0.05     & -0.21     & 13 & 0.11 &  0.04     &  0.04       & 13 &  0.12 & 0.03     & 0.38     & 10  & 0.10 &  0.04  \\
NGC~5904 & -1.34 & -0.51    & 14 & 0.10     & 0.02  &  0.12    & 14 & 0.12 & 0.05     & -0.08     & 14 & 0.10 &  0.03     &  0.17       & 13 &  0.10 & 0.03     & 0.64     &  8  & 0.10 &  0.03  \\
NGC~6218 & -1.33 & -0.36    & 11 & 0.10     & 0.02  &  0.06    & 11 & 0.11 & 0.04     &  0.05     & 11 & 0.10 &  0.04     &  0.15       & 11 &  0.10 & 0.02     & 0.42     & 10  & 0.10 &  0.02  \\
NGC~288  & -1.30 & -0.26    & 10 & 0.11     & 0.04  &  0.18    & 10 & 0.12 & 0.05     &  0.12     & 10 & 0.12 &  0.03     &  0.36       & 10 &  0.11 & 0.03     & 0.58     &  10 & 0.10 &  0.02  \\
NGC~6121 & -1.17 & -0.15    & 14 & 0.11     & 0.04  &  0.45    & 14 & 0.11 & 0.06     &  0.28     & 14 & 0.10 &  0.04     &  0.37       & 14 &  0.10 & 0.02     & 0.49     &  14 & 0.10 &  0.04  \\
NGC~2808 & -1.15 & -0.44    & 12 & 0.12     & 0.03  & -0.01    & 12 & 0.12 & 0.04     & -0.18     & 12 & 0.11 &  0.03     &  0.23       & 12 &  0.10 & 0.04     & 0.63     &  12 & 0.10 &  0.03  \\
NGC~6171 & -1.03 &  0.15    & 5  & 0.10     & 0.16  &  0.55    & 5  & 0.12 & 0.18     &  0.31     & 5  & 0.11 &  0.18     &  0.44       & 3  &  0.13 & 0.10     & 0.64     &  5  & 0.10 &  0.11  \\
NGC~6838 & -0.83 & -0.28    & 12 & 0.12     & 0.07  & -0.08    & 12 & 0.15 & 0.07     & -0.23     & 12 & 0.11 &  0.04     &  0.21       & 12 &  0.10 & 0.02     & 0.50     &  12 & 0.12 &  0.03  \\
NGC~104  & -0.77 & -0.23    & 11 & 0.12     & 0.07  & -0.15    & 11 & 0.11 & 0.07     & -0.30     & 11 & 0.11 &  0.04     & -0.06       & 11 &  0.12 & 0.02     & 0.37     &  11 & 0.10 &  0.02  \\
\hline
\hline
\end{tabular}
\end{table*}
\endgroup

\subsection{Non-LTE correction for Y}

Because our sample spans a large range of stellar parameters, non-LTE correction is a factor to take into consideration, especially due to their strong dependency on metallicity, which could lead to unreal abundance trends in our results. \citet{Storm2023} presented the Y non-LTE correction for a large range of stellar parameters in different Y lines. For the stars in our sample, the corrections range from $\sim$0.05 and $\sim$0.15 dex. Then, in the worst-case scenario, the maximum variation would be around 0.10 dex, which has a limited impact on the current analysis. Moreover, \citet{Guiglion2023} showed the Y spread along with the [Fe/H]; the results reveal that the spread did not change considerably ($\sim$0.02 dex), meaning that non-LTE corrections would not modify the potential spreads within a given cluster. Similar results were reported for Ba in the same article. Therefore, our results do not consider non-LTE corrections.

\subsection{Comparison with the literature}

\begin{figure}
        \centering
        \includegraphics[width=\columnwidth]{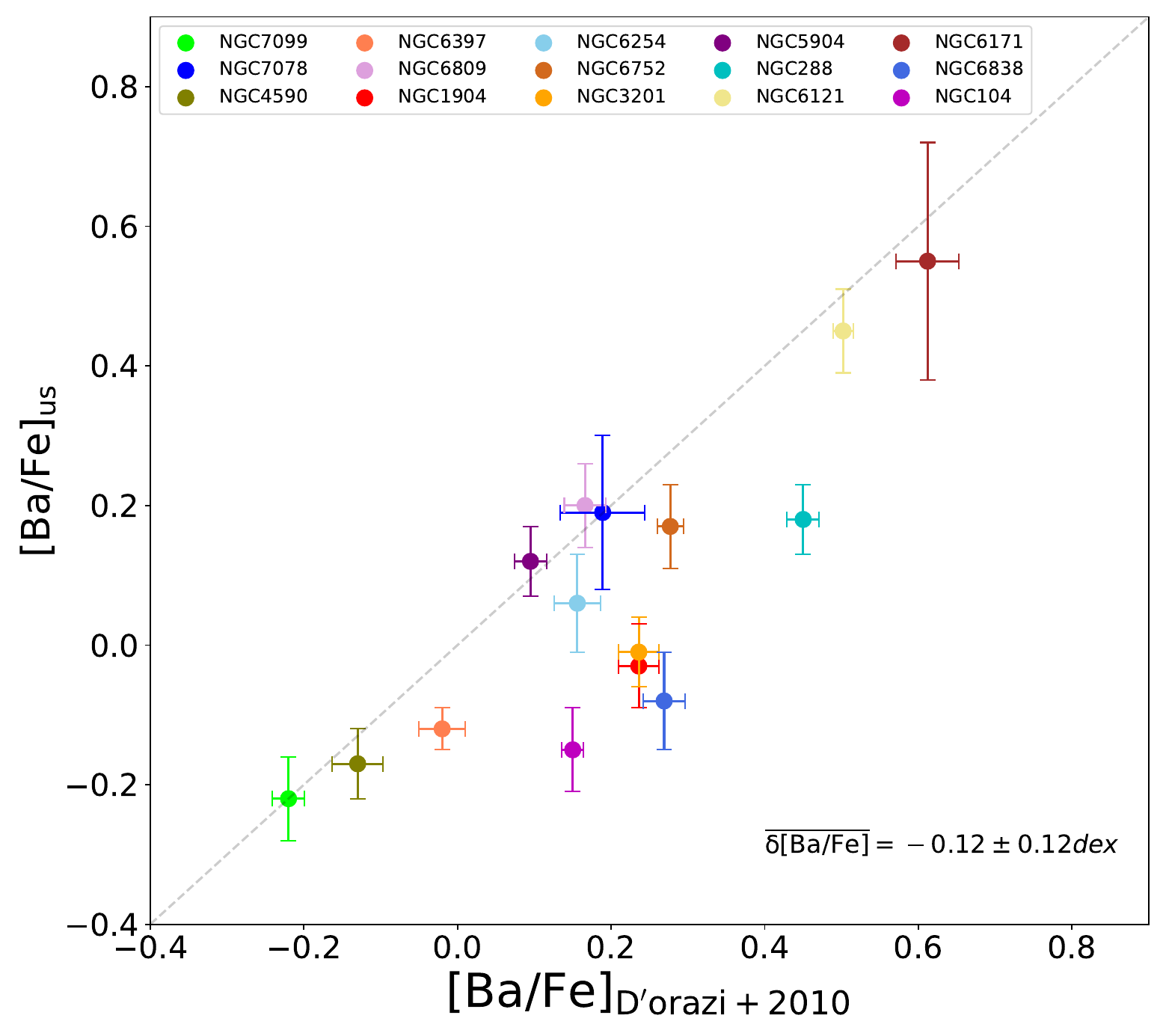}
        \caption{Comparison of Ba abundances obtained in the present article with \citet{dorazi2010_neutron} for in-common GCs. The average difference between our and their results, $\overline{\delta[Ba/Fe]}$, is also indicated.}
        \label{fig:dorazi_comp}
\end{figure}

\citet{dorazi2010_neutron} analysed the Ba abundances of 15 GCs included in our sample, for which we have 55 stars in common; however, using lower resolution GIRAFFE spectra of a larger number of stars per cluster. They used equivalent width to determine the chemical abundances and adopted stellar parameters derived identically from those used in the present article. Because the Ba abundances for individual stars were not published, Fig. \ref{fig:dorazi_comp} shows the comparison of our and their average Ba abundance for the 15 in-common GCs. 
We got constantly lower abundances for the whole sample. As shown in the figure, the average difference between our and their results is $\overline{\delta[Ba/Fe]}$= -0.12 $\pm$ 0.12 dex, probably due to the different methods used in the abundance determination and the lines considered. While we used Ba lines at 5853\AA, 6141\AA, and 6496\AA, \citet{dorazi2010_neutron} used only the second one.

\subsection{n-process elements: relation with O, Na, and Mg}

\begin{figure*}
     \centering
     \begin{subfigure}
         \centering
         \includegraphics[width=\textwidth]{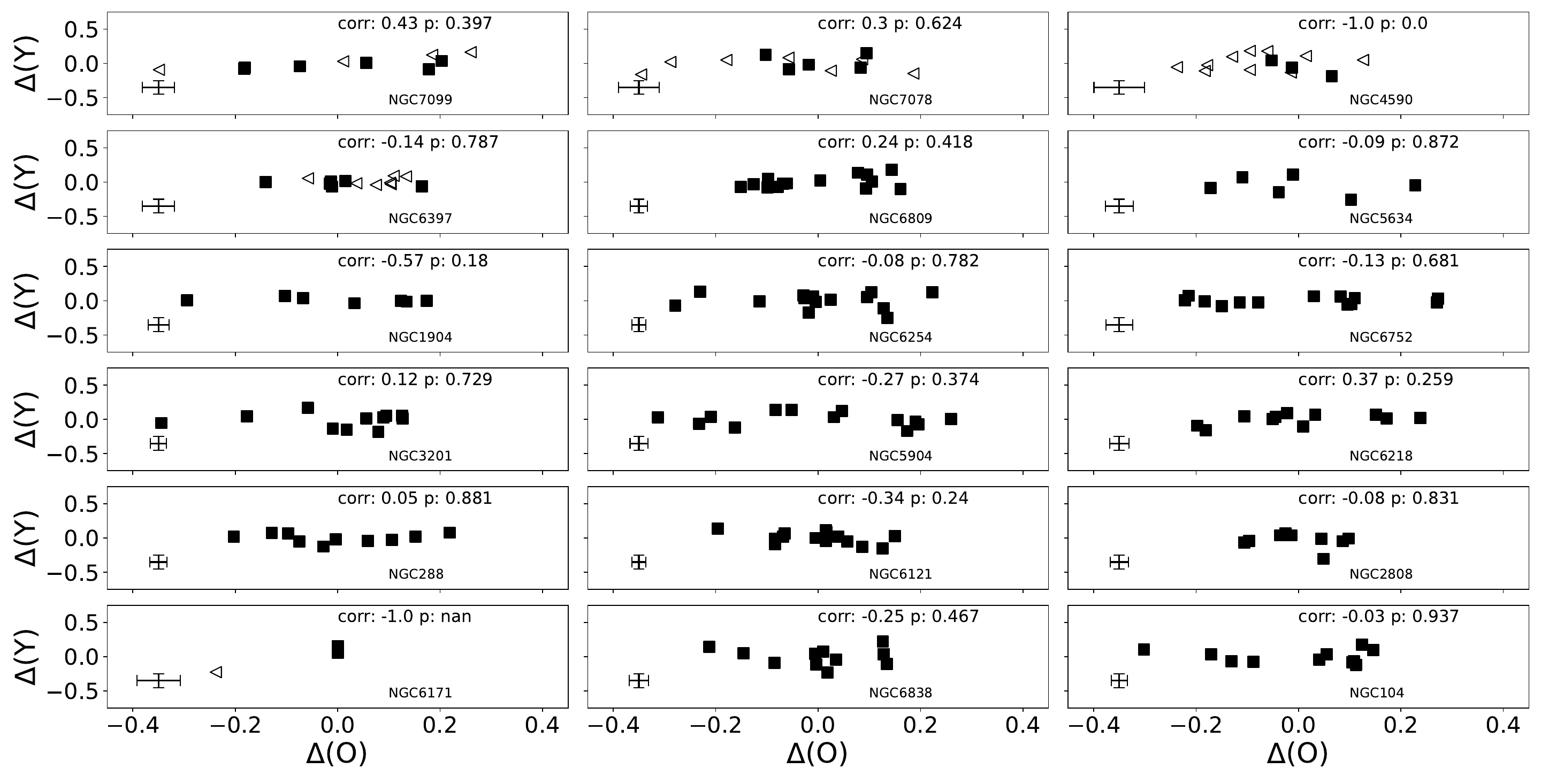}
     \end{subfigure}
     \vfill
     \begin{subfigure}
         \centering
         \includegraphics[width=\textwidth]{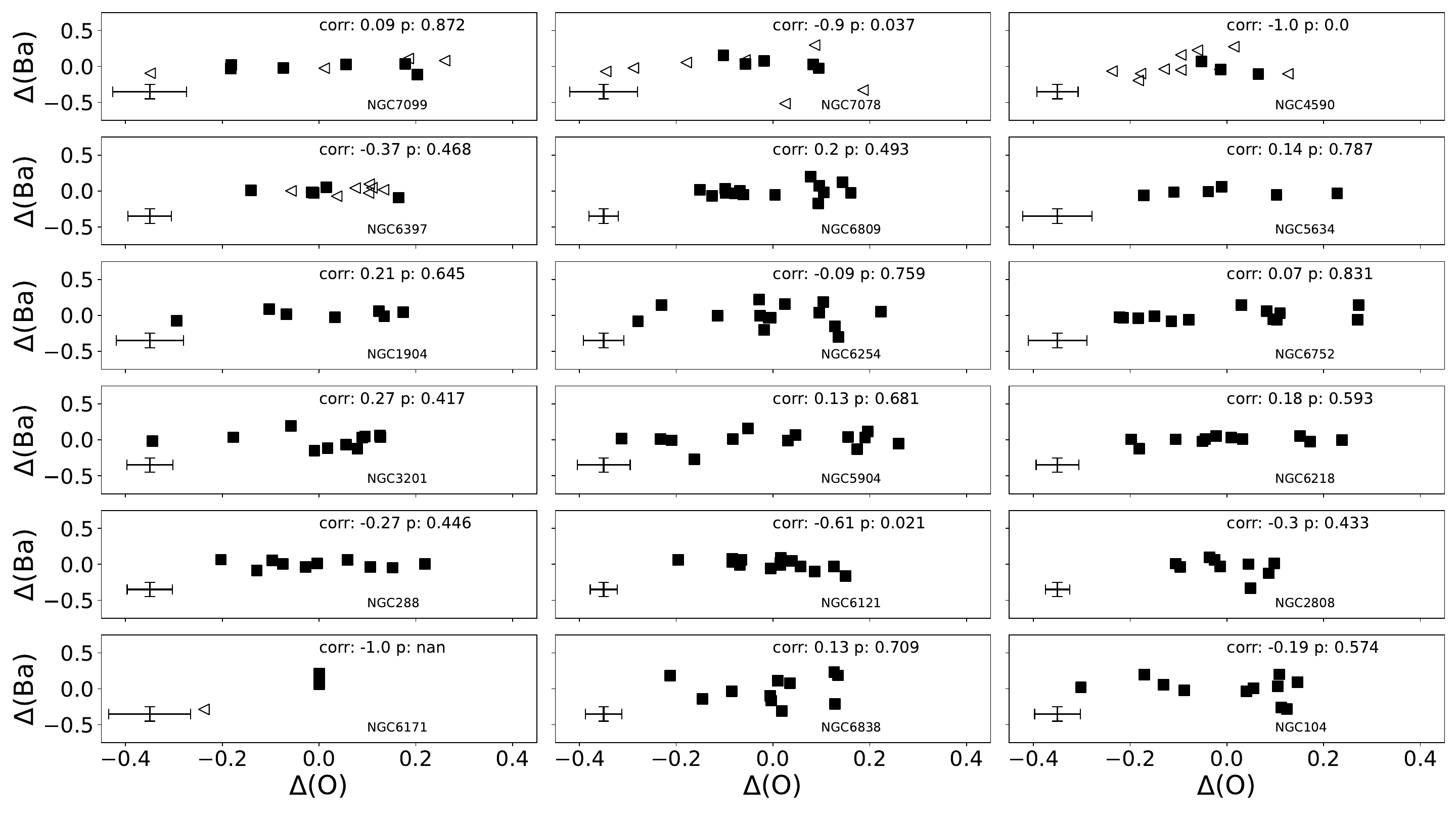}
     \end{subfigure}
     \vfill
        \caption{$\Delta$(Y) (upper) and $\Delta$(Ba) (lower) as a function of $\Delta$(Na) for each cluster of the sample. Filled squares and empty triangles represent actual measurement and upper limits, respectively.}
        \label{fig:DeltaYBa_O}
\end{figure*}

\begin{figure*}
     \centering
     \begin{subfigure}
         \centering
         \includegraphics[width=\textwidth]{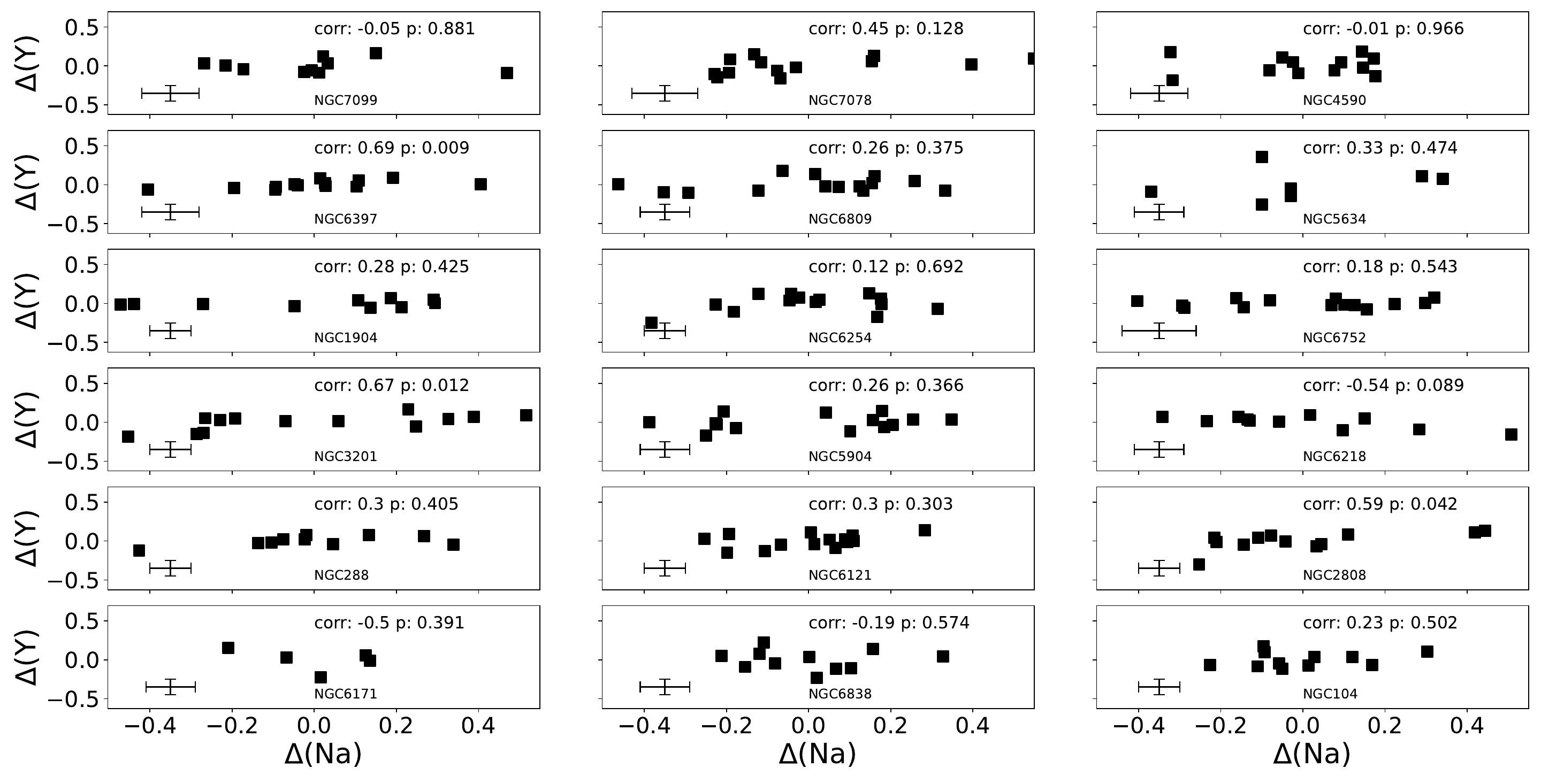}
     \end{subfigure}
     \vfill
     \begin{subfigure}
         \centering
         \includegraphics[width=\textwidth]{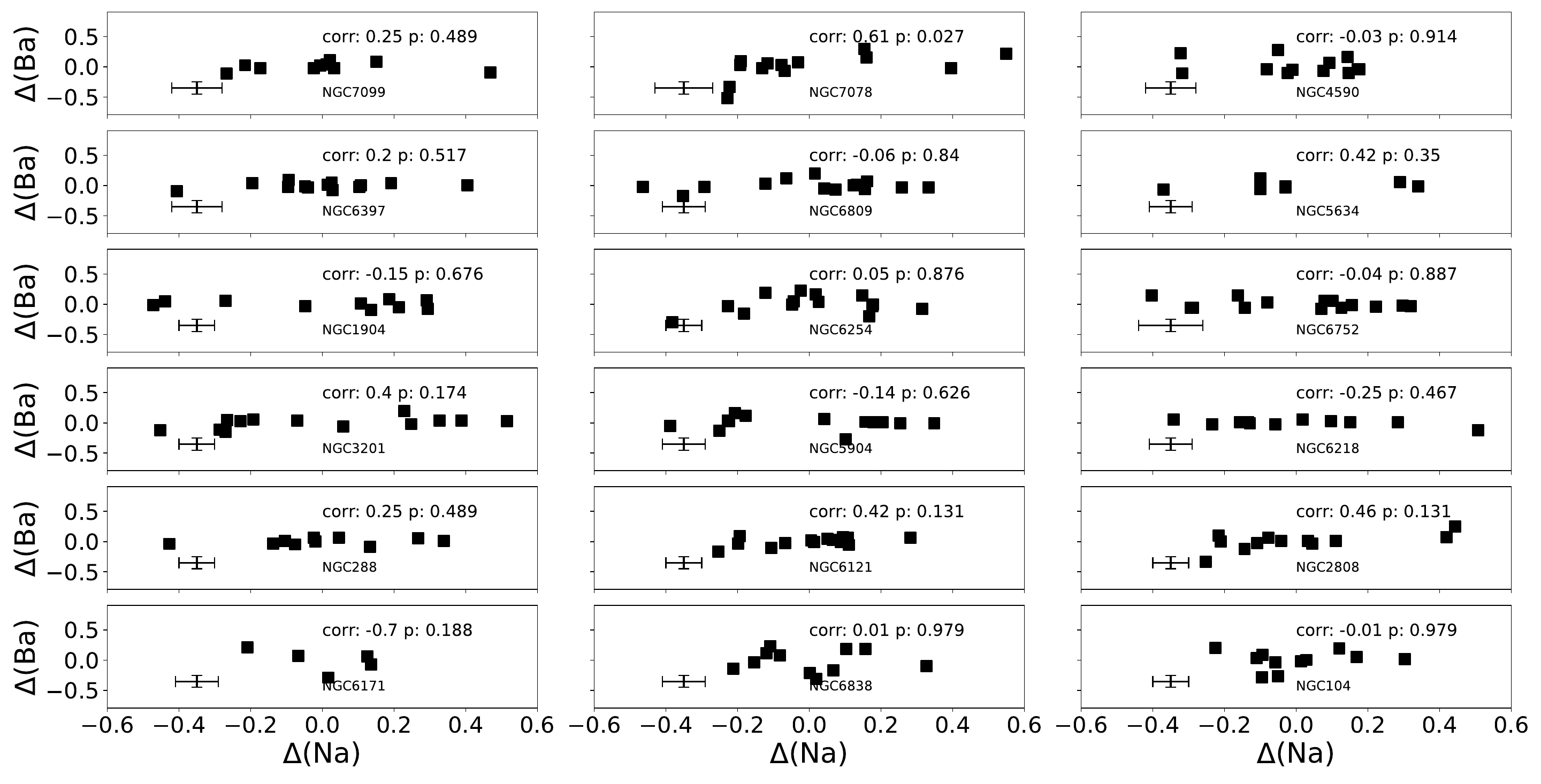}
     \end{subfigure}
     \vfill
        \caption{$\Delta$(Y) (upper) and $\Delta$(Ba) (lower) as a function of $\Delta$(Na) for each cluster of the sample. Filled squares and empty triangles represent actual measurement and upper limits, respectively.}
        \label{fig:DeltaYBa_Ind}
\end{figure*}

\begin{figure*}
     \centering
     \begin{subfigure}
         \centering
         \includegraphics[width=\textwidth]{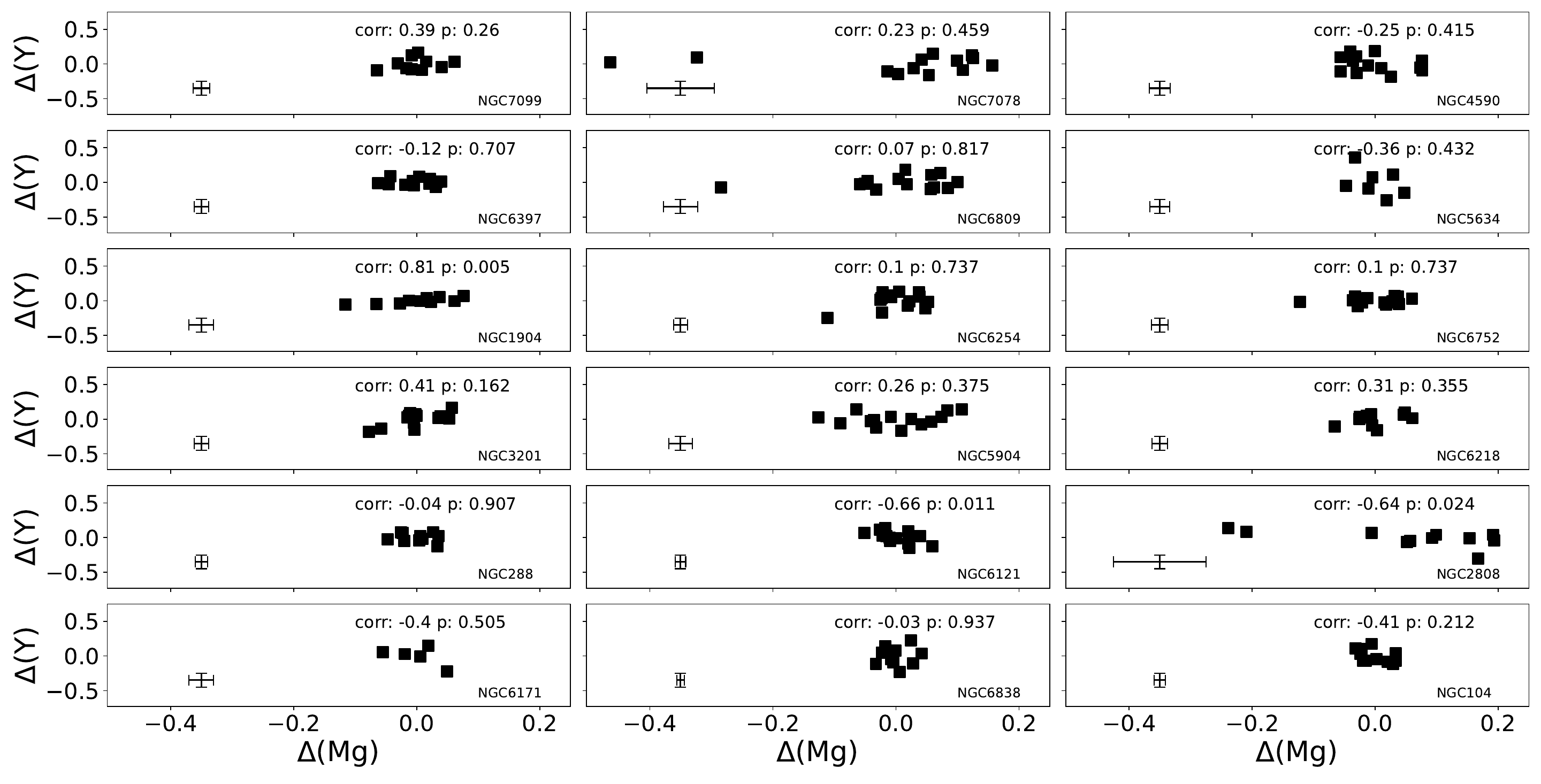}
     \end{subfigure}
     \vfill
     \begin{subfigure}
         \centering
         \includegraphics[width=\textwidth]{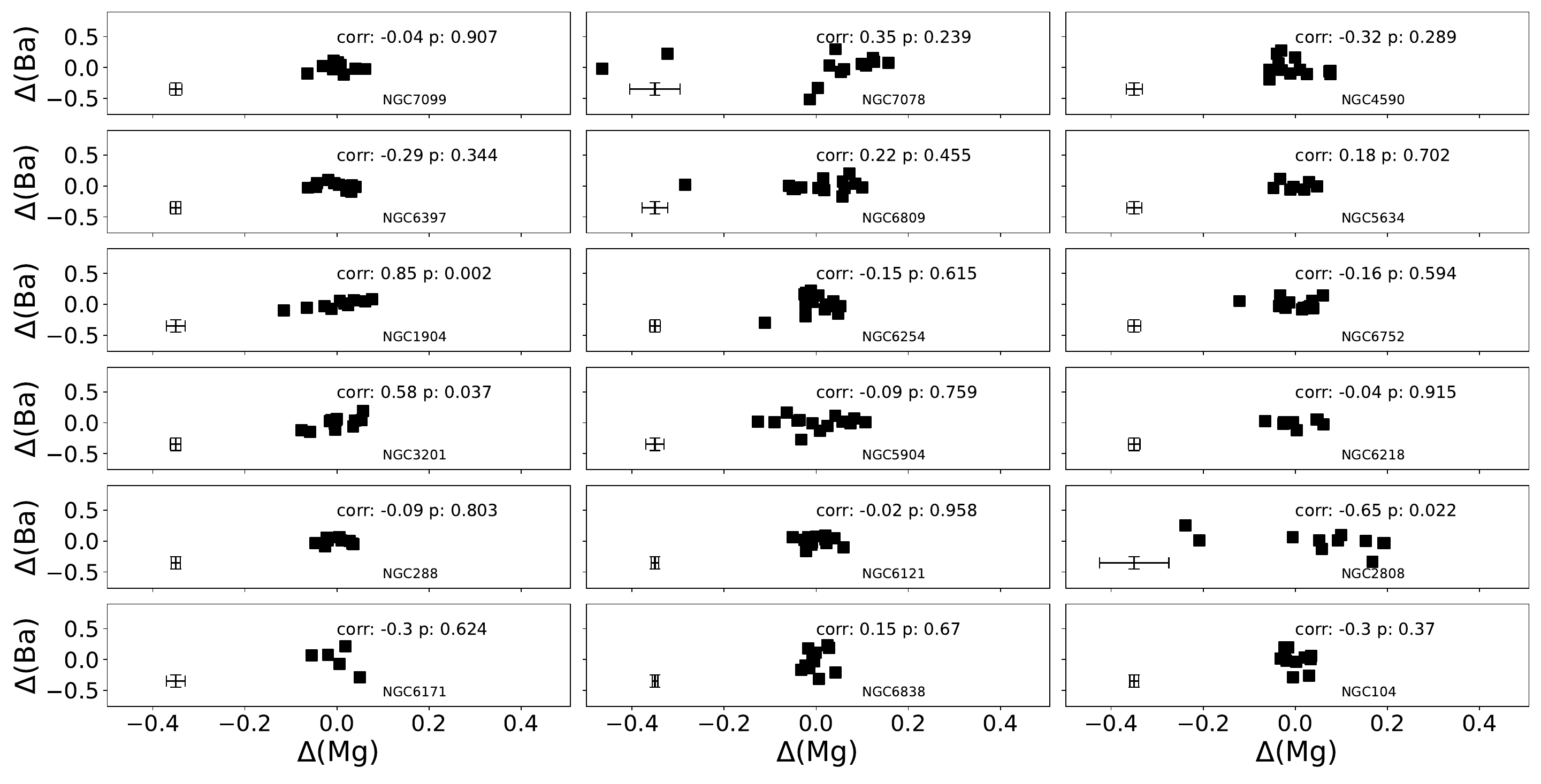}
     \end{subfigure}
     \vfill
        \caption{$\Delta$(Y) (upper) and $\Delta$(Ba) (lower) as a function of $\Delta$(Na) for each cluster of the sample. Filled squares and empty triangles represent actual measurement and upper limits, respectively.}
        \label{fig:DeltaYBa_Mg}
\end{figure*}

With a few exceptions, GCs have not shown heavy element variations among their different populations, and are not involved in the MSP phenomenon. Similarly to the analysis done by \citetalias{Schiappacasse-Ulloa2023} for NGC~6752, we explored the variation of $\Delta$Y and $\Delta$Ba with respect to Na. Figures \ref{fig:DeltaYBa_O}, \ref{fig:DeltaYBa_Ind}, and \ref{fig:DeltaYBa_Mg} show the results for the Y and Ba abundances as a function of the O, Na, and Mg, respectively, for each cluster of the sample. The Spearman correlation for actual measurements is reported in each panel. In the first figure, only weak or poorly significant correlations exist between Y and O in the whole sample. A few clusters (NGC~7078 and NGC~6121) show a negative correlation between Ba and O highly significant, which could be translated in a Ba decrease of about 0.07 and 0.05 dex for each 0.10 dex increment of O, respectively. In the second figure, the results for Y and Ba seem to display quite constant abundances within the associated errors along the different Na. However, there are GCs such as NGC~2808, NGC~6397, and NGC~3201 that display a positive correlation highly significant between Y and Na, which would produce an increment in Y of about 0.03, 0.015, and 0.02 dex for each increment in Na of 0.1 dex, respectively. Finally, the last figure shows the relation of Y-Mg and Ba-Mg. Most of the clusters display weak or non-significant Y-Mg correlations. The exception of it are the GCs NGC~1904, NGC~6121, and NGC~2808. Interestingly, the two latter clusters have a negative correlation, whereas the first has a positive one. In particular, NGC~1904 would have a Y increment of 0.06 dex for each 0.10 dex Mg, whereas NGC~6121 and NGC~2808 show a Y decrease of 0.17 and 0.03 dex for a Mg increment of 0.10 dex, respectively. Concerning the relation Ba-Mg, the GCs NGC~1904, NGC~3201, and NGC~2808 display a strong correlation, being the last GC the only one with a negative relation. While the latter displayed a Ba decrease of about 0.03 dex for each Mg increment of 0.10 dex, NGC~1904 and NGC~3201 showed an increment of 0.09 and 0.16 dex. Nevertheless, having s-process elements correlating with Na, without a corresponding negative correlation with Mg (or vice-versa) could indicate spurious occurrences due to the small number of statistics. On the other hand, because the proton-capture reactions produce intrinsically small Mg depletion (as opposed to large enhancements in Na), the Mg variations are difficult to observe. Then, these results should be taken with caution.

\begin{table*}
\footnotesize
\centering
\caption{Observational and rms error (excluding the \vm~ contribution) for each cluster. In addition, the IQRs of [Y/Fe], [Ba/Fe], [La/Fe], and [Eu/Fe] for each cluster are reported considering the \vm~ effects. Upper limits were not considered for the IQR computation.}
\label{tab:rms_deltas_IQR}
\begin{tabular}{ccccccccc}
\hline
\hline
Cluster & rms(Ba)$_{MEAN}$ & $\sigma$(Ba) & rms(Y)$_{MEAN}$  & $\sigma$(Y) & IQR$_{[Y/Fe]}$     & IQR$_{[Ba/Fe]}$     & IQR$_{[La/Fe]}$     & IQR$_{[Eu/Fe]}$ \\
\hline
NGC~7099 & 0.02   & 0.10  & 0.03    & 0.10 & 0.11  & 0.13        & ---         & ---   \\
NGC~7078 & 0.06   & 0.10  & 0.03    & 0.10 & 0.17  & 0.44        & ---         & 0.27  \\
NGC~4590 & 0.04   & 0.10  & 0.03    & 0.10 & 0.18  & 0.15        & ---         & ---   \\
NGC~6397 & 0.02   & 0.10  & 0.02    & 0.10 & 0.04  & 0.14        & ---         & 0.01  \\
NGC~6809 & 0.03   & 0.10  & 0.03    & 0.10 & 0.07  & 0.16        & 0.16        & 0.17  \\
NGC~5634 & 0.03   & 0.10  & 0.08    & 0.10 & 0.18  & 0.19        & 0.15        & 0.10  \\
NGC~1904 & 0.02   & 0.10  & 0.02    & 0.10 & 0.11  & 0.27        & 0.08        & 0.15  \\
NGC~6254 & 0.04   & 0.10  & 0.03    & 0.10 & 0.15  & 0.27        & 0.12        & 0.08  \\
NGC~6752 & 0.02   & 0.10  & 0.01    & 0.10 & 0.16  & 0.24        & 0.06        & 0.09  \\
NGC~3201 & 0.03   & 0.10  & 0.03    & 0.10 & 0.24  & 0.32        & 0.16        & 0.20  \\
NGC~5904 & 0.03   & 0.10  & 0.03    & 0.10 & 0.15  & 0.23        & 0.09        & 0.14  \\
NGC~6218 & 0.02   & 0.10  & 0.03    & 0.10 & 0.15  & 0.15        & 0.08        & 0.09  \\
NGC~288  & 0.02   & 0.10  & 0.02    & 0.10 & 0.14  & 0.14        & 0.10        & 0.09  \\
NGC~6121 & 0.02   & 0.10  & 0.03    & 0.10 & 0.23  & 0.39        & 0.06        & 0.18  \\
NGC~2808 & 0.04   & 0.11  & 0.04    & 0.12 & 0.11  & 0.11        & 0.12        & 0.15  \\
NGC~6171 & 0.09   & 0.10  & 0.07    & 0.10 & 0.53  & 0.41        & 0.15        & 0.22  \\
NGC~6838 & 0.06   & 0.10  & 0.04    & 0.10 & 0.17  & 0.41        & 0.08        & 0.26  \\
NGC~104  & 0.05   & 0.10  & 0.03    & 0.10 & 0.14  & 0.21        & 0.12        & 0.09  \\      
\hline
\hline
\end{tabular}
\end{table*}

\begin{figure*}
     \centering
     \begin{subfigure}
         \centering
         \includegraphics[width=\textwidth]{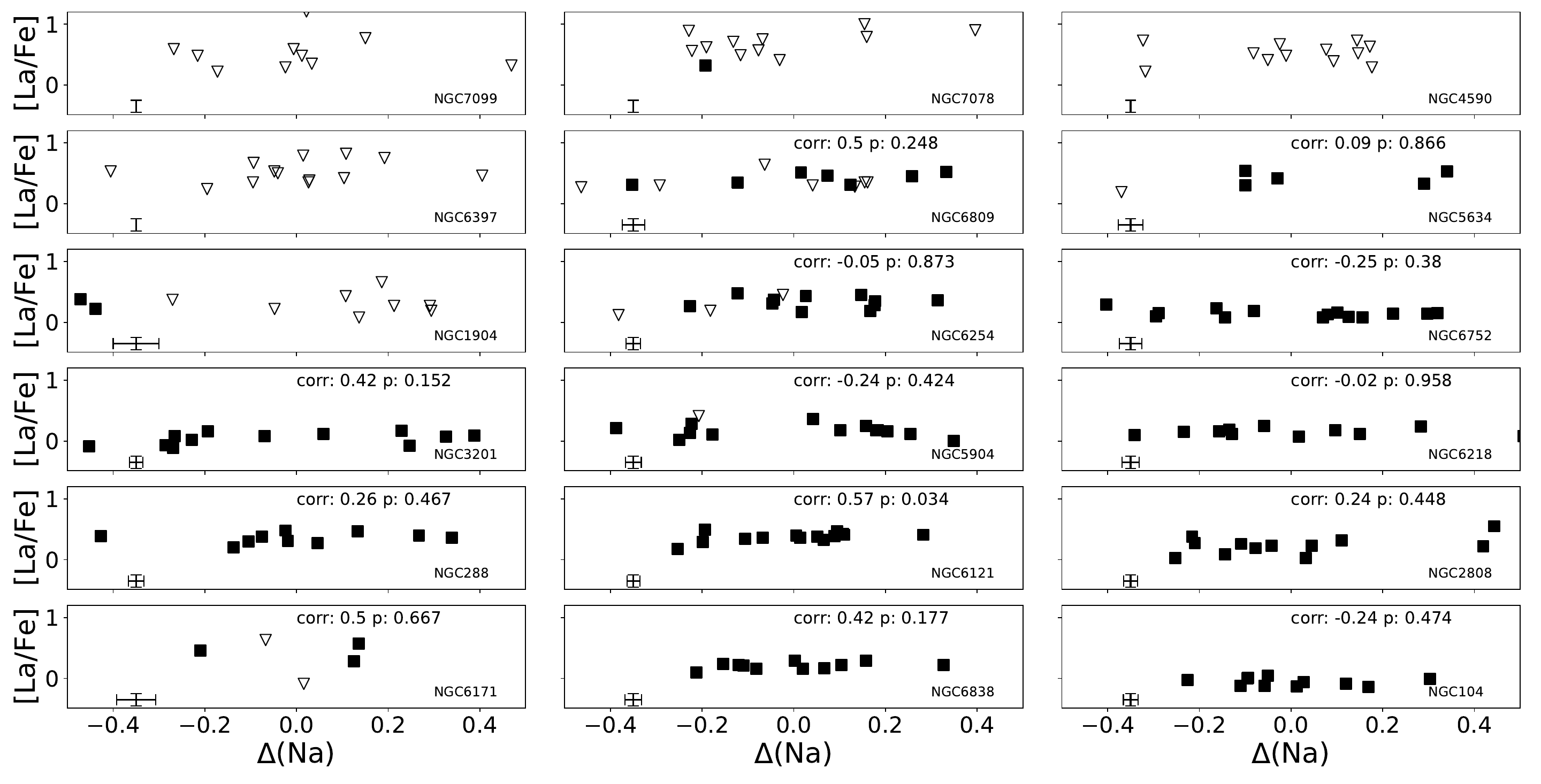}
     \end{subfigure}
     \vfill
     \begin{subfigure}
         \centering
         \includegraphics[width=\textwidth]{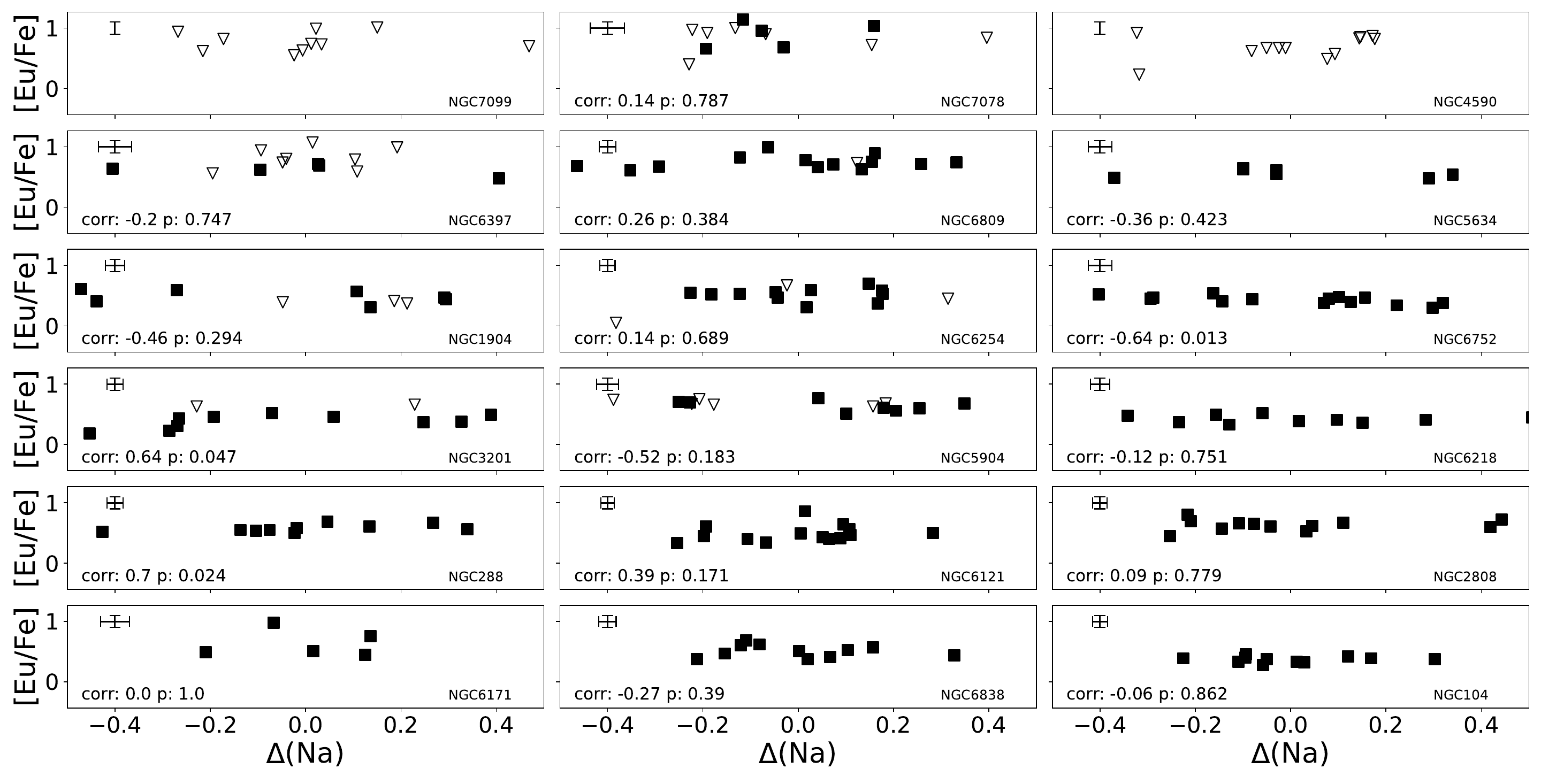}
     \end{subfigure}
     \vfill
        \caption{[La/Fe] (upper) and [Eu/Fe] (lower) as a function of $\Delta$ (Na) for each cluster of the sample. Filled squares and empty triangles represent actual measurement and upper limits, respectively.}
        \label{fig:DeltaEuLa_Ind}
\end{figure*}

Similarly, Fig. \ref{fig:DeltaEuLa_Ind} shows the results for La and Eu along with $\Delta$(Na). Although La and Eu abundances are dominated by upper limits in the more metal-poor clusters, the distribution of La and Eu does not display considerable spread. The only exceptions are the GCs NGC~7078 and NGC~6171, which display a larger Eu spread supporting the scenario of the r-process enrichment mentioned previously.
Moreover -in most clusters- the La and Eu results display a constant abundance along Na, showing the lack of correlation between these species. However, NGC~6121 showed a mildly significant correlation between La and Na. Similar results were found for NGC~3201, NGC288, and NGC~6752 for Eu and Na.

\begin{figure*}
        \centering
        \includegraphics[width=\textwidth]{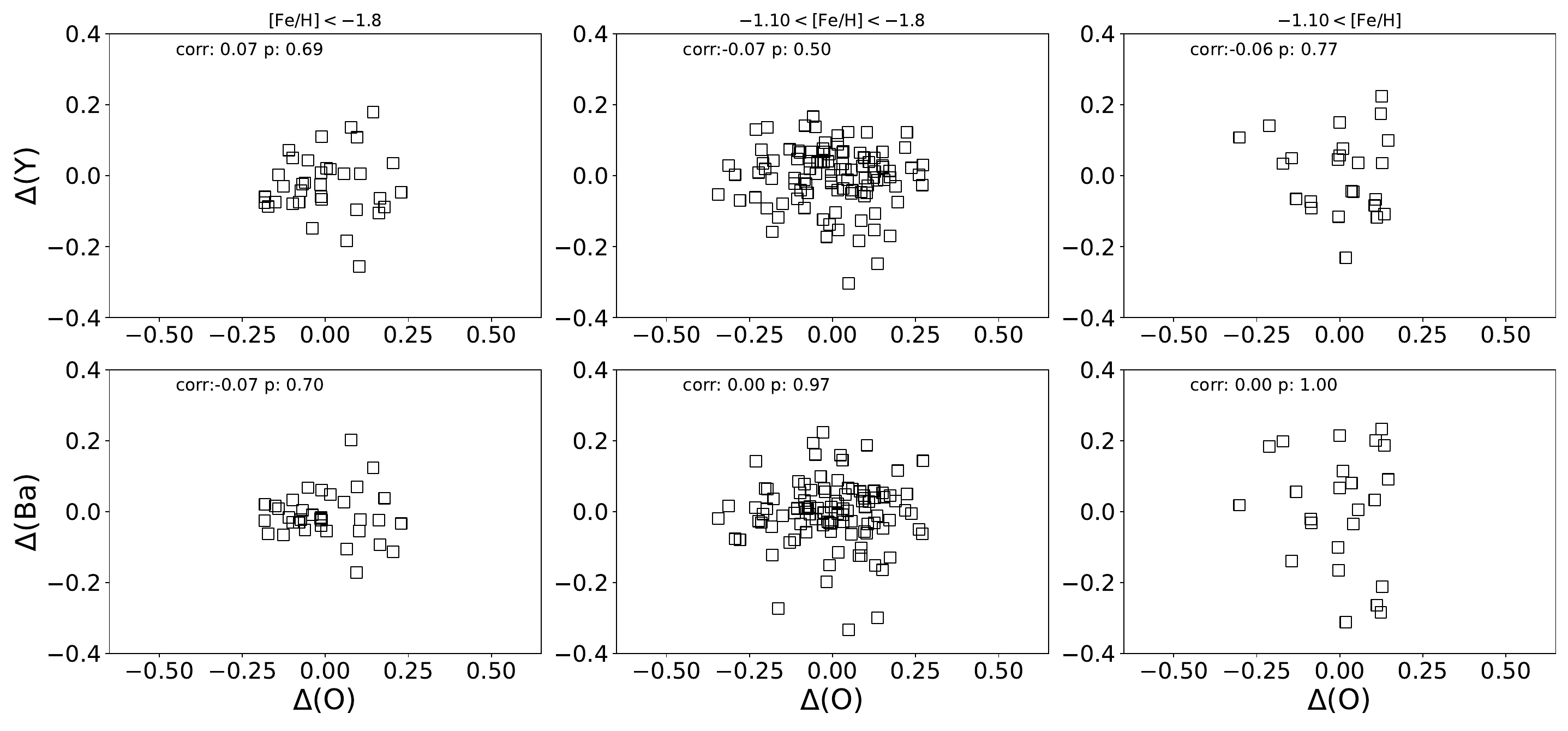}
        \caption{$\Delta$Y (upper panels) and $\Delta$Ba (lower panels) as a function of $\Delta$O for the whole survey sample (except for NGC~7078). The sample was divided into three metallicity bins: [Fe/H]<-1.80 dex (metal-poor; left panels), -1.80 dex <[Fe/H]<-1.10 dex (metal-mid; mid-panels), and [Fe/H]>-1.10 dex (metal-rich; right panels). The respective Spearman coefficient and p-value are reported on each panel.}
        \label{fig:DeltaBaYO}
\end{figure*}

\begin{figure*}
        \centering
        \includegraphics[width=\textwidth]{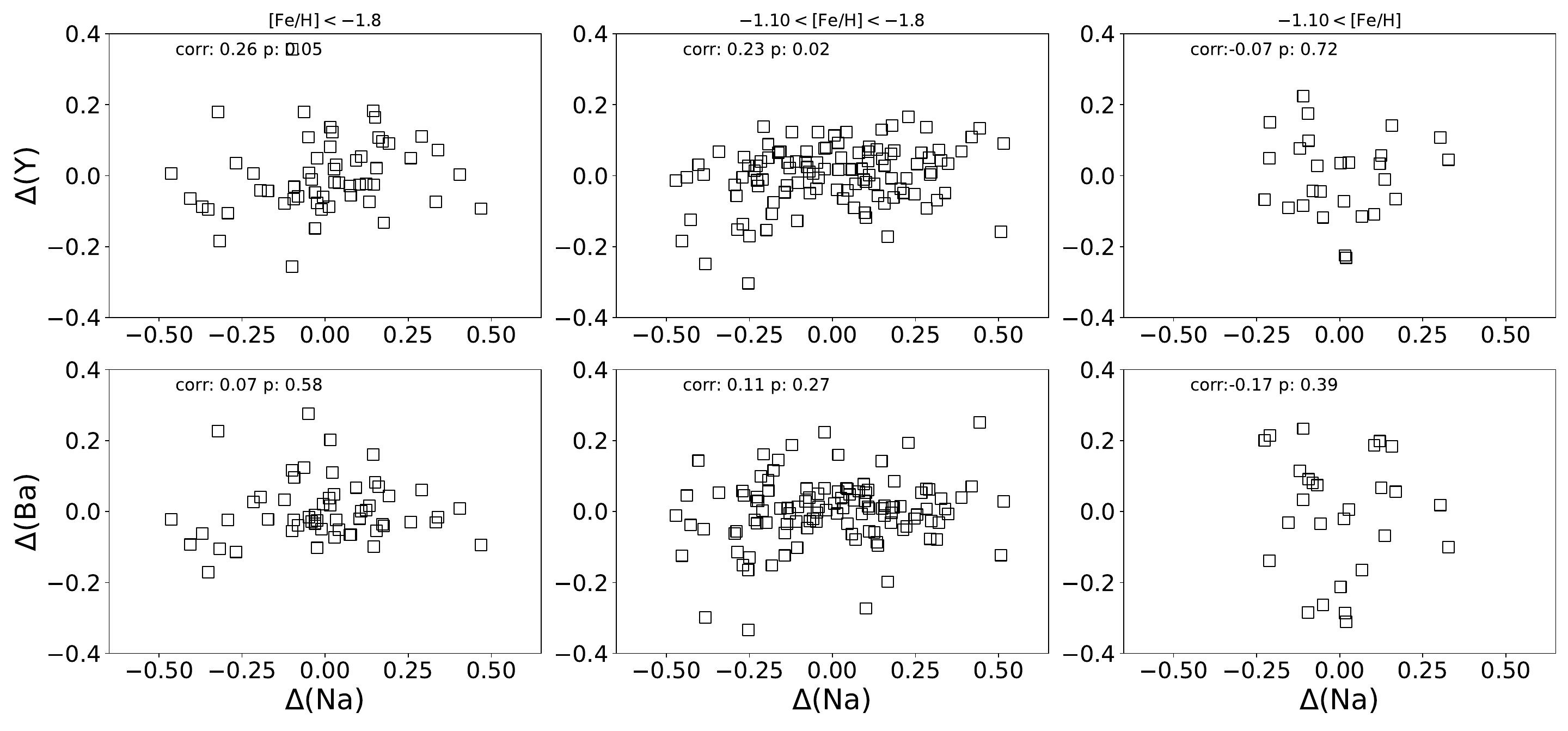}
        \caption{$\Delta$Y (upper panels) and $\Delta$Ba (lower panels) as a function of $\Delta$Na. It follows the same description as Fig.\ref{fig:DeltaBaYO}.}
        \label{fig:DeltaBaYNa}
\end{figure*}

\begin{figure*}
        \centering
        \includegraphics[width=\textwidth]{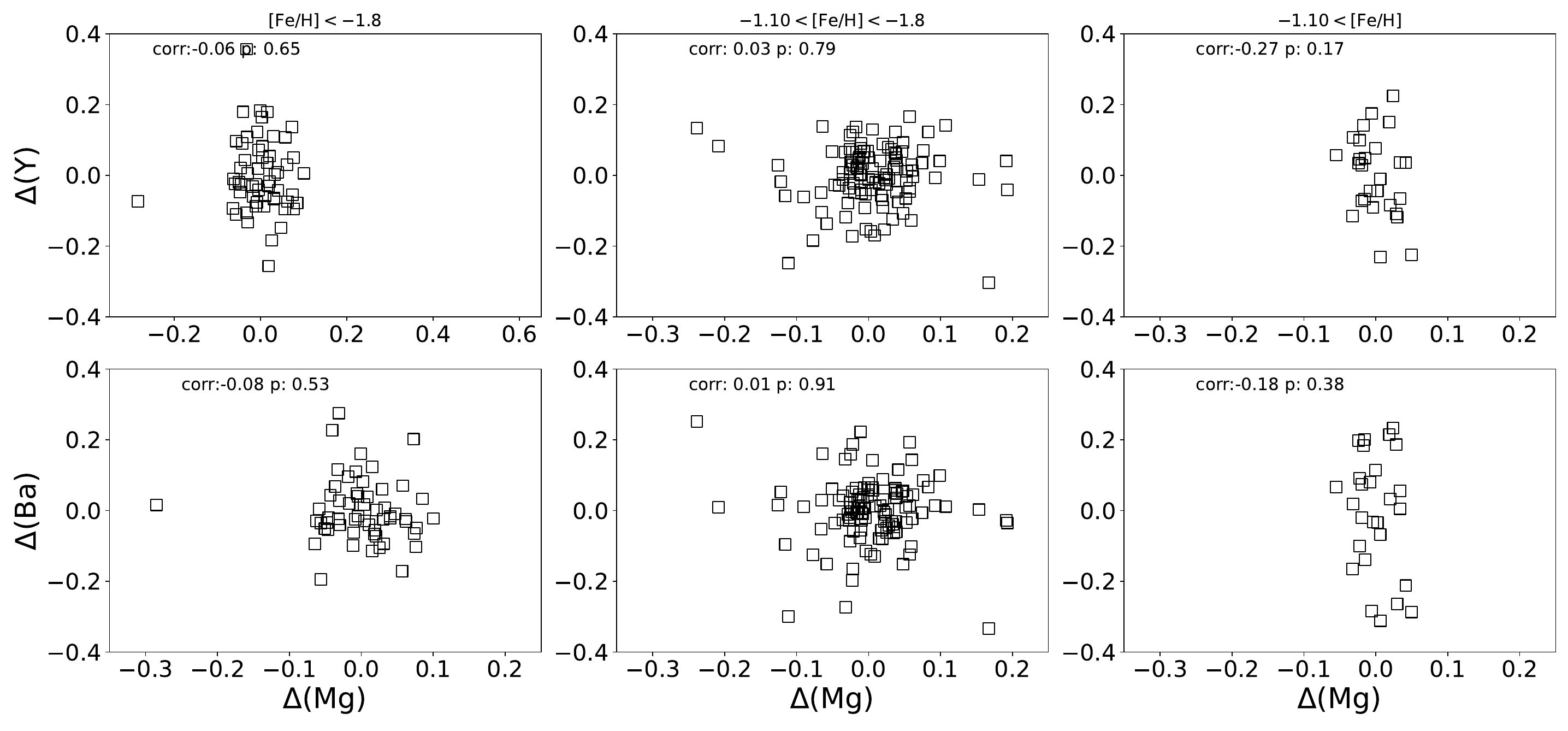}
        \caption{$\Delta$Y (upper panels) and $\Delta$Ba (lower panels) as a function of $\Delta$Mg. It follows the same description as Fig.\ref{fig:DeltaBaYO}.}
        \label{fig:DeltaBaYMg}
\end{figure*}

The trends were then examined on the combined sample, that is to say, on all the stars analysed in the present work, separated into groups according to their overall metallicity.
To do so, Figs. \ref{fig:DeltaBaYO},\ref{fig:DeltaBaYNa}, and \ref{fig:DeltaBaYMg} show the $\Delta$(Y) (upper row) and $\Delta$(Ba) (lower row) as a function of $\Delta$(O), $\Delta$(Na), and $\Delta$(Mg), respectively. This exercise aims to probe the variation of s-process elements along with the O, Na, and Mg abundance. Therefore, NGC~7078, known to display n-capture element spread attributable to the r-process, was excluded from the combined sample. The panels display the distribution for three metallicity bins: [Fe/H]<-1.80 dex (metal-poor; left panels), -1.80 dex <[Fe/H]<-1.10 dex (metal-mid; mid-panels), and [Fe/H]>-1.10 dex (metal-rich; right panels). Each figure indicates the corresponding Spearman coefficient and p-value for each metallicity bin. All the panels show quite flat distributions and weak correlations, which is valid for the whole sample and each metallicity bin. 

However, it is worth noticing that for the mid-metallicity regime (mid-panels), there is a mildly significant correlation between Y and Na. The correlation is similar in the low metallicity bin; however, its significance is lower than in the mid-metallicity regime, and in the high one, it disappears entirely.
We note, however, that in case of an actual correlation between those abundances, such a metallicity regime should be the most suitable one to detect it. In fact, in this regime, the lines are strong enough to be scarcely affected by noise but weak enough to have to be weakly affected by the \vm so that a linear fit can appropriately address its contribution. This correlation in the mid-metallicity regime would be translated in a Y increment of about 0.01 dex for each 0.1 dex increment of Na.

\subsection{Heavy elements distribution}

\begin{figure*}
        \centering
        \includegraphics[width=\textwidth]{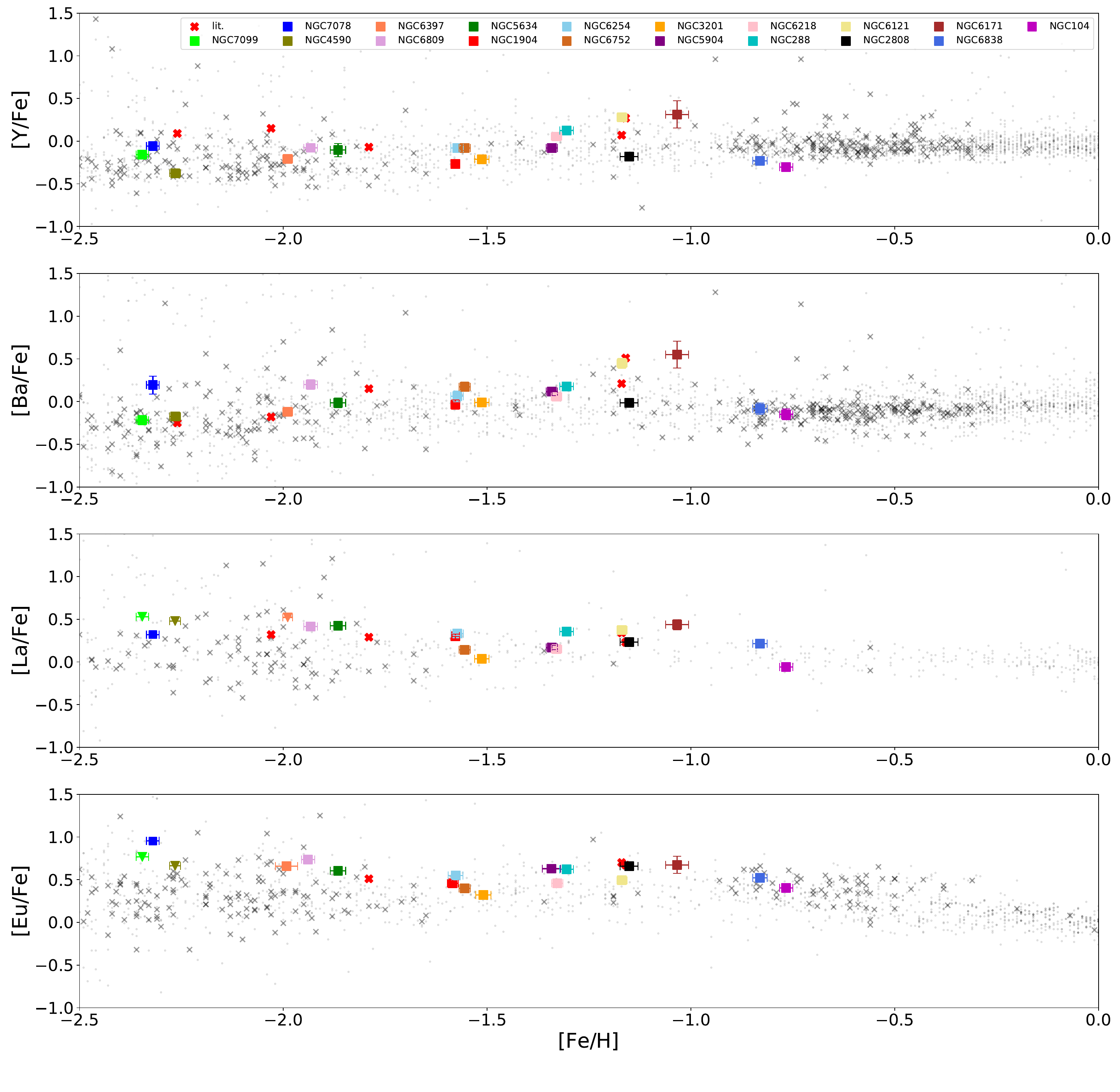}
        \caption{From the top to the bottom: [Y/Fe], [Ba/Fe], [La/Fe], and [Eu/Fe] as a function of the [Fe/H] for the whole sample. Coloured squares represent the GCs analysed in the present sample. Red-filled crosses represent GCs abundance from the literature as in Fig.\ref{fig:CuFe_Fe}. Grey dots show the field star abundances, and grey crosses represent bonafide Halo field stars ([Mg/Fe]>0.2 dex) taken from SAGA Database \citep{Suda2008}.}
        \label{fig:YBaEu_Fe}
\end{figure*}

Aiming to look at the overall content in n-capture elements, a comparison of the heavy elements analysed for the sample of GCs and the galactic field was performed. The Fig.\ref{fig:YBaEu_Fe} shows, from the top to the bottom, the results obtained for [Y/Fe], [Ba/Fe], [La/Fe], and [Eu/Fe] along with the [Fe/H]. The field star distribution (grey crosses) was taken from SAGA Database\footnote{Data compilation of Galactic abundances, including the vast majority of literature up to 2019 and composition studies from a large sample from 2019 to today: \url{http://sagadatabase.jp/}} \citet{Suda2008}. Each GC is represented with a different colour. Squares and triangles are actual measurements and upper limits, respectively. As was done in Fig. \ref{fig:CuFe_Fe}, we included --when possible-- literature results (in red crosses) from \citet{ngc1851} (NGC~1851), \citet{ngc362} (NGC~362), \citet{Terzan8} (Terzan~8), \citet{ngc4833} (NGC~4833), and \citet{ngc6093} (NGC~6093). In addition, Table \ref{tab:mean_spread}  displays the mean, spread, and the number of stars used to get the actual abundance for each element.

Field stars show a yttrium distribution, which increases with the metallicity having Y abundances ranging from $\sim$-0.60 dex at low metallicities up to solar abundances at high metallicities. In the upper panel, most of the GCs analysed follow closely the trend displayed by field stars at the correspondent metallicity. NGC~6121 and NGC~6171 are the only exceptions showing larger Y abundances than the field star counterparts.

Barium, at solar metallicity, has mostly an s-process origin \citep[85\%][]{Sneden2008}. Ba shows similar behaviour to Y along with [Fe/H]; however, the former displays slightly lower abundances than Y at [Fe/H]<-1.5 dex. In the second panel, similar to Y results, Ba abundances in almost all the GCs analysed follow the field stars trend. The GCs NGC~6121, NGC~6171, and NGC~7078 display higher abundance than expected for stars at that metallicity. 

Field stars display a lanthanum distribution slightly super-solar at [Fe/H]<-1 dex, which becomes solar for richer metallicities. The third panel shows that the GCs surveyed fit the field stars trend. It is worth noticing that only the upper limit was set for the metal-poor GCs NGC~7099, NGC~4590, and NGC~6397 because the La lines became too weak. For NGC~7078, La abundance was determined in only one star, so the result should be taken cautiously. 

Europium is known to be a pure r-process element \citep[97\% at solar metallicity][]{Simmerer2004}. In the lowest panel, the Eu distribution in the field displays a quite constant over-abundant at about [Fe/H]<-0.7 dex, which constantly decreases toward higher metallicities, showing the iron production by SN Ia after 0.1-1.0 Gyr, which agrees with both observations and models \citep{Cescutti2006}. All the GCs analysed seem to follow closely the upper envelope of the distribution drawn by the field stars. It is worth noticing, that in the GCs NGC~4590 and NGC~7099, the Eu detection was not possible. Moreover, the GC NGC~7078 displays a slight Eu over-abundance with respect to the field stars at the same metallicity.

In general, most of the surveyed GCs closely follow the field distribution\footnote{We remind the reader that SAGA is a compilation of data from different sources. While these results are scaled to our solar abundances, there are still some differences, such as in the method used for the abundance determination, model atmosphere, $\log$ gf values, etc.}, meaning there is no evidence of a peculiar n-capture enrichment. Our results are in good agreement with literature GCs of similar metallicity. 

On the other hand, Table \ref{tab:rms_deltas_IQR} reports the IQRs of [Y/Fe], [Ba/Fe], [La/Fe], and [Eu/Fe]. Upper limits were not considered in the IQR computation for La and Eu. It is worth noticing that off-the-trend GCs display (NGC~7078, NGC~6171, and NGC~6121) also a larger internal dispersion. NGC~7078 has been reported as GCs with the largest spread in both Ba and Eu. The present analysis reports a [Ba/Fe] abundance ranging from -0.29 dex to 1.02 dex. Previous studies have reported a difference of $\sim$0.45 dex \citep{Otsuki2006} and $\sim$0.55 dex \citep{Sobeck2011}. The larger Ba spread found in the present analysis can be related to the larger \vm~ range compared to the cited articles. For comparison, when the Ba intrinsic spread (without considering the effect of \vm) is considered, it decreases to $\sim$0.80 dex. In a larger sample of 63 stars \citet{Worley2013} reported bimodal distribution for both Ba and Eu, finding a difference of up to 1.25 dex for the first one and about 0.80 dex for the second one. In the case of the present article, the [Eu/Fe] difference is at least 0.59 dex (upper limits could enlarge this difference), which is similar to the difference reported by \citet{Otsuki2006} ($\sim$0.55 dex) and \citet{Sobeck2011} (0.57 dex) in their sample of 3 RGB stars. The large dispersion reported in both Ba and Eu, presented in our results and the literature, agrees with a peculiar r-process element enrichment. 

On the other hand, NGC~6171 displays a large IQR in all the n-capture elements measured. \citet{OConnell2011} analysed the La and Eu abundances in 13 stars of the cluster, which showed a good agreement with the present article ($\langle$[La/Fe]$\rangle$=0.41$\pm$0.12 and $\langle$[Eu/Fe]$\rangle$=0.73$\pm$0.13). Moreover, they reported a large difference in the Eu ($\sim$0.50) and La ($\sim$0.40) content in their sample, which agrees with the large IQR mentioned before arguing in favour of an early r-process enrichment. Finally, the GC NGC~6121 was found to show a Y bimodal distribution \citep{Villanova2011}, which was later challenged by \citet{Dorazi2013}, whose results are consistent with the present analysis. The cluster was found to display an intrinsic high s-process enrichment due to a particular higher concentration of these species in the protocluster cloud \citep{Yong2008b}, which agrees with the [Y/Fe]=0.44 dex and [Ba/Fe]=0.50 dex found by \citep{Dorazi2013} and \citep{dorazi2010_neutron}, respectively. Moreover, the La (0.48 dex) and Eu (0.40 dex) results from \citep{Yong2008b} are in good agreement with the ones presented here. 

Further discussion of internal spread is beyond the aims of the present paper and will be addressed in an upcoming work, which is currently in preparation.

\subsection{[Ba/Eu] and [Ba/Y] ratios}

\begin{figure*}
        \centering
        \includegraphics[width=\textwidth]{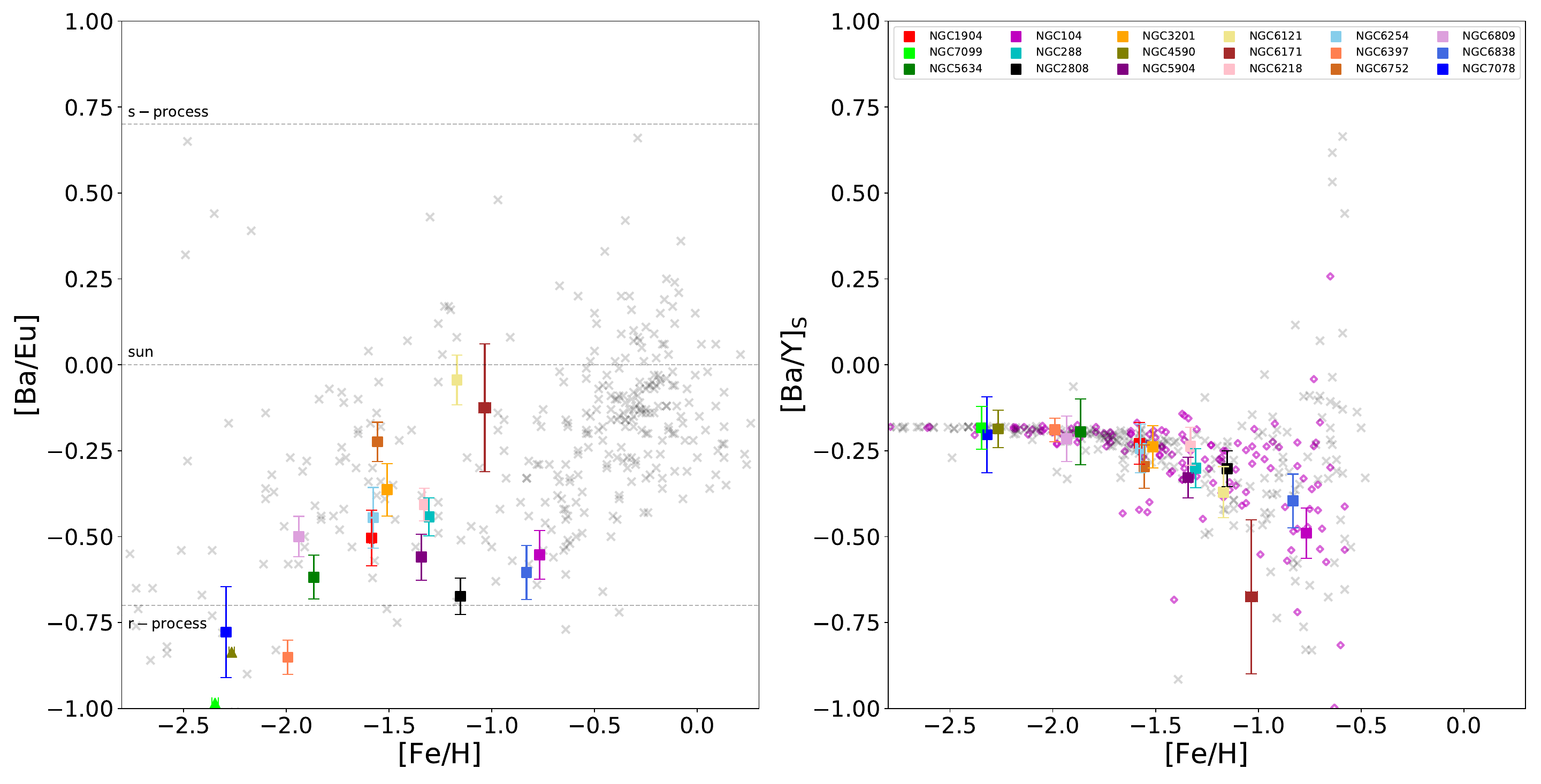}
        \caption{Left and right panels display the abundance [Ba/Eu] and [Ba/Y]$_s$ ratios as a function of [Fe/H], respectively. Dashed lines at [Ba/Eu] 0.70 dex and -0.70 dex indicate the ratio for a full s-process and full r-process enrichment. The dashed line at [Ba/Eu] 0.00 dex displays the solar ratio. Magenta diamonds in the right-hand panel show dwarf galaxies results taken from \citet{Suda2008}. Other symbols and colours follow the description given in Fig. \ref{fig:YBaEu_Fe}.}
        \label{fig:BaEu_Fe}
\end{figure*}

Figure \ref{fig:BaEu_Fe} shows the ratio between the Ba and Y (right panel) and Ba over Eu (left panel) as a function of [Fe/H]. The ratio of these elements can provide means to disentangle the contribution of the r- and s-process to the heavy element content in the cluster. The symbols follow the same description as the previous figures. In addition, on the right-hand panel, we included in magenta diamonds dwarf galaxies results from \citet{Suda2008} to compare their behaviour and the one for GCs.

The [Ba/Eu] distribution as a function of [Fe/H] provides insight into the process by which our Galaxy was enriched. The dotted horizontal lines at [Ba/Eu] -0.70 dex and 0.70 dex, reflect a pure enrichment from r-process and s-process species, respectively. The [Ba/Eu] pattern followed by the field stars goes from a pure r-process enrichment at low metallicities to a continuous contribution of s-process at solar metallicity. Although there are GCs with similar metallicities but discrepant [Ba/Eu] (e.g., NGC~2808 and NGC~6121), the results for most of the GCs display a similar behaviour as field stars. In addition, if [Fe/H] is considered a proxy of time --with more metal-poor stars being older than the ones with higher metallicity-- it is possible to see the rise of the s-process elements along the time. The results are compatible with pure r-process abundances for the more metal-poor cluster, meaning that their abundances are influenced by explosive events like SNe type II or merging neutron stars. As field stars, in GCs, the contribution of the s-process enrichment increases with metallicity; however, it remains dominated by the r-process. It is worth noticing that for the GCs NGC~4590 and NGC~7099, we reported lower limits for the [Ba/Eu] ratios.

In s-process production, Y and Ba are part of the first and second peaks of s-process elements, respectively. Consequently, their ratio investigates the contribution of l$_s$ and h$_s$ elements. In the case of AGB stars, their nucleosynthesis is linked to the stars' mass and metallicity. Specifically, the [h$_s$/l$_s$] ratio tends to decrease as the star mass increases, which can vary depending on the star's metallicity. Nevertheless, when dealing with low metallicities, such as those observed in metal-poor globular clusters (GCs), the r-process contributes to the synthesis of Ba and Y. This complicates the direct use of the [Ba/Y] ratio in this scenario. To address the r-process contribution of these elements, we adopted values for the r-process contribution to Ba and Y from the solar system r-pattern derived by \citep{Simmerer2004}. These values are scaled to align with the europium (Eu) abundance measured in both GC and field stars. The notation [Ba/Y]$_s$ denotes the Ba over Y ratio, considering only the contribution from the s-process. The results are shown in the right panel of Fig.\ref{fig:BaEu_Fe}, which reveals that the [Ba/Y]$_s$ ratio in the GC sample remains constant at low metallicity, but decreases at metallicities higher than -1.5 dex. The decrement is also seen in a fraction of field and dwarf galaxy stars; however, the behaviour of GCs regarding the [Ba/Y]$_s$ ratio seems better defined. The increase [Fe/H] prompts a shift in the s-process pattern of GC towards Y instead of Ba. This shift, from the h$_s$ to the l$_s$ elements, suggests an augmented contribution from lower-mass AGB stars at later stages of the Galaxy.

\subsection{Clusters Comparison: cluster-to-cluster difference}

\begin{figure*}
    \centering
     \begin{subfigure}
         \centering
         \includegraphics[width=\textwidth]{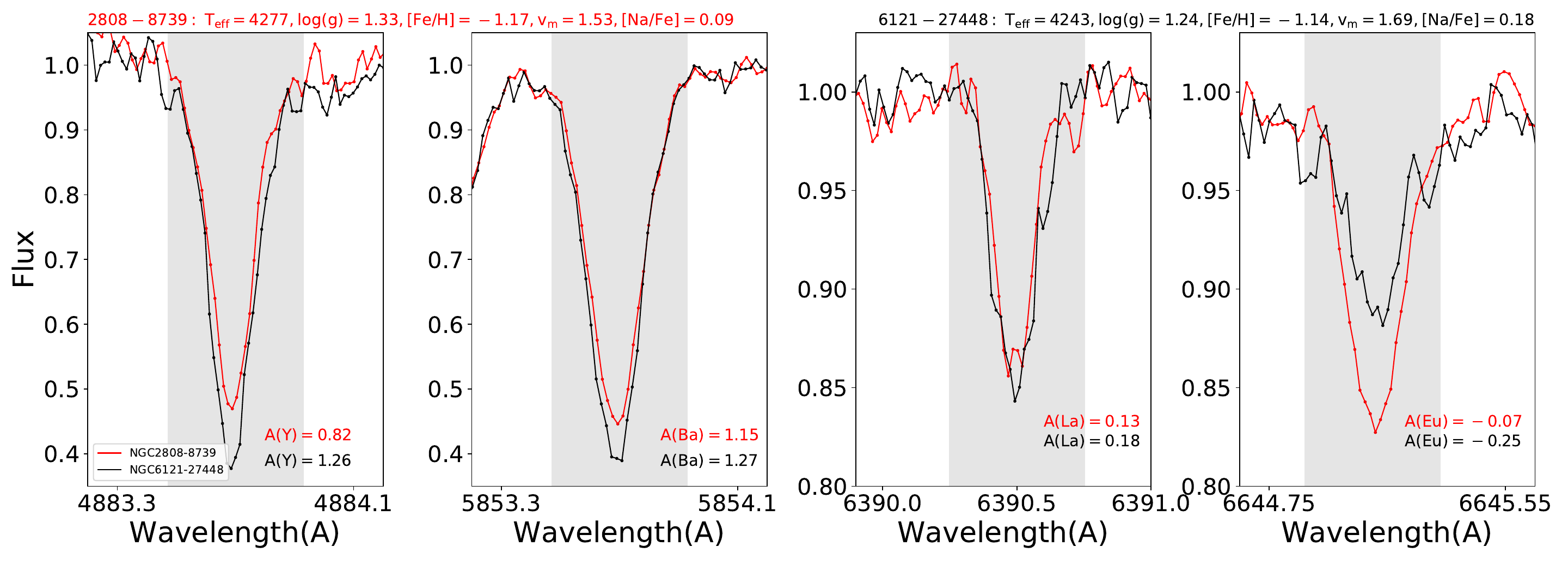}
     \end{subfigure}
     \vfill
     \begin{subfigure}
         \centering
         \includegraphics[width=\textwidth]{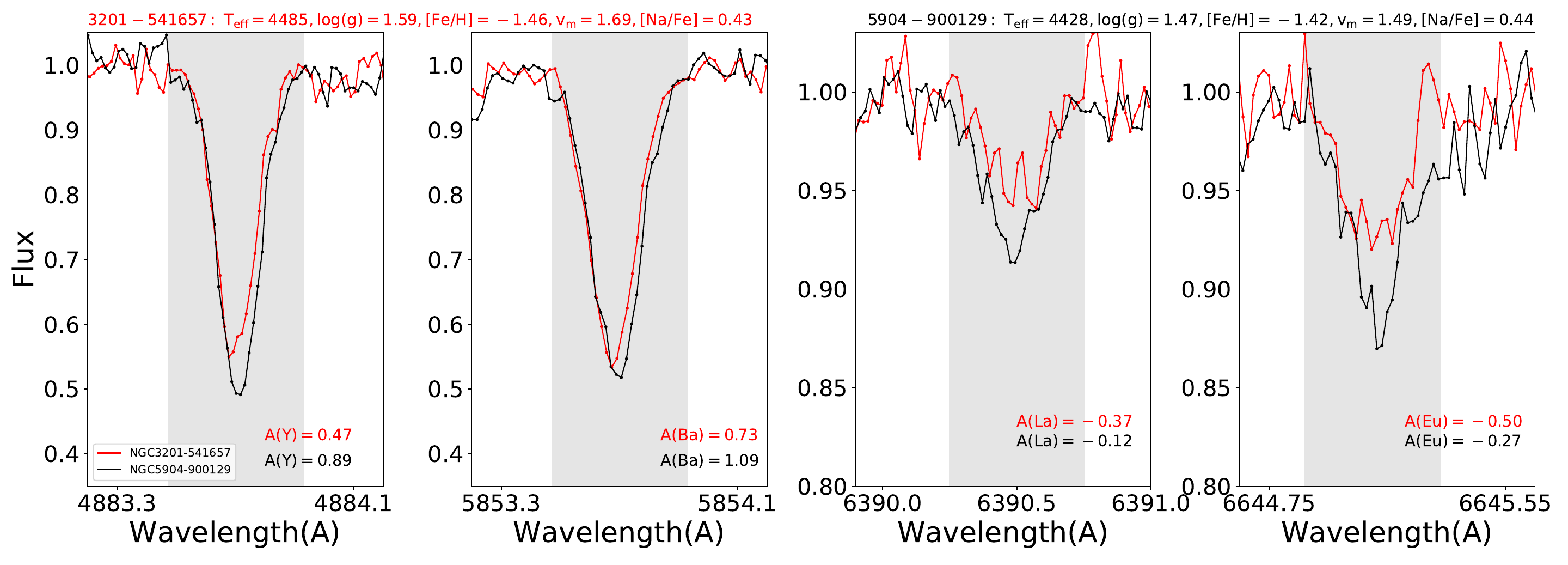}
     \end{subfigure}
   \vfill
        \caption{Pair of stars with similar stellar parameters and Na abundances as was reported by \citet{Carretta2009u}, but different n-capture abundances. From left to right: Y line at 4883\AA, Ba line at 5853\AA, La line at 6391\AA, and Eu line at 6645\AA.}
        \label{fig:comp_lines}
\end{figure*}

To assess the discrepant Ba/Eu ratios among GCs with similar metallicities, we have compared pairs of stars with similar stellar parameters in different clusters. The comparison is shown in Fig. \ref{fig:comp_lines}. The pairs also share similar Na abundance as reported by \citet{Carretta2009u}. In the first row of the figure, the comparison between the stars of NGC~6121 (ID=27448) and NGC~2808 (ID=8739), two clusters with similar metallicities ([Fe/H]$\sim$-1.2 dex), but quite different n-capture abundance. As the spectra comparison shows, there is higher abundance in their s-process elements (Y, Ba and La), however, this behaviour changes for the r-process elements. Because the stars have only slightly different \vm, its effect cannot explain such a difference in abundance. This comparison suggests that the large difference($\sim$0.70 dex) shown in Fig. \ref{fig:BaEu_Fe} is real, meaning the NGC~6121 has a higher enrichment of s-process elements than NGC~2808 and the latter has a higher r-process enrichment. The second row compares a star pair, in GCs NGC~3201 (ID=541657) and NGC~5904 (ID=900129). The two stars with similar stellar parameters and Na abundance show a systematic overabundance in favour of the second one for all the elements analysed, suggesting an overall different n-capture enrichment, but still slightly more shifted to the r-process.

\subsection{Comparison with chemical evolution models}

\begin{figure*}
    \centering
     \begin{subfigure}
         \centering
         \includegraphics[width=0.49\textwidth]{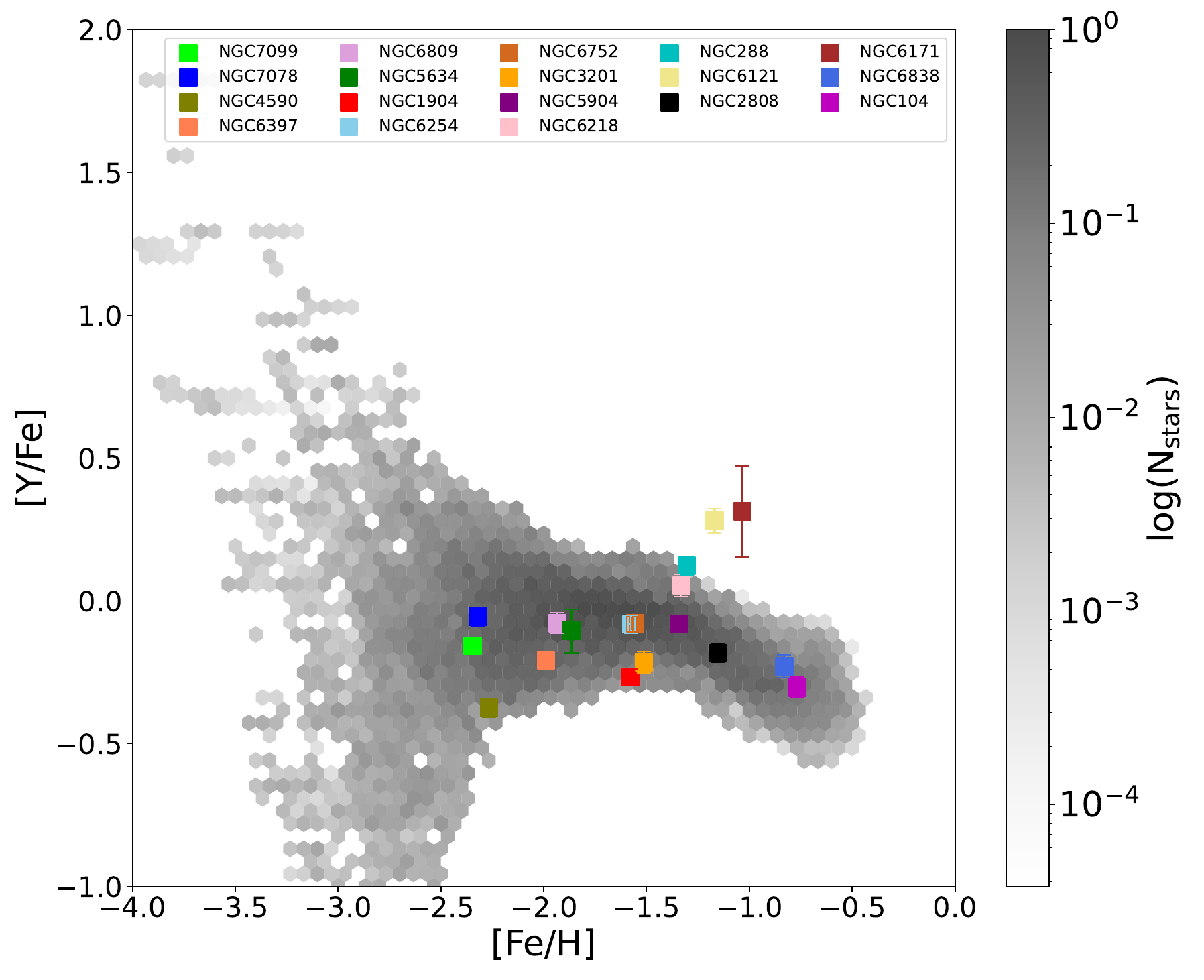}
     \end{subfigure}
     \hfill
     \begin{subfigure}
         \centering
         \includegraphics[width=0.49\textwidth]{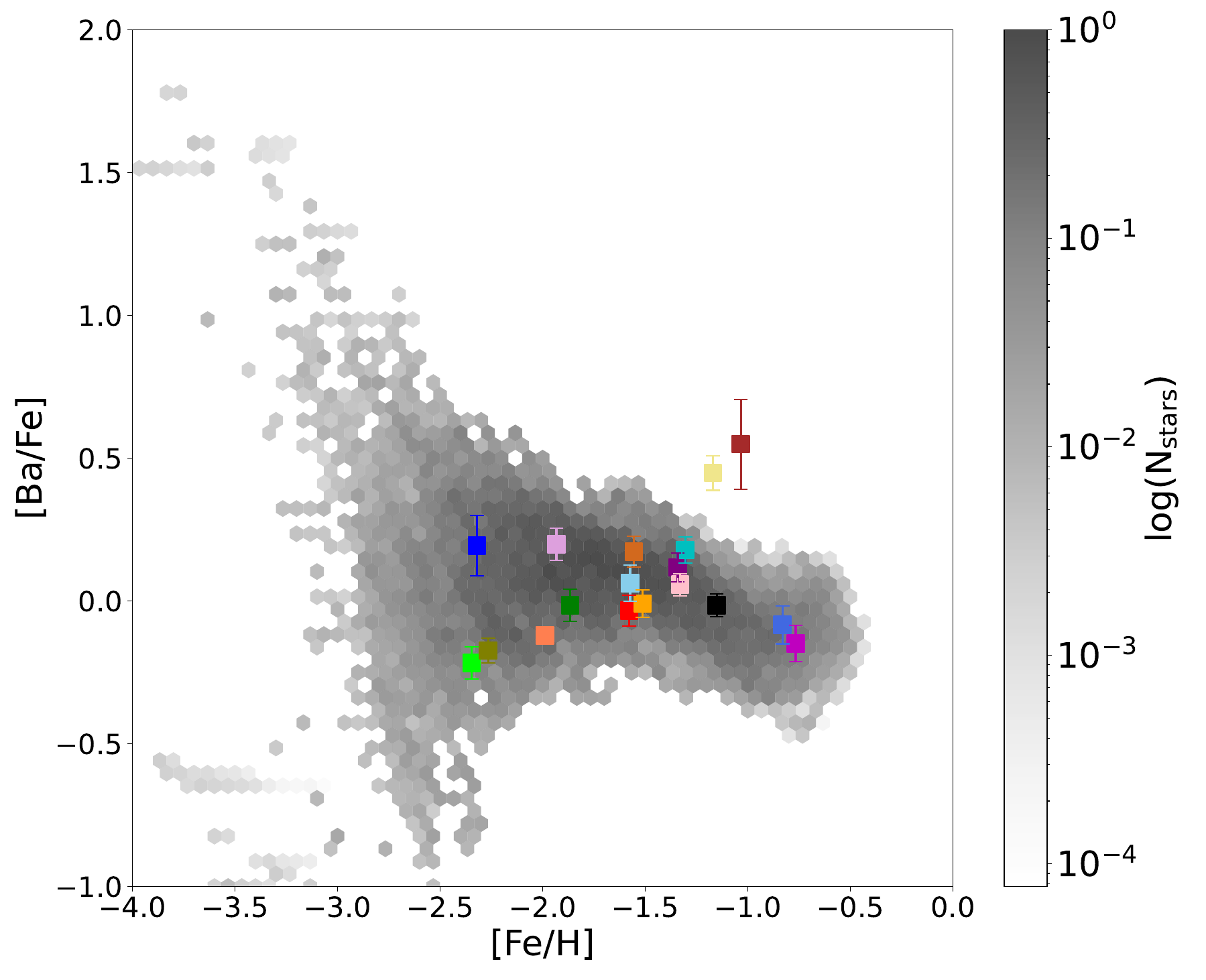}
     \end{subfigure}
   \vfill
     \begin{subfigure}
         \centering
         \includegraphics[width=0.49\textwidth]{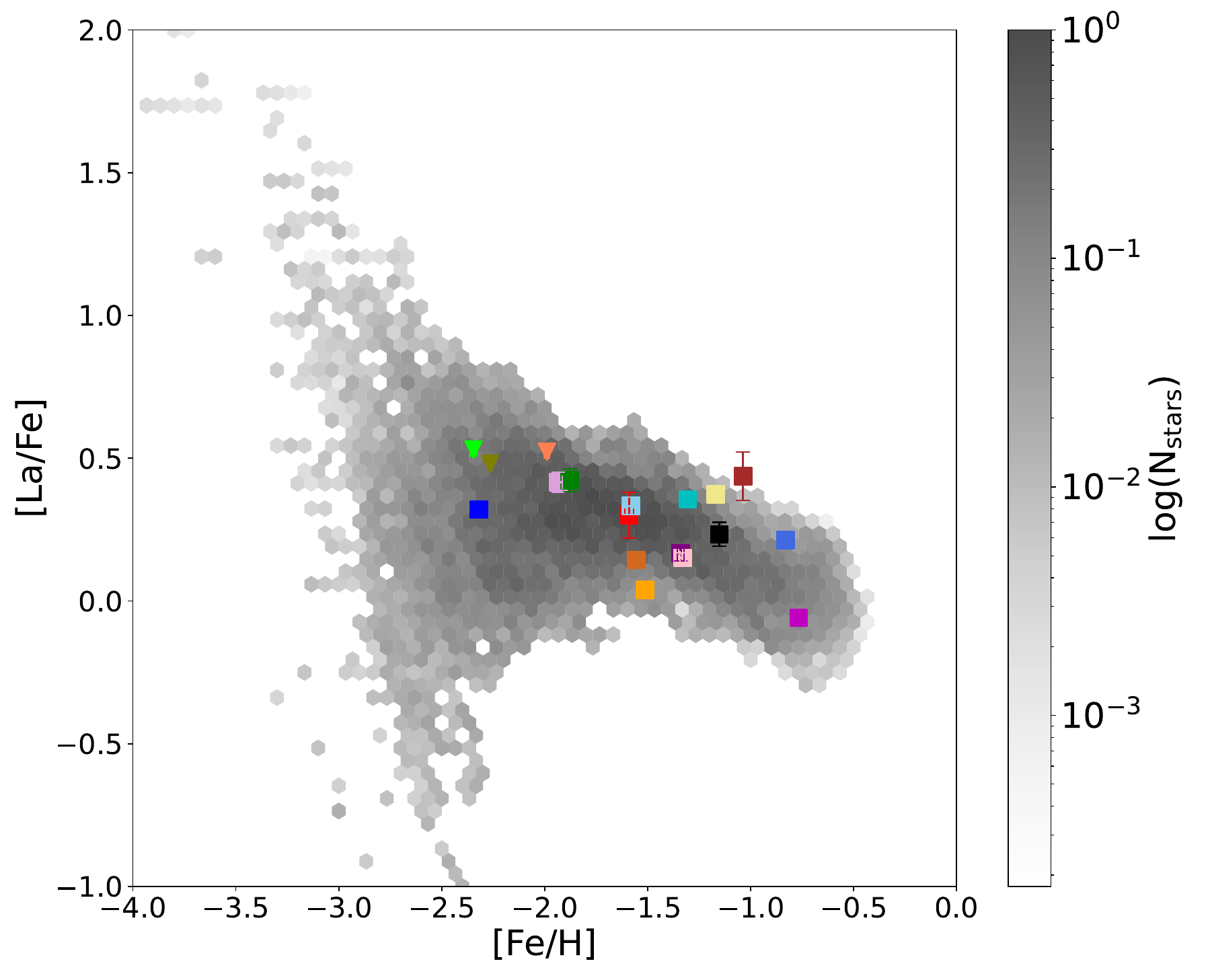}
     \end{subfigure}
   \hfill
     \begin{subfigure}
         \centering
         \includegraphics[width=0.49\textwidth]{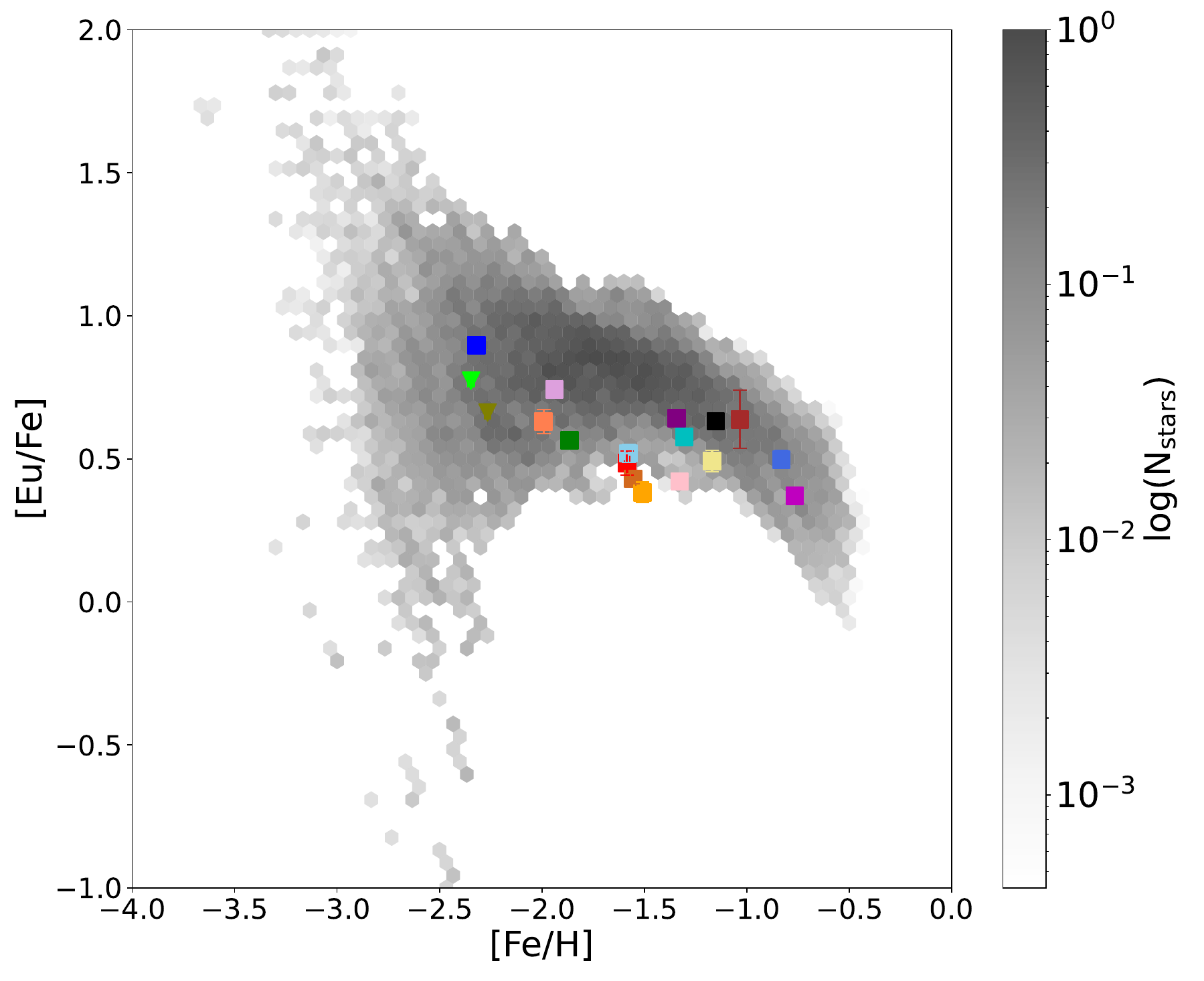}
     \end{subfigure}
   \caption{Comparison between the n-capture element abundances measured in the present work and modelled abundances from \citet{Cescutti2014} coloured-coded by the logarithmic number of stars in each bin.}
        \label{fig:models}
\end{figure*}

As mentioned in Sec. \ref{introduction}, the main nucleosynthetic sites for the s- and r-processes are mainly AGB stars -- with some contribution of FRMS-- and neutron star mergers and magneto-rotational driven (MRD) SNe, respectively. \citet{Cescutti2014} proposed a model for the chemical enrichment of the halo considering different sources of heavy elements (for details about the model, we refer the reader to the cited article). In particular, they tested the models with electron capture (EC) SNe or/and MRD SNe with/without an early enrichment of s-process elements from FRMS. According to \citet{Cescutti2014}, to better reproduce the observed n-capture element distribution in the Galactic halo, the model should take into account a mix of pollution coming from FRMS and MRD for the s- and r-process enrichment, respectively. Fig. \ref{fig:models} shows the comparison of our results (coloured symbols) and the predictions for Y, Ba, La, and Eu from the mentioned model. The figure is coloured by $\log_{stars}$, which reflects the probability of finding a long-living star. As shown in the present figure, the model closely reproduces the observations except for the [Y/Fe] and [Ba/Fe] abundances in NGC~6121 and NGC~6171, which --as was discussed previously-- show a particularly high content in these elements. In addition, although those clusters show a La abundance in good agreement with the models, it is worth noticing that they are placed in the upper envelope of the model's distribution.

As commented by \citet{Cescutti2014}, both r-process sources analysed in their models (EC SNe and MRD) reproduced quite well the halo distribution of Eu, showing good agreement among these sources in the most metal-poor regime ([Fe/H]<-2 dex), however with some slight discrepancies at intermediate metallicities (-2 dex <[Fe/H]<-1 dex). The model used in comparison with our results reflects good agreement with the metallicity of our sample being no discrepant with the MRD + FRMS scenario. We hope in the future, with the present and other observational constraints, we could shed light on the contributor sources of n-capture elements of the halo.

\section{In-situ and Ex-situ GCs}
\label{Sec:Origin}

Different authors have tried to determine the origin of GCs and related it to both their dynamical and chemical features. We used the results published by \citet{Massari2019} to identify potential relations with the n-capture element abundances obtained in the present article. Among the GCs in our sample, there are 16 analysed by \citet{Massari2019} for which 7 (NGC~104, NGC~6171, NGC~6218, NGC~6397, NGC~6752, NGC~6838, and NGC~7078) were identified as in-situ, 6 (NGC~288, NGC~2808, NGC~3201, NGC~4590, NGC~5904, and NGC~7099) as ex-situ (most associated with the Gaia-Enceladus Stream), and 3 GCs (NGC~6121, NGC~6809, and NGC~6254) with uncertain origin.

We analysed if in-situ and ex-situ GCs behave in different fashions. All the GCs identified as in-situ showed slightly higher Y and Ba abundances with respect to the ex-situ GCs; moreover, the latter displayed smaller spreads than the first ones in all the n-capture analyses. The mean and standard deviation for in-situ (ex-situ) GCs are $\overline{[Y/Fe]}$=-0.07 and $\sigma$([Y/Fe])=0.21 ($\overline{[Y/Fe]}$=-0.15 and $\sigma$([Y/Fe])=0.16), $\overline{[Ba/Fe]}$=0.09 and $\sigma$([Ba/Fe])=0.24 ($\overline{[Ba/Fe]}$=-0.02 and $\sigma$([Ba/Fe])=0.16), $\overline{[La/Fe]}$=0.20 and $\sigma$([La/Fe])=0.17 ($\overline{[La/Fe]}$=0.20 and $\sigma$([La/Fe])=0.13), and $\overline{[Eu/Fe]}$=0.56 and $\sigma$([Eu/Fe])=0.18 ($\overline{[Eu/Fe]}$=0.56 and $\sigma$([Eu/Fe])=0.12) dex. Note that GCs with upper limits were not considered in that calculation. 

Firstly, we investigated whether these slight discrepancies in the average abundances were linked to differences in metallicity, age, or \vm. We found that for Ba and Y, there seems to be a trend with metallicity, as the most metal-rich ones have systematically higher abundances of these elements. As each cluster spans approximately the same evolutionary range, the Ba and Y transitions in metal-rich clusters are more likely to be saturated, and the measurement of abundances from them might be affected by systematic errors. Nevertheless, it is worth noticing that this effect would not affect the internal spread of each cluster and --in any case-- the overall pattern as a function of metallicity is in good agreement with the field. Secondly, because s-process production changes over time, we investigated any potential relation with the cluster age. To this aim, we used the results determined by \citet{VandenBerg2013}, however, there were not any trends for GCs with different origins. Finally, after analysing the maximum difference of \vm~ within each cluster, we note that this difference seems quite homogeneous along the clusters (with only two outliers), meaning that \vm~ affects all these clusters similarly. Then, this behaviour is not caused by different ranges in \vm. Nevertheless, a Student's t-test comparing the mean abundances of these two groups showed that their differences were not significant. Therefore, there is no evidence of different chemical evolution among them.

\section{Chemical abundances and cluster mass}
\label{Sec:Cl_Mass}

\begin{figure*}
        \centering
        \includegraphics[width=\textwidth]{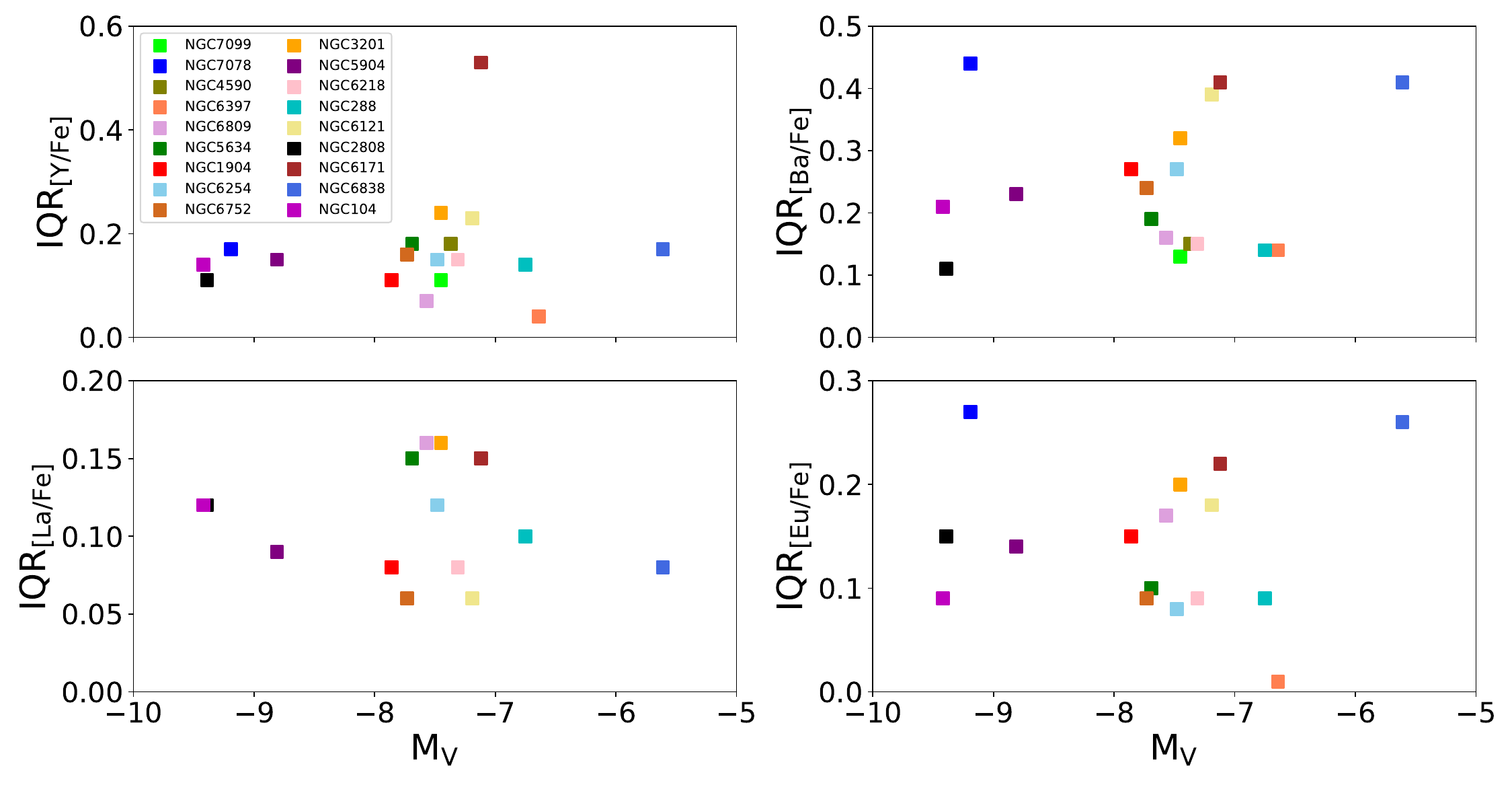}
        \caption{IQRs for Y, Ba, La, and Eu abundances as a function of the absolute visual magnitude for the whole sample. The M$_{V}$ were taken from the Harris Catalogue \citep{Harris2010}.}
        \label{fig:YBaLaEu_MVt}
\end{figure*}

Several studies have compared the abundance patterns of GCs with global properties such as cluster mass. Using part of the sample presented here, \citet{Carretta2009u} related the Mg-Al anti-correlation with the mass and the metallicity of the GCs, which was later confirmed by \citet{Pancino2017}. Similarly, \citet{Masseron2019} analysed a sample of 885 GC stars and found evidence of a correlation of the Al spread present in GCs with the cluster mass. The latter suggested that the Mg-Al reaction decreases its importance in more massive GCs. It is interesting to perform a similar analysis using the n-capture element abundances. A comparison was performed against the absolute magnitude (M$_{V}$), a proxy for the cluster mass. The relation between M$_{V}$ \citep[from Harris Catalogue;][]{Harris2010} and the spread reported (represented by the IQR) for Y, Ba, La, and Eu can be seen in Fig. \ref{fig:YBaLaEu_MVt}. All the mentioned IQRs display a quite flat distribution with a quite constant spread along the M$_{V}$, meaning there is no evidence of any trends with cluster mass neither in s-process species nor Eu abundances. Hence, we find no evidence that cluster mass does play a role in retaining n-capture-enriched material.

\section{Discussion}

In general terms, insofar as heavy elements are concerned, the GCs in our sample behave similarly to field stars at the same metallicity. Nevertheless, some present peculiarities, such as significant spreads or correlations between elements. Those cases are briefly discussed below.

\textit{NGC~7078:} It displays considerable n-capture element dispersion, explained by peculiar chemical enrichment from an r-process. Moreover, we found highly significant correlations between Ba and O and Na, being negative and positive, respectively. These correlations suggest that the nucleosynthetic sites destroying (producing) O (Na) were also contributing to the Ba production.

\textit{NGC~1904:} Although it has shown quite similar behaviour to field stars at its metallicity, NGC~1904 displays a significant correlation between both Y and Ba with Mg. The dispersion in all these elements is modest, however, the nucleosynthetic site responsible for the little Mg destruction would also deplete a small amount of s-process elements.

\textit{NGC~2808:} It shows a quite constant n-capture distribution, however, correlating with Na. In addition, it presents a negative correlation --highly significant-- correlation between both Y and Ba with Mg. According to \citetalias{Carretta2009u}, NGC~2808 presents a large dispersion in Mg (up to 0.7 dex), which seems to be bimodal. Because the Mg-poor group has higher abundances of Y and Ba, the responsible for the Mg production in NGC~2808 should be able to destroy s-process species. The fact that a positive correlation between Y and Na, coupled with a negative correlation between Y and Mg, could reflect a real effect in NGC~2808. 

\textit{NGC~6121:} It shows slightly overall higher s-process abundances with respect to the field stars at its metallicity and presents a large dispersion within their members. Interestingly, it displays significant negative correlations between Y-Mg and Ba-O. The spread reported by \citetalias{Carretta2009u} for O and Mg is modest. Then, the nucleosynthesis site of s-process elements either differs from the one destroying O and Mg or is destroyed in the process.

\section{Summary and Conclusions}

We then analysed 210 UVES spectra of RGB stars belonging to 18 GCs with a large range of metallicities. The sample previously studied by \citealt{Carretta2009u} and \citet{Carretta2017} mainly focused on determining hot H-burning elements. For homogeneity, the present article used the same stellar parameters as in the mentioned ones to extend the analysis to Cu, Y, Ba, La, and Eu, aiming to study the overall behaviour of n-capture elements in GCs. We aimed to analyse the potential trend in producing the enriched hot H-burning and s-process elements. Y, Ba, La, and Eu abundances are generally quite constant within all the GCs in our sample. 

Heavy elements in GCs display the same distribution as field stars, meaning that GCs have the same chemical enrichment and do not show considerable spread in the elements considered. A special case was found for NGC~7078, which displays the largest spread in heavy elements. The latter is in good agreement with the literature and has been attributed to an initial spread in r-process enrichment. The distribution with respect to the field, two GCs (NGC~6121 and NGC~6171), had a Y and Ba abundance over the field star patterns. A further examination revealed that the spread in their Y and Ba abundances is at least partially due to the \vm. However, a line-to-line comparison of stars with similar stellar parameters revealed a real spread in the abundances reported in both clusters. In the same fashion as field stars, the [Ba/Eu] ratio in GCs shows a continuous s-process enrichment over time, revealing that at the beginning (low metallicities), both field stars and GCs were mainly enriched by r-process sources, while at higher metallicities, the contribution of s-process sources (like AGB of different masses) becomes more important. In addition, we analysed the Y and Ba abundances along with the Na abundances for the whole sample to study their overall behaviour in GCs. To do so, the sample was divided into three metallicity bins. In the intermediate-metallicity regime (-1.10 dex < [Fe/H] < -1.80 dex), there is a mildly significant correlation between Y and Na. Although the trend is low, it could imply a modest production of light s-process elements with the same nucleosynthetic site where Na is produced. On the other hand, the halo chemical evolution model derived by \citet{Cescutti2014} --which considers s- and r-process enrichment from FRMS and MRD, respectively-- reproduces closely the observations reported for the whole sample of this article. The only exceptions are NGC~6121 and NGC~6171, which display higher abundances in Y and Ba.

We compared the n-capture element abundances of GCs as a function of their origin according to the classification given by \citet{Massari2019}. We did not find significant differences between in-situ and ex-situ ones in the n-capture elements analysed. Therefore, no strong evidence exists of a different chemical evolution among these groups.

\begin{acknowledgements}
J.S-U and his work were supported by the National Agency for Research and Development (ANID)/Programa de Becas de Doctorado en el extranjero/DOCTORADO BECASCHILE/2019-72200126. This work was partially funded by the PRIN INAF 2019 grant ObFu 1.05.01.85.14 (\emph{'Building up the halo: chemo-dynamical tagging in the age of large surveys'}, PI. S. Lucatello. G.C. acknowledges the supported by the European Union (ChETEC-INFRA, project no. 101008324).)      
\end{acknowledgements}

%
%

\bibliographystyle{aa}
\bibliography{bibliography}

\begin{thebibliography}{80}
\expandafter\ifx\csname natexlab\endcsname\relax\def\natexlab#1{#1}\fi

\bibitem[{{Alonso} {et~al.}(1999){Alonso}, {Arribas}, \& {Mart{\'\i}nez-Roger}}]{Alonso1999}
{Alonso}, A., {Arribas}, S., \& {Mart{\'\i}nez-Roger}, C. 1999, \aaps, 140, 261

\bibitem[{{Arakelyan} {et~al.}(2020){Arakelyan}, {Pilipenko}, \& {Sharina}}]{Arakelyan2020}
{Arakelyan}, N.~R., {Pilipenko}, S.~V., \& {Sharina}, M.~E. 2020, Astrophysical Bulletin, 75, 394

\bibitem[{{Arcones} {et~al.}(2007){Arcones}, {Janka}, \& {Scheck}}]{Arcones2007}
{Arcones}, A., {Janka}, H.~T., \& {Scheck}, L. 2007, \aap, 467, 1227

\bibitem[{{Asplund} {et~al.}(2009){Asplund}, {Grevesse}, {Sauval}, \& {Scott}}]{Asplund_isoratios}
{Asplund}, M., {Grevesse}, N., {Sauval}, A.~J., \& {Scott}, P. 2009, \araa, 47, 481

\bibitem[{{Bastian} \& {Lardo}(2018)}]{Bastian2018}
{Bastian}, N. \& {Lardo}, C. 2018, \araa, 56, 83

\bibitem[{{Busso} {et~al.}(2001){Busso}, {Gallino}, {Lambert}, {Travaglio}, \& {Smith}}]{Busso2001}
{Busso}, M., {Gallino}, R., {Lambert}, D.~L., {Travaglio}, C., \& {Smith}, V.~V. 2001, \apj, 557, 802

\bibitem[{{Carretta} \& {Bragaglia}(2022)}]{Carretta2022}
{Carretta}, E. \& {Bragaglia}, A. 2022, \aap, 660, L1

\bibitem[{{Carretta} {et~al.}(2009){Carretta}, {Bragaglia}, {Gratton}, \& {Lucatello}}]{Carretta2009u}
{Carretta}, E., {Bragaglia}, A., {Gratton}, R., \& {Lucatello}, S. 2009, \aap, 505, 139

\bibitem[{{Carretta} {et~al.}(2014{\natexlab{a}}){Carretta}, {Bragaglia}, {Gratton}, {D'Orazi}, {Lucatello}, {Momany}, {Sollima}, {Bellazzini}, {Catanzaro}, \& {Leone}}]{ngc4833}
{Carretta}, E., {Bragaglia}, A., {Gratton}, R.~G., {et~al.} 2014{\natexlab{a}}, \aap, 564, A60

\bibitem[{{Carretta} {et~al.}(2014{\natexlab{b}}){Carretta}, {Bragaglia}, {Gratton}, {D'Orazi}, {Lucatello}, \& {Sollima}}]{Terzan8}
{Carretta}, E., {Bragaglia}, A., {Gratton}, R.~G., {et~al.} 2014{\natexlab{b}}, \aap, 561, A87

\bibitem[{{Carretta} {et~al.}(2015){Carretta}, {Bragaglia}, {Gratton}, {D'Orazi}, {Lucatello}, {Sollima}, {Momany}, {Catanzaro}, \& {Leone}}]{ngc6093}
{Carretta}, E., {Bragaglia}, A., {Gratton}, R.~G., {et~al.} 2015, \aap, 578, A116

\bibitem[{{Carretta} {et~al.}(2013){Carretta}, {Bragaglia}, {Gratton}, {Lucatello}, {D'Orazi}, {Bellazzini}, {Catanzaro}, {Leone}, {Momany}, \& {Sollima}}]{ngc362}
{Carretta}, E., {Bragaglia}, A., {Gratton}, R.~G., {et~al.} 2013, \aap, 557, A138

\bibitem[{{Carretta} {et~al.}(2017){Carretta}, {Bragaglia}, {Lucatello}, {D'Orazi}, {Gratton}, {Donati}, {Sollima}, \& {Sneden}}]{Carretta2017}
{Carretta}, E., {Bragaglia}, A., {Lucatello}, S., {et~al.} 2017, \aap, 600, A118

\bibitem[{{Carretta} {et~al.}(2011){Carretta}, {Lucatello}, {Gratton}, {Bragaglia}, \& {D'Orazi}}]{ngc1851}
{Carretta}, E., {Lucatello}, S., {Gratton}, R.~G., {Bragaglia}, A., \& {D'Orazi}, V. 2011, \aap, 533, A69

\bibitem[{{Cavallo} {et~al.}(2023){Cavallo}, {Cescutti}, \& {Matteucci}}]{Cavallo2023}
{Cavallo}, L., {Cescutti}, G., \& {Matteucci}, F. 2023, \aap, 674, A130

\bibitem[{Cayrel {et~al.}(1988)Cayrel, Burkhart, Van’t~Veer, Michaud, \& Tutukov}]{cayrel1988proc}
Cayrel, R., Burkhart, C., Van’t~Veer, C., Michaud, G., \& Tutukov, A. 1988

\bibitem[{{Cescutti} \& {Chiappini}(2014)}]{Cescutti2014}
{Cescutti}, G. \& {Chiappini}, C. 2014, \aap, 565, A51

\bibitem[{{Cescutti} {et~al.}(2006){Cescutti}, {Fran{\c{c}}ois}, {Matteucci}, {Cayrel}, \& {Spite}}]{Cescutti2006}
{Cescutti}, G., {Fran{\c{c}}ois}, P., {Matteucci}, F., {Cayrel}, R., \& {Spite}, M. 2006, \aap, 448, 557

\bibitem[{{Cescutti} \& {Matteucci}(2022)}]{Cescutti2022}
{Cescutti}, G. \& {Matteucci}, F. 2022, Universe, 8, 173

\bibitem[{{Cohen}(2011)}]{Cohen2011}
{Cohen}, J.~G. 2011, \apjl, 740, L38

\bibitem[{Cowan {et~al.}(1991)Cowan, Thielemann, \& Truran}]{Cowan1991}
Cowan, J.~J., Thielemann, F.-K., \& Truran, J.~W. 1991, Physics Reports, 208, 267

\bibitem[{{Cseh} {et~al.}(2018){Cseh}, {Lugaro}, {D'Orazi}, {de Castro}, {Pereira}, {Karakas}, {Moln{\'a}r}, {Plachy}, {Szab{\'o}}, {Pignatari}, \& {Cristallo}}]{Cseh2018}
{Cseh}, B., {Lugaro}, M., {D'Orazi}, V., {et~al.} 2018, \aap, 620, A146

\bibitem[{{Cunha} {et~al.}(2002){Cunha}, {Smith}, {Suntzeff}, {Norris}, {Da Costa}, \& {Plez}}]{Cunha2002}
{Cunha}, K., {Smith}, V.~V., {Suntzeff}, N.~B., {et~al.} 2002, \aj, 124, 379

\bibitem[{{de Mink} {et~al.}(2009){de Mink}, {Pols}, {Langer}, \& {Izzard}}]{DeMink2009}
{de Mink}, S.~E., {Pols}, O.~R., {Langer}, N., \& {Izzard}, R.~G. 2009, \aap, 507, L1

\bibitem[{{Decressin} {et~al.}(2007){Decressin}, {Charbonnel}, \& {Meynet}}]{Decressin2007}
{Decressin}, T., {Charbonnel}, C., \& {Meynet}, G. 2007, \aap, 475, 859

\bibitem[{{D'Orazi} {et~al.}(2013){D'Orazi}, {Campbell}, {Lugaro}, {Lattanzio}, {Pignatari}, \& {Carretta}}]{Dorazi2013}
{D'Orazi}, V., {Campbell}, S.~W., {Lugaro}, M., {et~al.} 2013, \mnras, 433, 366

\bibitem[{{D'Orazi} {et~al.}(2010){D'Orazi}, {Gratton}, {Lucatello}, {Carretta}, {Bragaglia}, \& {Marino}}]{dorazi2010_neutron}
{D'Orazi}, V., {Gratton}, R., {Lucatello}, S., {et~al.} 2010, \apjl, 719, L213

\bibitem[{{Fern{\'a}ndez-Alvar} {et~al.}(2018){Fern{\'a}ndez-Alvar}, {Carigi}, {Schuster}, {Hayes}, {{\'A}vila-Vergara}, {Majewski}, {Allende Prieto}, {Beers}, {S{\'a}nchez}, {Zamora}, {Garc{\'\i}a-Hern{\'a}ndez}, {Tang}, {Fern{\'a}ndez-Trincado}, {Tissera}, {Geisler}, \& {Villanova}}]{Fernandez-Alvar2018}
{Fern{\'a}ndez-Alvar}, E., {Carigi}, L., {Schuster}, W.~J., {et~al.} 2018, \apj, 852, 50

\bibitem[{{Freeman} \& {Bland-Hawthorn}(2002)}]{Freeman-Bland2002}
{Freeman}, K. \& {Bland-Hawthorn}, J. 2002, \araa, 40, 487

\bibitem[{{Frischknecht} {et~al.}(2016){Frischknecht}, {Hirschi}, {Pignatari}, {Maeder}, {Meynet}, {Chiappini}, {Thielemann}, {Rauscher}, {Georgy}, \& {Ekstr{\"o}m}}]{Frischknecht2016}
{Frischknecht}, U., {Hirschi}, R., {Pignatari}, M., {et~al.} 2016, \mnras, 456, 1803

\bibitem[{{Gaia Collaboration} {et~al.}(2022){Gaia Collaboration}, {Vallenari}, {Brown}, {Prusti}, {de Bruijne}, {Arenou}, {Babusiaux}, {Biermann}, {Creevey}, {Ducourant}, {Evans}, {Eyer}, {Guerra}, {Hutton}, {Jordi}, {Klioner}, {Lammers}, {Lindegren}, {Luri}, {Mignard}, {Panem}, {Pourbaix}, {Randich}, {Sartoretti}, {Soubiran}, {Tanga}, {Walton}, {Bailer-Jones}, {Bastian}, {Drimmel}, {Jansen}, {Katz}, {Lattanzi}, {van Leeuwen}, {Bakker}, {Cacciari}, {Casta{\~n}eda}, {De Angeli}, {Fabricius}, {Fouesneau}, {Fr{\'e}mat}, {Galluccio}, {Guerrier}, {Heiter}, {Masana}, {Messineo}, {Mowlavi}, {Nicolas}, {Nienartowicz}, {Pailler}, {Panuzzo}, {Riclet}, {Roux}, {Seabroke}, {Sordo{\o}rcit}, {Th{\'e}venin}, {Gracia-Abril}, {Portell}, {Teyssier}, {Altmann}, {Andrae}, {Audard}, {Bellas-Velidis}, {Benson}, {Berthier}, {Blomme}, {Burgess}, {Busonero}, {Busso}, {C{\'a}novas}, {Carry}, {Cellino}, {Cheek}, {Clementini}, {Damerdji}, {Davidson}, {de Teodoro}, {Nu{\~n}ez Campos}, {Delchambre}, {Dell'Oro}, {Esquej},
  {Fern{\'a}ndez-Hern{\'a}ndez}, {Fraile}, {Garabato}, {Garc{\'\i}a-Lario}, {Gosset}, {Haigron}, {Halbwachs}, {Hambly}, {Harrison}, {Hern{\'a}ndez}, {Hestroffer}, {Hodgkin}, {Holl}, {Jan{\ss}en}, {Jevardat de Fombelle}, {Jordan}, {Krone-Martins}, {Lanzafame}, {L{\"o}ffler}, {Marchal}, {Marrese}, {Moitinho}, {Muinonen}, {Osborne}, {Pancino}, {Pauwels}, {Recio-Blanco}, {Reyl{\'e}}, {Riello}, {Rimoldini}, {Roegiers}, {Rybizki}, {Sarro}, {Siopis}, {Smith}, {Sozzetti}, {Utrilla}, {van Leeuwen}, {Abbas}, {{\'A}brah{\'a}m}, {Abreu Aramburu}, {Aerts}, {Aguado}, {Ajaj}, {Aldea-Montero}, {Altavilla}, {{\'A}lvarez}, {Alves}, {Anders}, {Anderson}, {Anglada Varela}, {Antoja}, {Baines}, {Baker}, {Balaguer-N{\'u}{\~n}ez}, {Balbinot}, {Balog}, {Barache}, {Barbato}, {Barros}, {Barstow}, {Bartolom{\'e}}, {Bassilana}, {Bauchet}, {Becciani}, {Bellazzini}, {Berihuete}, {Bernet}, {Bertone}, {Bianchi}, {Binnenfeld}, {Blanco-Cuaresma}, {Blazere}, {Boch}, {Bombrun}, {Bossini}, {Bouquillon}, {Bragaglia}, {Bramante}, {Breedt},
  {Bressan}, {Brouillet}, {Brugaletta}, {Bucciarelli}, {Burlacu}, {Butkevich}, {Buzzi}, {Caffau}, {Cancelliere}, {Cantat-Gaudin}, {Carballo}, {Carlucci}, {Carnerero}, {Carrasco}, {Casamiquela}, {Castellani}, {Castro-Ginard}, {Chaoul}, {Charlot}, {Chemin}, {Chiaramida}, {Chiavassa}, {Chornay}, {Comoretto}, {Contursi}, {Cooper}, {Cornez}, {Cowell}, {Crifo}, {Cropper}, {Crosta}, {Crowley}, {Dafonte}, {Dapergolas}, {David}, {David}, {de Laverny}, {De Luise}, {De March}, {De Ridder}, {de Souza}, {de Torres}, {del Peloso}, {del Pozo}, {Delbo}, {Delgado}, {Delisle}, {Demouchy}, {Dharmawardena}, {Di Matteo}, {Diakite}, {Diener}, {Distefano}, {Dolding}, {Edvardsson}, {Enke}, {Fabre}, {Fabrizio}, {Faigler}, {Fedorets}, {Fernique}, {Fienga}, {Figueras}, {Fournier}, {Fouron}, {Fragkoudi}, {Gai}, {Garcia-Gutierrez}, {Garcia-Reinaldos}, {Garc{\'\i}a-Torres}, {Garofalo}, {Gavel}, {Gavras}, {Gerlach}, {Geyer}, {Giacobbe}, {Gilmore}, {Girona}, {Giuffrida}, {Gomel}, {Gomez}, {Gonz{\'a}lez-N{\'u}{\~n}ez},
  {Gonz{\'a}lez-Santamar{\'\i}a}, {Gonz{\'a}lez-Vidal}, {Granvik}, {Guillout}, {Guiraud}, {Guti{\'e}rrez-S{\'a}nchez}, {Guy}, {Hatzidimitriou}, {Hauser}, {Haywood}, {Helmer}, {Helmi}, {Sarmiento}, {Hidalgo}, {Hilger}, {H{\l}adczuk}, {Hobbs}, {Holland}, {Huckle}, {Jardine}, {Jasniewicz}, {Jean-Antoine Piccolo}, {Jim{\'e}nez-Arranz}, {Jorissen}, {Juaristi Campillo}, {Julbe}, {Karbevska}, {Kervella}, {Khanna}, {Kontizas}, {Kordopatis}, {Korn}, {K{\'o}sp{\'a}l}, {Kostrzewa-Rutkowska}, {Kruszy{\'n}ska}, {Kun}, {Laizeau}, {Lambert}, {Lanza}, {Lasne}, {Le Campion}, {Lebreton}, {Lebzelter}, {Leccia}, {Leclerc}, {Lecoeur-Taibi}, {Liao}, {Licata}, {Lindstr{\o}m}, {Lister}, {Livanou}, {Lobel}, {Lorca}, {Loup}, {Madrero Pardo}, {Magdaleno Romeo}, {Managau}, {Mann}, {Manteiga}, {Marchant}, {Marconi}, {Marcos}, {Marcos Santos}, {Mar{\'\i}n Pina}, {Marinoni}, {Marocco}, {Marshall}, {Polo}, {Mart{\'\i}n-Fleitas}, {Marton}, {Mary}, {Masip}, {Massari}, {Mastrobuono-Battisti}, {Mazeh}, {McMillan}, {Messina}, {Michalik},
  {Millar}, {Mints}, {Molina}, {Molinaro}, {Moln{\'a}r}, {Monari}, {Mongui{\'o}}, {Montegriffo}, {Montero}, {Mor}, {Mora}, {Morbidelli}, {Morel}, {Morris}, {Muraveva}, {Murphy}, {Musella}, {Nagy}, {Noval}, {Oca{\~n}a}, {Ogden}, {Ordenovic}, {Osinde}, {Pagani}, {Pagano}, {Palaversa}, {Palicio}, {Pallas-Quintela}, {Panahi}, {Payne-Wardenaar}, {Pe{\~n}alosa Esteller}, {Penttil{\"a}}, {Pichon}, {Piersimoni}, {Pineau}, {Plachy}, {Plum}, {Poggio}, {Pr{\v{s}}a}, {Pulone}, {Racero}, {Ragaini}, {Rainer}, {Raiteri}, {Rambaux}, {Ramos}, {Ramos-Lerate}, {Re Fiorentin}, {Regibo}, {Richards}, {Rios Diaz}, {Ripepi}, {Riva}, {Rix}, {Rixon}, {Robichon}, {Robin}, {Robin}, {Roelens}, {Rogues}, {Rohrbasser}, {Romero-G{\'o}mez}, {Rowell}, {Royer}, {Ruz Mieres}, {Rybicki}, {Sadowski}, {S{\'a}ez N{\'u}{\~n}ez}, {Sagrist{\`a} Sell{\'e}s}, {Sahlmann}, {Salguero}, {Samaras}, {Sanchez Gimenez}, {Sanna}, {Santove{\~n}a}, {Sarasso}, {Schultheis}, {Sciacca}, {Segol}, {Segovia}, {S{\'e}gransan}, {Semeux}, {Shahaf}, {Siddiqui}, {Siebert},
  {Siltala}, {Silvelo}, {Slezak}, {Slezak}, {Smart}, {Snaith}, {Solano}, {Solitro}, {Souami}, {Souchay}, {Spagna}, {Spina}, {Spoto}, {Steele}, {Steidelm{\"u}ller}, {Stephenson}, {S{\"u}veges}, {Surdej}, {Szabados}, {Szegedi-Elek}, {Taris}, {Taylo}, {Teixeira}, {Tolomei}, {Tonello}, {Torra}, {Torra}, {Torralba Elipe}, {Trabucchi}, {Tsounis}, {Turon}, {Ulla}, {Unger}, {Vaillant}, {van Dillen}, {van Reeven}, {Vanel}, {Vecchiato}, {Viala}, {Vicente}, {Voutsinas}, {Weiler}, {Wevers}, {Wyrzykowski}, {Yoldas}, {Yvard}, {Zhao}, {Zorec}, {Zucker}, \& {Zwitter}}]{GaiaDR3_2022}
{Gaia Collaboration}, {Vallenari}, A., {Brown}, A.~G.~A., {et~al.} 2022, arXiv e-prints, arXiv:2208.00211

\bibitem[{{Gallagher}(1967)}]{Gallagher1967}
{Gallagher}, A. 1967, Physical Review, 157, 24

\bibitem[{{Geisler} {et~al.}(2007){Geisler}, {Wallerstein}, {Smith}, \& {Casetti-Dinescu}}]{Geisler2007}
{Geisler}, D., {Wallerstein}, G., {Smith}, V.~V., \& {Casetti-Dinescu}, D.~I. 2007, \pasp, 119, 939

\bibitem[{{Gratton} {et~al.}(2019){Gratton}, {Bragaglia}, {Carretta}, {D'Orazi}, {Lucatello}, \& {Sollima}}]{Gratton2019}
{Gratton}, R., {Bragaglia}, A., {Carretta}, E., {et~al.} 2019, \aapr, 27, 8

\bibitem[{Gratton {et~al.}(2004)Gratton, Sneden, \& Carretta}]{Gratton2004}
Gratton, R., Sneden, C., \& Carretta, E. 2004, Annual Review of Astronomy and Astrophysics, 42, 385

\bibitem[{{Gratton} {et~al.}(2012){Gratton}, {Carretta}, \& {Bragaglia}}]{Gratton2012}
{Gratton}, R.~G., {Carretta}, E., \& {Bragaglia}, A. 2012, \aapr, 20, 50

\bibitem[{{Guiglion} {et~al.}(2023){Guiglion}, {Bergemann}, {Storm}, {Lian}, {Cescutti}, \& {Serenelli}}]{Guiglion2023}
{Guiglion}, G., {Bergemann}, M., {Storm}, N., {et~al.} 2023, arXiv e-prints, arXiv:2311.05459

\bibitem[{{Harris}(2010)}]{Harris2010}
{Harris}, W.~E. 2010, arXiv e-prints, arXiv:1012.3224

\bibitem[{{H{\"o}hle} {et~al.}(1982){H{\"o}hle}, {H{\"u}hnermann}, \& {Wagner}}]{Hohle1982}
{H{\"o}hle}, C., {H{\"u}hnermann}, H., \& {Wagner}, H. 1982, Zeitschrift fur Physik A Hadrons and Nuclei, 304, 279

\bibitem[{{Horta} {et~al.}(2020){Horta}, {Schiavon}, {Mackereth}, {Beers}, {Fern{\'a}ndez-Trincado}, {Frinchaboy}, {Garc{\'\i}a-Hern{\'a}ndez}, {Geisler}, {Hasselquist}, {J{\"o}nsson}, {Lane}, {Majewski}, {M{\'e}sz{\'a}ros}, {Bidin}, {Nataf}, {Roman-Lopes}, {Nitschelm}, {Vargas-Gonz{\'a}lez}, \& {Zasowski}}]{Horta2020}
{Horta}, D., {Schiavon}, R.~P., {Mackereth}, J.~T., {et~al.} 2020, \mnras, 493, 3363

\bibitem[{{Ishigaki} {et~al.}(2013){Ishigaki}, {Aoki}, \& {Chiba}}]{Ishigaki2013}
{Ishigaki}, M.~N., {Aoki}, W., \& {Chiba}, M. 2013, \apj, 771, 67

\bibitem[{{James} {et~al.}(2004){James}, {Fran{\c{c}}ois}, {Bonifacio}, {Carretta}, {Gratton}, \& {Spite}}]{James2004}
{James}, G., {Fran{\c{c}}ois}, P., {Bonifacio}, P., {et~al.} 2004, \aap, 427, 825

\bibitem[{{Kasen} {et~al.}(2017){Kasen}, {Metzger}, {Barnes}, {Quataert}, \& {Ramirez-Ruiz}}]{Kasen2017}
{Kasen}, D., {Metzger}, B., {Barnes}, J., {Quataert}, E., \& {Ramirez-Ruiz}, E. 2017, \nat, 551, 80

\bibitem[{{Kirby} {et~al.}(2020){Kirby}, {Duggan}, {Ramirez-Ruiz}, \& {Macias}}]{Kirby2020}
{Kirby}, E.~N., {Duggan}, G., {Ramirez-Ruiz}, E., \& {Macias}, P. 2020, \apjl, 891, L13

\bibitem[{{Kraft}(1994)}]{Kraft1994}
{Kraft}, R.~P. 1994, \pasp, 106, 553

\bibitem[{{Kurucz}(1993)}]{Kurucz1993}
{Kurucz}, R. 1993, ATLAS9 Stellar Atmosphere Programs and 2 km/s grid. Kurucz CD-ROM No. 13. Cambridge, 13

\bibitem[{{Kurucz} \& {Bell}(1995)}]{Kurucz1995}
{Kurucz}, R.~L. \& {Bell}, B. 1995, {Atomic line list} (Cambridge: SAO)

\bibitem[{{Lawler} {et~al.}(2001){Lawler}, {Wickliffe}, {den Hartog}, \& {Sneden}}]{Lawler2001}
{Lawler}, J.~E., {Wickliffe}, M.~E., {den Hartog}, E.~A., \& {Sneden}, C. 2001, \apj, 563, 1075

\bibitem[{{Lee} {et~al.}(2019){Lee}, {Kim}, {Johnson}, {Chung}, {Jang}, {Lim}, \& {Kang}}]{Lee2019}
{Lee}, Y.-W., {Kim}, J.~J., {Johnson}, C.~I., {et~al.} 2019, \apjl, 878, L2

\bibitem[{{Limongi} \& {Chieffi}(2018)}]{Limongi2018}
{Limongi}, M. \& {Chieffi}, A. 2018, \apjs, 237, 13

\bibitem[{{Martell} {et~al.}(2011){Martell}, {Smolinski}, {Beers}, \& {Grebel}}]{Martell2011}
{Martell}, S.~L., {Smolinski}, J.~P., {Beers}, T.~C., \& {Grebel}, E.~K. 2011, \aap, 534, A136

\bibitem[{{Massari} {et~al.}(2019){Massari}, {Koppelman}, \& {Helmi}}]{Massari2019}
{Massari}, D., {Koppelman}, H.~H., \& {Helmi}, A. 2019, \aap, 630, L4

\bibitem[{{Masseron} {et~al.}(2019){Masseron}, {Garc{\'\i}a-Hern{\'a}ndez}, {M{\'e}sz{\'a}ros}, {Zamora}, {Dell'Agli}, {Allende Prieto}, {Edvardsson}, {Shetrone}, {Plez}, {Fern{\'a}ndez-Trincado}, {Cunha}, {J{\"o}nsson}, {Geisler}, {Beers}, \& {Cohen}}]{Masseron2019}
{Masseron}, T., {Garc{\'\i}a-Hern{\'a}ndez}, D.~A., {M{\'e}sz{\'a}ros}, S., {et~al.} 2019, \aap, 622, A191

\bibitem[{{Mucciarelli}(2011)}]{Mucciarelli2011}
{Mucciarelli}, A. 2011, \aap, 528, A44

\bibitem[{{Nishimura} {et~al.}(2015){Nishimura}, {Takiwaki}, \& {Thielemann}}]{Nishimura2015}
{Nishimura}, N., {Takiwaki}, T., \& {Thielemann}, F.-K. 2015, \apj, 810, 109

\bibitem[{{O'Connell} {et~al.}(2011){O'Connell}, {Johnson}, {Pilachowski}, \& {Burks}}]{OConnell2011}
{O'Connell}, J.~E., {Johnson}, C.~I., {Pilachowski}, C.~A., \& {Burks}, G. 2011, \pasp, 123, 1139

\bibitem[{{Otsuki} {et~al.}(2006){Otsuki}, {Honda}, {Aoki}, {Kajino}, \& {Mathews}}]{Otsuki2006}
{Otsuki}, K., {Honda}, S., {Aoki}, W., {Kajino}, T., \& {Mathews}, G.~J. 2006, \apjl, 641, L117

\bibitem[{{Pancino} {et~al.}(2017){Pancino}, {Romano}, {Tang}, {Tautvai{\v{s}}ien{\.{e}}}, {Casey}, {Gruyters}, {Geisler}, {San Roman}, {Randich}, {Alfaro}, {Bragaglia}, {Flaccomio}, {Korn}, {Recio-Blanco}, {Smiljanic}, {Carraro}, {Bayo}, {Costado}, {Damiani}, {Jofr{\'e}}, {Lardo}, {de Laverny}, {Monaco}, {Morbidelli}, {Sbordone}, {Sousa}, \& {Villanova}}]{Pancino2017}
{Pancino}, E., {Romano}, D., {Tang}, B., {et~al.} 2017, \aap, 601, A112

\bibitem[{{Perego} {et~al.}(2021){Perego}, {Thielemann}, \& {Cescutti}}]{Perego2021}
{Perego}, A., {Thielemann}, F.~K., \& {Cescutti}, G. 2021, in Handbook of Gravitational Wave Astronomy, 13

\bibitem[{{Pignatari} {et~al.}(2010){Pignatari}, {Gallino}, {Heil}, {Wiescher}, {K{\"a}ppeler}, {Herwig}, \& {Bisterzo}}]{Pignatari2010}
{Pignatari}, M., {Gallino}, R., {Heil}, M., {et~al.} 2010, \apj, 710, 1557

\bibitem[{{Placco} {et~al.}(2021){Placco}, {Sneden}, {Roederer}, {Lawler}, {Den Hartog}, {Hejazi}, {Maas}, \& {Bernath}}]{Placco2021}
{Placco}, V.~M., {Sneden}, C., {Roederer}, I.~U., {et~al.} 2021, Research Notes of the American Astronomical Society, 5, 92

\bibitem[{Prantzos {et~al.}(2020)Prantzos, Abia, Cristallo, Limongi, \& Chieffi}]{Prantzos2020}
Prantzos, N., Abia, C., Cristallo, S., Limongi, M., \& Chieffi, A. 2020, Monthly Notices of the Royal Astronomical Society, 491, 1832

\bibitem[{{Recio-Blanco}(2018)}]{RecioBlanco2018}
{Recio-Blanco}, A. 2018, \aap, 620, A194

\bibitem[{{Schiappacasse-Ulloa} \& {Lucatello}(2023)}]{Schiappacasse-Ulloa2023}
{Schiappacasse-Ulloa}, J. \& {Lucatello}, S. 2023, \mnras, 520, 5938

\bibitem[{{Siegel} {et~al.}(2019){Siegel}, {Barnes}, \& {Metzger}}]{Siegel2019}
{Siegel}, D.~M., {Barnes}, J., \& {Metzger}, B.~D. 2019, \nat, 569, 241

\bibitem[{{Simmerer} {et~al.}(2004){Simmerer}, {Sneden}, {Cowan}, {Collier}, {Woolf}, \& {Lawler}}]{Simmerer2004}
{Simmerer}, J., {Sneden}, C., {Cowan}, J.~J., {et~al.} 2004, \apj, 617, 1091

\bibitem[{{Simmerer} {et~al.}(2003){Simmerer}, {Sneden}, {Ivans}, {Kraft}, {Shetrone}, \& {Smith}}]{Simmerer2003}
{Simmerer}, J., {Sneden}, C., {Ivans}, I.~I., {et~al.} 2003, \aj, 125, 2018

\bibitem[{{Skrutskie} {et~al.}(2006){Skrutskie}, {Cutri}, {Stiening}, {Weinberg}, {Schneider}, {Carpenter}, {Beichman}, {Capps}, {Chester}, {Elias}, {Huchra}, {Liebert}, {Lonsdale}, {Monet}, {Price}, {Seitzer}, {Jarrett}, {Kirkpatrick}, {Gizis}, {Howard}, {Evans}, {Fowler}, {Fullmer}, {Hurt}, {Light}, {Kopan}, {Marsh}, {McCallon}, {Tam}, {Van Dyk}, \& {Wheelock}}]{Skrutskie2006}
{Skrutskie}, M.~F., {Cutri}, R.~M., {Stiening}, R., {et~al.} 2006, \aj, 131, 1163

\bibitem[{{Smith}(1987)}]{Smith1987}
{Smith}, G.~H. 1987, \pasp, 99, 67

\bibitem[{{Sneden} {et~al.}(2008){Sneden}, {Cowan}, \& {Gallino}}]{Sneden2008}
{Sneden}, C., {Cowan}, J.~J., \& {Gallino}, R. 2008, \araa, 46, 241

\bibitem[{{Sobeck} {et~al.}(2011){Sobeck}, {Kraft}, {Sneden}, {Preston}, {Cowan}, {Smith}, {Thompson}, {Shectman}, \& {Burley}}]{Sobeck2011}
{Sobeck}, J.~S., {Kraft}, R.~P., {Sneden}, C., {et~al.} 2011, \aj, 141, 175

\bibitem[{Spearman(1904)}]{Spearman}
Spearman, C. 1904, The American Journal of Psychology, 15, 72

\bibitem[{{Storm} \& {Bergemann}(2023)}]{Storm2023}
{Storm}, N. \& {Bergemann}, M. 2023, \mnras, 525, 3718

\bibitem[{{Suda} {et~al.}(2008){Suda}, {Katsuta}, {Yamada}, {Suwa}, {Ishizuka}, {Komiya}, {Sorai}, {Aikawa}, \& {Fujimoto}}]{Suda2008}
{Suda}, T., {Katsuta}, Y., {Yamada}, S., {et~al.} 2008, \pasj, 60, 1159

\bibitem[{{VandenBerg} {et~al.}(2013){VandenBerg}, {Brogaard}, {Leaman}, \& {Casagrande}}]{VandenBerg2013}
{VandenBerg}, D.~A., {Brogaard}, K., {Leaman}, R., \& {Casagrande}, L. 2013, \apj, 775, 134

\bibitem[{{Ventura} {et~al.}(2001){Ventura}, {D'Antona}, {Mazzitelli}, \& {Gratton}}]{Ventura2001}
{Ventura}, P., {D'Antona}, F., {Mazzitelli}, I., \& {Gratton}, R. 2001, \apjl, 550, L65

\bibitem[{{Villanova} \& {Geisler}(2011)}]{Villanova2011}
{Villanova}, S. \& {Geisler}, D. 2011, \aap, 535, A31

\bibitem[{{Wehrhahn}(2021)}]{Wehrhahn2021}
{Wehrhahn}, A. 2021, in The 20.5th Cambridge Workshop on Cool Stars, Stellar Systems, and the Sun (CS20.5), Cambridge Workshop on Cool Stars, Stellar Systems, and the Sun, 1

\bibitem[{{Worley} {et~al.}(2013){Worley}, {Hill}, {Sobeck}, \& {Carretta}}]{Worley2013}
{Worley}, C.~C., {Hill}, V., {Sobeck}, J., \& {Carretta}, E. 2013, \aap, 553, A47

\bibitem[{{Yong} {et~al.}(2008){Yong}, {Karakas}, {Lambert}, {Chieffi}, \& {Limongi}}]{Yong2008b}
{Yong}, D., {Karakas}, A.~I., {Lambert}, D.~L., {Chieffi}, A., \& {Limongi}, M. 2008, \apj, 689, 1031

\end{thebibliography}

\end{document}